\documentclass[twocolumn,english,aps,prb,nofootinbib]{revtex4-1}
\usepackage{CJK}

\usepackage{amsfonts}
\usepackage{amsmath}
\usepackage{ dsfont }
\usepackage{hyperref}
\usepackage{amssymb}
\usepackage{array,multirow}
\usepackage{longtable}
\usepackage{xcolor}
\usepackage{relsize}%\mathlarger command 
\usepackage[bottom]{footmisc}

\usepackage{tikz}
\usetikzlibrary{calc} 
\newcommand{\be}{\begin{equation}}
\newcommand{\ee}{\end{equation}}
\newcommand{\bea}{\begin{equation} \begin{aligned}}
\newcommand{\eea}{\end{aligned} \end{equation} }
\newcommand{\bi}{\begin{itemize}}
\newcommand{\ei}{\end{itemize}}

\newcolumntype{C}[1]{>{\centering\arraybackslash}p{#1}}

\newcommand{\la}{\lambda}
\renewcommand{\be}{\beta}
\newcommand{\al}{\alpha}
\newcommand{\eps}{\epsilon}

\newcommand{\lp}{\left(}
\newcommand{\rp}{\right)}
\newcommand{\bpm}{\begin{pmatrix}}
\newcommand{\epm}{\end{pmatrix}}

\newcommand{\del}{\partial}

\newcommand{\mbf}[1]{\mathbf{#1}}
 
\newcommand{\dff}{\delta_1^{0}}
\newcommand{\dFf}{\delta_1^{\Phi/2}}
\newcommand{\dfs}{\delta_2^{0}}
\newcommand{\dFs}{\delta_2^{\Phi/2}}
\newcommand{\dft}{\delta_3^{0}}
\newcommand{\dFt}{\delta_3^{\Phi/2}}

\usepackage{xspace}
\usepackage{color}
\usepackage{graphicx}
\usepackage[space]{grffile}
\usepackage{verbatim}
\usepackage{amsmath}
\usepackage{amssymb}
\usepackage{mathrsfs}
\usepackage{wasysym}
\usepackage[caption=false]{subfig}
\usepackage{url}
\usepackage{bbold}
\usepackage{slashed}
\usepackage{epstopdf}
\usepackage{braket}
\usepackage{float}
\usepackage[percent]{overpic}

\DeclareRobustCommand{\Sec}[1]{Sec.~\ref{#1}}

\DeclareRobustCommand{\App}[1]{App.~\ref{#1}}

\DeclareRobustCommand{\Tab}[1]{Table~\ref{#1}}

\DeclareRobustCommand{\Fig}[1]{Fig.~\ref{#1}}

\DeclareRobustCommand{\Eq}[1]{Eq.~(\ref{#1})}
\DeclareRobustCommand{\Eqs}[2]{Eqs.~(\ref{#1}) and (\ref{#2})}
\DeclareRobustCommand{\Ref}[1]{Ref.~\cite{#1}}

\newcommand{\boundellipse}[3]% center, xdim, ydim
{[black,fill=blue!30] (#1) ellipse (#2 and #3)
}
\newcommand{\boundellipseW}[3]% center, xdim, ydim
{[white,fill=white] (#1) ellipse (#2 and #3)
}

\begin{document}
\date{\today}
\title{Hofstadter Topology with Real Space Invariants and Reentrant Projective Symmetries
}

\author{Jonah Herzog-Arbeitman$^{1}$}
%\thanks{These authors contributed equally.}
\author{Zhi-Da Song$^{1,2}$}
\author{Luis Elcoro$^{3,4}$}
%\thanks{These authors contributed equally.}
%\author{Nicolas Regnault$^{1,2}$}
\author{B. Andrei Bernevig$^{1,5,6}$}

\affiliation{$^1$Department of Physics, Princeton University, Princeton, NJ 08544}
\affiliation{$^2$International Center for Quantum Materials, School of Physics, Peking University, Beijing 100871, China}
\affiliation{$^3$Department of Physics, University of the Basque Country UPV/EHU, Apdo. 644, 48080 Bilbao, Spain}
\affiliation{$^4$EHU Quantum Center, University of the Basque Country UPV/EHU}
\affiliation{$^5$Donostia International Physics Center, P. Manuel de Lardizabal 4, 20018 Donostia-San Sebastian, Spain}
\affiliation{$^6$IKERBASQUE, Basque Foundation for Science, Bilbao, Spain}
\date{\today}

%2021arXiv210804831S HOf SC

\begin{abstract}
Adding magnetic flux to a band structure breaks Bloch's theorem by realizing a projective representation of the translation group. The resulting Hofstadter spectrum encodes the non-perturbative response of the bands to flux. Depending on their topology, adding flux can enforce a bulk gap closing (a Hofstadter semimetal) or boundary state pumping (a Hofstadter topological insulator). In this work, we present a real-space classification of these Hofstadter phases. We give topological indices in terms of symmetry-protected Real Space Invariants (RSIs) which encode bulk and boundary responses of fragile topological states to flux. In fact, we find that the flux periodicity in tight-binding models causes the symmetries which are broken by the magnetic field to reenter at strong flux where they form projective point group representations. We completely classify the reentrant projective point groups and find that the Schur multipliers which define them are Arahanov-Bohm phases calculated along the bonds of the crystal. We find that a nontrivial Schur multiplier is enough to predict and protect the Hofstadter response with only zero-flux topology. 
\end{abstract}
\maketitle

\emph{Introduction}. The magnetic translation group \cite{PhysRev.134.A1602} is a famous example of how symmetries acquire projective representations in quantum mechanics, dramatically altering the behavior of the system. When magnetic flux is threaded through a two-dimensional crystal, the translation operators $T_i$ along the $i$th lattice vector no longer commute and instead form a \emph{projective} group
\bea
\label{eq:mtg}
T_1 T_2 = e^{i \phi} T_2 T_1, \qquad \phi = \frac{e}{\hbar} \oint \mbf{A} \cdot d\mbf{r},
\eea
where $\mbf{A}$ is the vector potential of the magnetic field, $e$ is the electron charge, and the flux $\phi$ is the Aharanov-Bohm phase difference between the two paths around the unit cell \cite{PhysRev.115.485}. The square lattice \cite{PhysRevB.14.2239} provides a concrete realization of \Eq{eq:mtg}, which results in a fractal energy spectrum known as the Hofstadter Butterfly. In the present work, we find that point group symmetries can also be projective in strong flux. Their representations constrain the Hofstadter butterfly of a generic model in the paradigm of Hofstadter topology \cite{firstpaper}, where magnetic flux acts as a pumping parameter (a third dimension) \cite{PhysRevB.27.6083,2018arXiv181002373W,2022PhRvL.128x6602W}. We classify Hofstadter semimetals (SMs) with protected gap closings and Hofstadter higher order topological insulators (HOTIs) with boundary state pumping in all magnetic point groups. Their topological indices are written in terms of real space invariants (RSIs) \cite{song2019real} calculated with and without flux. Hofstadter physics has recently garnered attention from many directions \cite{2021arXiv211213837P,2022arXiv220111586M,2021NatCo..12.6433H,2022arXiv220405320Z,2022arXiv220405320Z,2022arXiv220411737S,2022arXiv220413116S,2022arXiv220501406F,2022PhyE..14215311A,2022arXiv220508545M,2022arXiv220113062G,2022arXiv220602810Y,2022arXiv220604543M,2022arXiv220703028G,2021arXiv211014570W,2021arXiv210810353G,2021PhRvB.104c5161M,2021PhRvB.104c5305M,PhysRevLett.126.056401,2021PhRvB.103x5107L,2021JPhD...54O4004Z,2021arXiv210710393X,2019PhRvB.100x5108O,2021arXiv210803196A,2021PhRvB.104r4501S,2020PhRvB.102o5102A,2021PhRvB.103s5105T,2021arXiv210912919L,2019arXiv190810976O,2022arXiv220611891B}, especially with the profusion of experiments on moir\'e materials \cite{yu2021correlated,huber2021brownzak,finney2021unusual,polshyn2021topological,Shen_2021,Park_2021,Lu_2020,2020arXiv200614000B,2020arXiv200613963L,2020arXiv200713390D} up to $2\pi$ flux \cite{PhysRevLett.128.217701}. Our results offer a unified, symmetry-based approach\cite{2017Natur.547..298B,2020arXiv201000598E} to the problem while shedding new light on 2D topology.

\begin{figure}[h]
 \centering
\includegraphics[width=6.5cm]{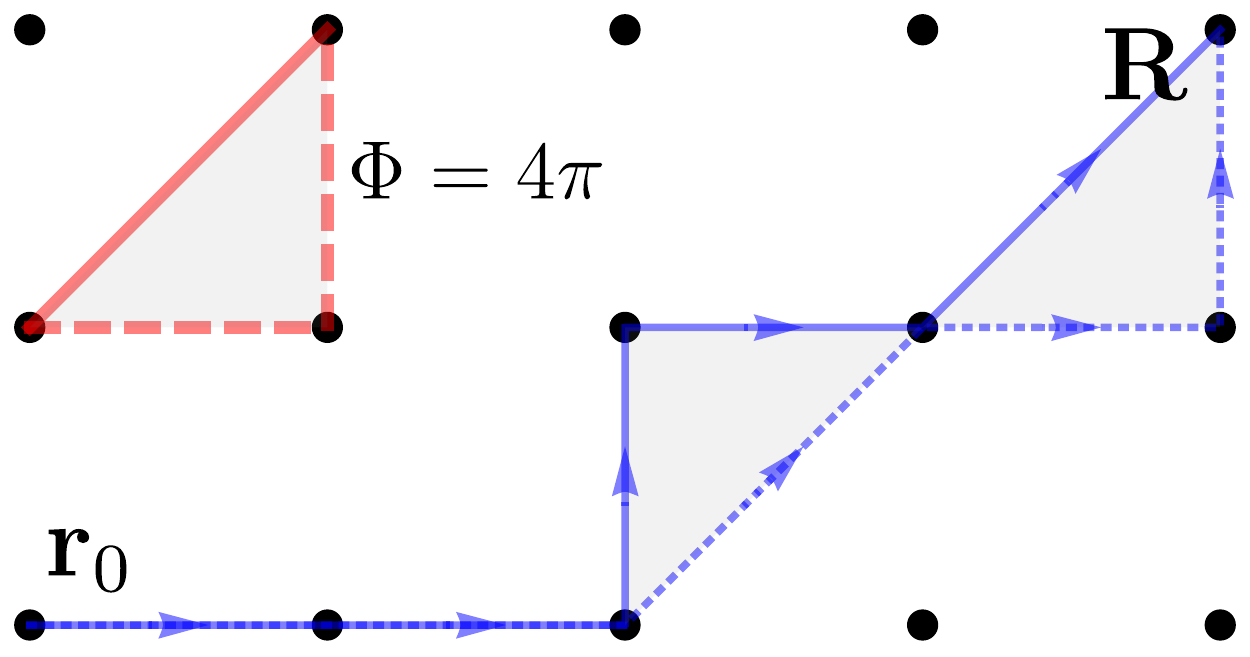} 
\caption{Square lattice with nearest neighbor (dashed red) and diagonal (solid red) hoppings. With straight line Peierls paths, $\Phi = 2\pi \times 2$ because the minimal loop (shaded red) encloses area $1/2$, shaded gray. In blue, we show examples of the flux-$\Phi$ string created by $U$ from $\mbf{r}_0$ to an orbital at $\mbf{R}$. The path of the flux string is unobservable and may be deformed arbitrarily along Peierls paths because the difference in flux (shaded blue) is a multiple of $2\pi$.} 
\label{fig_PPU}
\end{figure}

\emph{Hamiltonians.}
The Hofstadter Hamiltonian $H^\phi$ describes electrons in a constant perpendicular magnetic field via the Peierls substitution. We choose units where $\hbar, e, $ and the unit cell area equal one, so $\phi = \pmb{\nabla} \times \mbf{A}$. Consider a hopping term $\braket{\mbf{r}'|H|\mbf{r}} = t_{\mbf{r}\mbf{r}'}$, where $\ket{\mbf{r}}$ an electron state at $\mbf{r}$. Under the Peierls substitution, $\braket{\mbf{r}'|H^{\phi}|\mbf{r}} = t_{\mbf{r}\mbf{r}'} \exp i \int_{\mbf{r}}^{\mbf{r}'} \mbf{A} \cdot d\mbf{r}$. The ``Peierls path" along which the integral is taken can be determined by the Wannier functions of the zero-field groundstate \cite{2018arXiv181111786L} and in the simplest case is a straight line. Because the Peierls paths are determined by the ground state electron density, the paths themselves respect the lattice symmetries. The spectrum of $H^\phi$ is gauge-invariant but depends on the Peierls paths. Importantly, $H^{\phi + \Phi} = U H^\phi U^\dag$ has a nontrivial periodicity in flux \cite{firstpaper}, where
\bea
\label{eq:Umain}
U \ket{\mbf{r}} = \exp \lp i \int_{\mbf{r}_0}^{\mbf{r}} \mbf{A}^{\Phi} \cdot d\mbf{r} \rp \ket{\mbf{r}}, \quad \pmb{\nabla} \times \mbf{A}^{\Phi} = \Phi
\eea
where $\mbf{r}_0$ is the position of an arbitrary but fixed orbital, and the integral is taken along (any) sequence of Peierls paths. Thus the spectrum is periodic in $\Phi \in 2\pi \mathbb{N}$, defined such that all closed integrals along Peierls paths enclose a multiple of $2\pi$ flux and $U$ is single-valued (\Fig{fig_PPU}). For nearest neighbor hoppings on the square lattice, $\Phi = 2\pi$. 

\begin{figure}[h]
 \centering
\includegraphics[width=7cm]{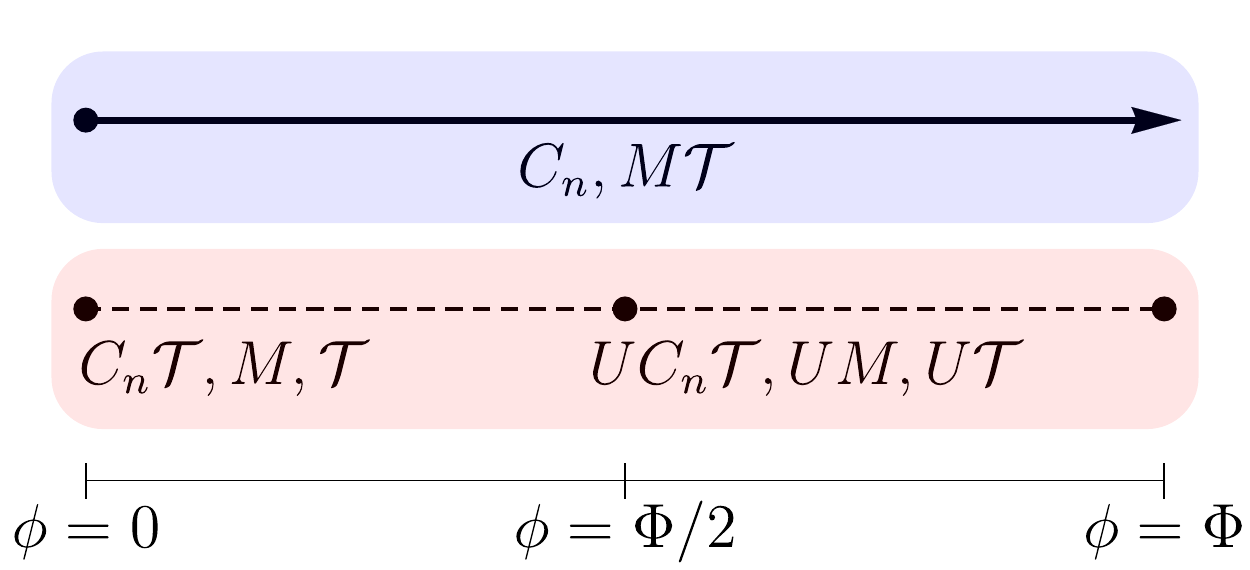} 
\caption{PG symmetries preserved (blue) and broken (red) in flux. At multiples of $\Phi/2$, the broken symmetries are reentrant as implemented by the flux periodicity operator $U$. } 
\label{fig:symmetrybehavior}
\end{figure}

\emph{Symmetries.} We now discuss the symmetries of crystalline systems composed of $n$-fold rotations $C_n$, mirrors $M = M_x,M_y$, and anti-unitary time reversal $\mathcal{T}$. In nonzero flux, the symmetries divide into two categories (see \Fig{fig:symmetrybehavior}). $M$ and $\mathcal{T}$ flip the sign of $\phi$ while rotations preserve it, so $C_n\mathcal{T}$ and $M$ are broken but $C_n$ and $M\mathcal{T}$ remain in flux (see \App{app:chicalc}). The symmetries broken in flux play a crucial role in the Hofstadter spectrum. In fact, these symmetries are \emph{reentrant} at strong flux $\phi = \Phi/2$. If $[C_n\mathcal{T}, H^{\phi=0}] = 0$, then
\bea
UC_n\mathcal{T} H^{\Phi/2} (UC_n\mathcal{T})^{-1} &= U H^{-\Phi/2} U^\dag = H^{\Phi/2} \\
\eea
so $UC_n\mathcal{T}$ is a symmetry of $H^{\Phi/2}$. The same is true for $UM$ and $U \mathcal{T}$. Since $U$ is a diagonal unitary matrix in the orbital basis and $\mathcal{T}$ acts locally on the orbitals, we have $(U \mathcal{T})^2 = \mathcal{T}^2$ \cite{firstpaper}. 
These reentrant symmetries can form a projective representation of the point group $G_x$. 

\begin{figure}[h]
 \centering
\includegraphics[width=4cm]{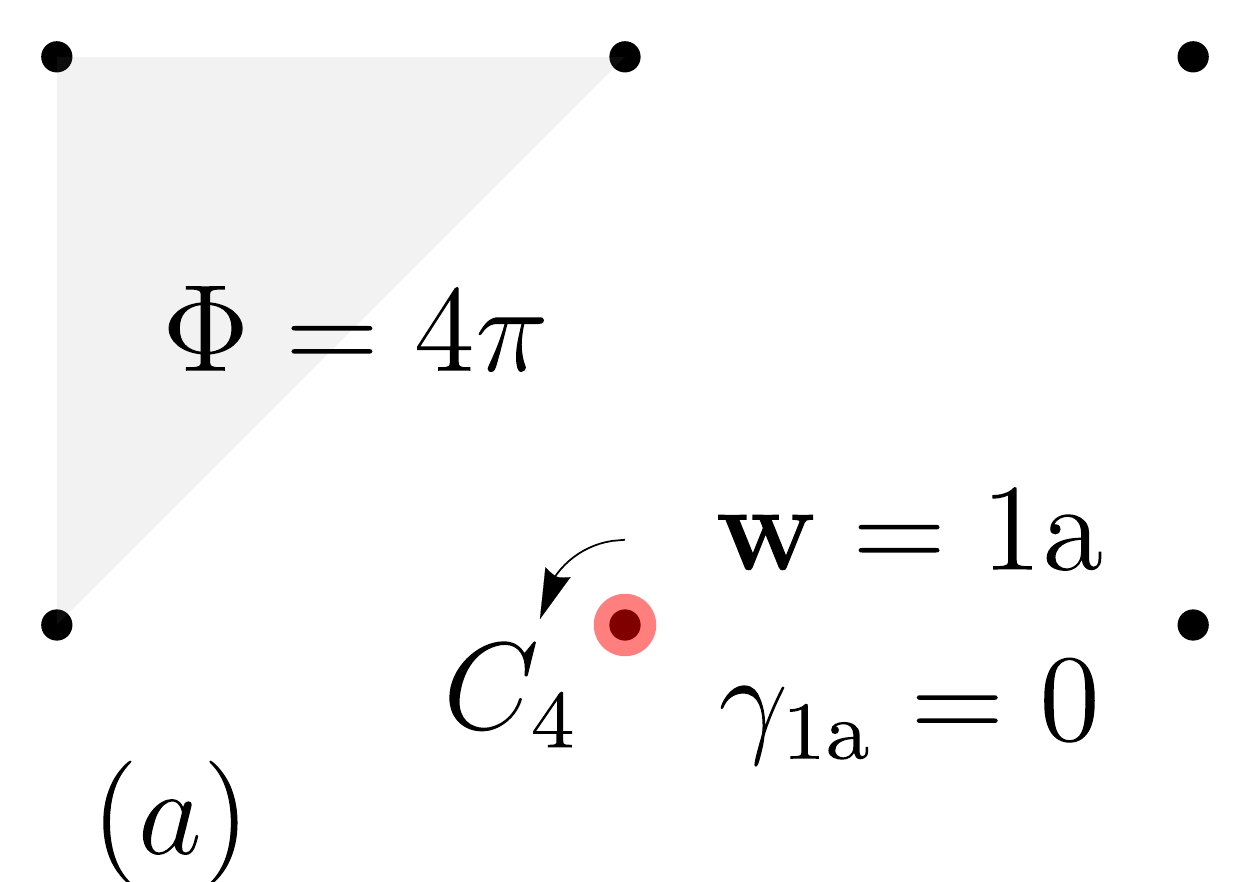} 
\includegraphics[width=4cm]{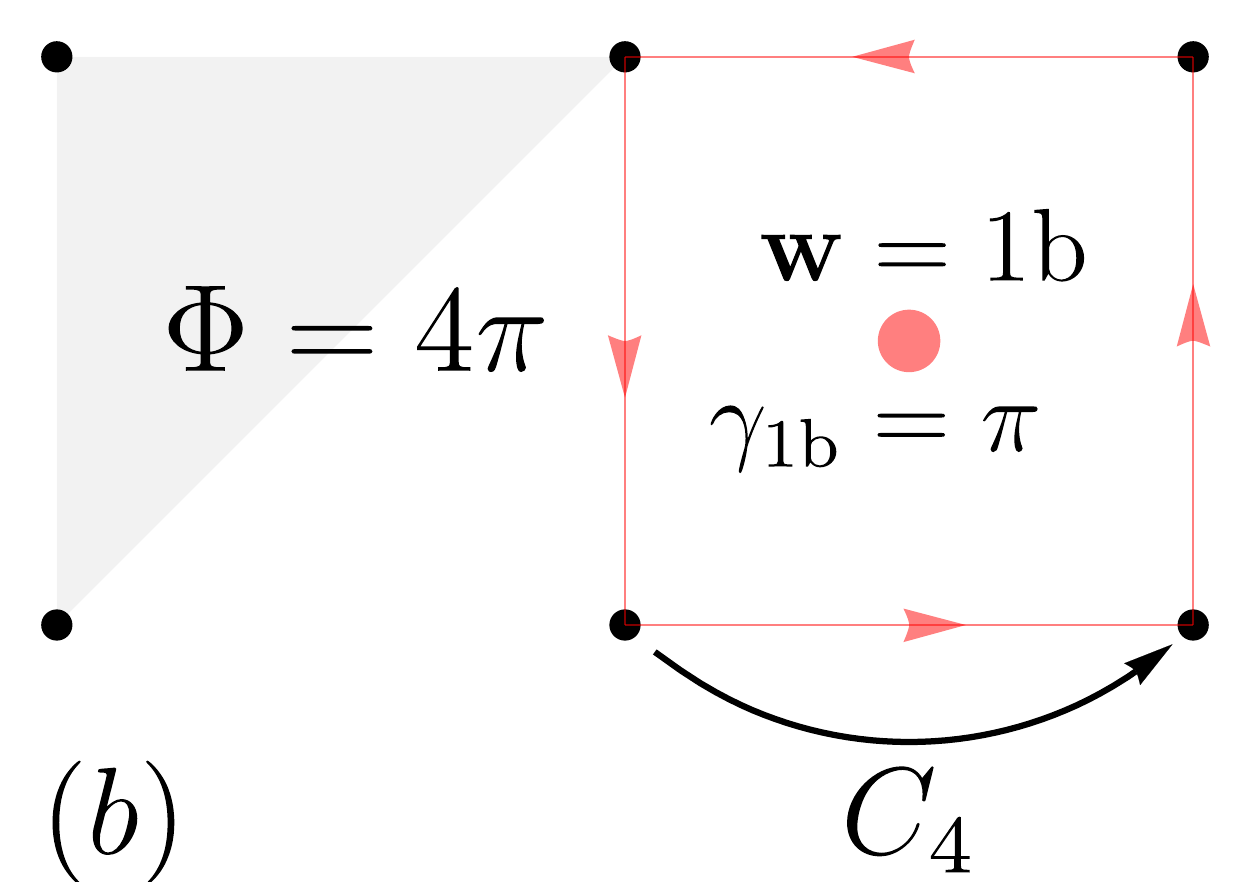} 
\caption{Calculating $\gamma_{\mbf{x}}$ at $\mbf{x} = 1a,1b$  on the square lattice with Peierls paths given in \Fig{fig_PPU} which enforce $\Phi = 4\pi$. $(a)$ At the 1a position (black), $\gamma_{1a} = 0$. $(b)$ At the 1b position, $\gamma_{1b} = \pi$, which we calculate using \Eq{eq:gammadef} by choosing a $C_4$-symmetric path around the unit cell on nearest-neighbor Peierls paths. If we considered an alternate model without the diagonal Peierls path, then $\Phi = 2\pi$ and $\gamma_{1b} = \pi/2$.}
\label{fig_gammaa}
\end{figure}

Consider a Wyckoff position $\mbf{x}$ and fix $\mbf{A}(\mbf{r}) = \frac{\phi}{2} \hat{z} \times (\mbf{r} - \mbf{x})$ to be in the symmetric gauge centered at $\mbf{x}$ so the $C_n \in G_x$ operator remains unchanged in flux (see \App{app:Cnchi}). To determine the group structure of the symmetries at $\phi = \Phi/2$, we derive in \App{app:gammaprop} the commutation relation
\bea
\label{eq:gammadef}
C_n^\dag U C_n = e^{i \gamma_{\mbf{x}}}  U, \quad \gamma_{\mbf{x}} = \frac{1}{n} \int_{\mathcal{C}_{\mbf{x}}} \mbf{A}^{\Phi} \cdot d \mbf{r} \mod 2\pi  \\
\eea
where $\mathcal{C}_{\mbf{x}}$ is a $C_n$-symmetric loop \emph{taken along Peierls paths} enclosing $\mbf{x}$. We prove that $\gamma_{\mbf{x}}$ is independent of the choice of loop (\App{app:gammaprop}), but we emphasize that $\gamma_{\mbf{x}}$ depends on the Wyckoff position $\mbf{x}$. Note that $\gamma_{\mbf{x}} \in \frac{2\pi}{n} \mathds{Z}_n$ is quantized because all closed loops along Peierls paths enclose multiples of $2\pi$ flux. If $\ket{\la}$ is an eigenstate of $H^{\phi=0}$ with $C_n$ eigenvalue $\la$, then $U \ket{\la}$ is an eigenstate of $H^\Phi$ since $H^{\Phi} = U H^{\phi=0} U^\dag$. The eigenvalue of $U \ket{\la}$ is
\bea
\label{eq:irrepflow}
C_n U \ket{\la} =  e^{-i \gamma_{\mbf{x}}}   U C_n \ket{\la} = \la e^{-i \gamma_{\mbf{x}}}  U \ket{\la}
\eea
so $\gamma_\mbf{x} \neq 0$ indicates angular momentum is transferred with flux, indicating irrep flow. Finally, if there is an orbital of the Hamiltonian located at $\mbf{x}$, then we can shrink the loop $\mathcal{C}_{\mbf{x}}$ to be a single point, and hence $\gamma_{\mbf{x}} = 0$ (see \Fig{fig_gammaa}a). Conventional straight-line Peierls paths can have nontrivial $\gamma_{\mbf{x}}$, as shown in \Fig{fig_gammaa}b where $\gamma_{1a} = 0$ but $\gamma_{1b} = \pi$. When referring to a fixed PG and Wyckoff position, we will drop the $\mbf{x}$ subscript. 

The reentrant symmetries $UC_n\mathcal{T}$ and $UM$ can form nontrivial central extensions of the conventional PGs at $\Phi/2$ flux when $\gamma \neq 0$, leading to projective representations which we call \emph{non-crystalline}. In the context of group theory, $\gamma$ is referred to as the Schur multiplier or 2-cocyle of the central extension. For instance, consider the PG $G_x=41'$ which is generated by $C_4$ and $\mathcal{T}$. Let us now consider $\phi=\Phi/2$ where the point symmetries generating $G^{\Phi/2}_x$ are $C_4$ and $U\mathcal{T}$. These symmetries can generate a projective  representation of $41'$ which we denote by $4_\gamma 1', \gamma = \pi/2, \pi, 3\pi/2$.  We build their irreps from the $C_4$ eigenstates $\ket{\la}$ in \Eq{eq:irrepflow}. Using \Eq{eq:gammadef}, $C_4 U\mathcal{T}\ket{\la} = e^{- i \gamma} \la^* U\mathcal{T} \ket{\la}$. If $\la \neq e^{- i \gamma} \la^*$, then $\ket{\la}$ and $U \mathcal{T} \ket{\la}$ must be distinct states which carry a 2D irrep since they are transformed to each other by $C_4$. If $\gamma =\pi/2$, there are two 2D irreps which we denote by $^1\!EA$ and $^2\!EB$ (see \Tab{tab:4irreps}). If $\gamma =\pi$, we find a 2D irrep $AB$ and two 1D irreps ${}^1\!E, {}^2\!E$. From the group theory perspective, $4_\pi 1'$ is actually not a nontrivial central extension: it can be lifted to the non-projective group $41'$ by taking $C_4 \to i C_4$. However, $4_\pi 1'$ is physically distinguished from $41'$ (and can still protect Hofstadter topology) because the overall phase of $C_4$ is fixed according to the $\phi=0$ convention. We enumerate all of the non-crystalline PGs and their irreps in \App{app:Luis}. All 51 of the non-crystalline PGs (as well as the 31 crystalline PGs) appear on the \href{www.cryst.ehu.es/cryst/projectiverepres}{Bilbao Crystallographic Server}\cite{aroyo2006bilbao}.

\begin{table}
    \centering
\begin{tabular}{c|ccc}
$41'$ & 1 & $C_4$ & $C_2$  \\
\hline
$A$ & 1 & 1 & 1 \\
$B$ & 1 & $-1$ & 1 \\
$^1\!E^2\!E$ & 2 & $0$ & $-2$ \\
\end{tabular} \
\begin{tabular}{c|ccc}
\rule{0pt}{-2.5ex} $4_{\pi/2}1'$ & 1 & $C_4$ & $C_2$ \\
\hline
\rule{0pt}{2.5ex} \!\!$^1\!EA$ & 2 & $1-i$ & 0 \\
$^2\!EB$ & 2 & $-1+i$ & 0\\
\end{tabular} \
\begin{tabular}{c|ccc}
$4_\pi1'$ & 1 & $C_4$ & $C_2$  \\
\hline
$AB$ & 2 & 0 & 2 \\
$^1\!E$ & 1 & $-i$ & $-1$ \\
$^2\!E$ & 1 & $i$ & $-1$ \\
\end{tabular} 
\caption{We list the (partial) character tables for the irreps of $41'$ and two of its projective representations without SOC. (The irreps of $4_{3\pi/2}1'$ are the complex conjugates of $4_{\pi/2}1'$.) We name the irreps according to their $C_4$ eigenvalues where $A,B,^1\!E,^2\!E$ correspond to $+1,-1,-i,i$ respectively. 
 \label{tab:4irreps} }
\end{table}

\emph{Hofstadter Response of an Obstructed Atomic State.} The symmetry and topology of $H^{\phi=0}$ determine the flux response and hence have a fundamental effect on the Hofstadter spectrum. A nonzero mirror Chern number enforces a bulk gap closing in finite flux \cite{firstpaper,2020arXiv200614000B,2015NatSR...513277Z,2014PhRvB..90k5305Z,andreibook} and a nonzero Kane-Mele index enforces edge state pumping in flux for $\Phi/2\pi$ odd, while a trivial atomic state remains gapped \cite{firstpaper}. We now show that symmetry-protected analogues of these phases also exist. Their topological invariants may be found in \App{app:tables}. 
%An overview of our classification is given in \Tab{tab:fullclassmain} organized by the zero-flux PGs. 

We first study the Hofstadter SM state which is defined by an enforced gap closing at finite flux $\phi \in (0,\Phi)$. We consider a fixed Wyckoff position with PG $G_x$ at zero flux, $G_x^{\Phi/2}$ at $\Phi/2$ flux, and $G^\phi_x \subseteq G_x$ at generic flux. In a Hofstadter SM, the bulk gap is closed in $\phi$ when a level crossing occurs. To avoid eigenvalue repulsion, the crossing states must be different irreps of $G^\phi_x$, and thus the ground states before and after the crossing have different irreps. Before showing formally how RSIs detect this irrep exchange, we give a simple example. 

\begin{figure}
 \centering
\includegraphics[width=8cm]{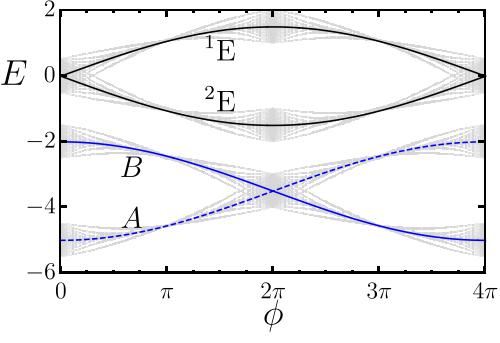} 
\caption{Hofstadter spectrum of an obstructed atomic state whose Peierls paths give $\Phi = 4\pi, \gamma_{1b} = \pi$. The flat band limit is analytically solvable (see \App{app:flatbandham}) and is shown in blue (valence bands) and black (conduction bands). Small terms can be added to broaden the bands, and the Hofstadter spectrum in this case is shown in gray. In the flat band limit where the valence bands are compact Wannier functions, each is labeled by its $C_4$ irrep with a crossing at $\Phi/2$. Note that the ${}^1E{}^2E$ irrep protected by $\mathcal{T}$ is broken by flux. }
\label{fig_OALcross}
\end{figure}

Consider a Hamiltonian on open boundary conditions with $s$ orbitals at corners of a square (four 1a sites). The center is the $x=$1b position and has $G_x = 41'$. We include nearest and second nearest neighbors with straight-line Peierls paths in flux, exactly as in \Fig{fig_gammaa}b. In the symmetric gauge centered at 1b, we obtain (see \App{app:square})
\bea
\label{eq:Hplaq}
H^\phi_{4} = \bpm
0 & t e^{-i \phi/4} & t' &   t e^{i \phi/4} \\
 t e^{i \phi/4}  & 0 &  t e^{-i \phi/4}  & t' \\
 t' &   t e^{i \phi/4} & 0 &   t e^{-i \phi/4} \\
 t e^{-i \phi/4}     & t' &   t e^{i \phi/4}  & 0 \\
\epm \ .
\eea
The diagonal hopping $t'$ sets $\Phi = 4\pi$ and $\gamma_{1b} = \pi$ (\Fig{fig_gammaa}b). The symmetries at $\phi = 0$ are $C_4$ which permutes the sites around 1b and $\mathcal{T} = K$ which is complex conjugation. $C_4$ remains a symmetry for all $\phi$, but $\mathcal{T}$ is broken in flux. Although the spectrum is $4\pi$-periodic, the eigenstates are not. \Eq{eq:irrepflow} shows there is irrep flow, e.g. if the lowest energy eigenstate of $H_4$ at $\phi = 0$ is an $A$ irrep, then at $\phi = \Phi$ the lowest energy eigenstate is a $B$ irrep because $\gamma_{1b} = \pi$. In this case, the ground state has changed irreps although $C_4$ symmetry is never broken, but this is only possible if there is a gap closing in flux. This gap closing can also be predicted entirely from the projective symmetries at $\phi = \Phi/2 = 2\pi$. There $U\mathcal{T}$ reenters as a symmetry with $U = \text{diag}(1,-1,1,-1)$ computed from \Eq{eq:Umain}.
 \Tab{tab:4irreps} shows that at $\phi = \Phi/2$, the PG is $4_\pi 1'$ which has the irrep $AB$, so the level crossing occurs at exactly $\phi = \Phi/2$ where the $A$ and $B$ irreps are degenerate (see \Fig{fig_OALcross}). In this example, the Hofstadter SM phase was deduced from only zero-flux data: a nonzero Schur multiplier and the irreps of the ground state of $H^{\phi =0}$ enforced irrep flow at $\Phi$ and a gap closing exactly at $\Phi/2$. We call such phases ``Peierls-indicated."

In this decoupled example, the multiplicities of each irrep at $x=1b$ characterized the whole ground state. In a general Hamiltonian with nontrivial bands where the number of irreps at a Wyckoff position is not adiabatically well-defined, we use RSIs to study irrep flow. Defined in \Ref{song2019real}, RSIs are computed in real space from the Wannier states at $x$ and are invariant under $G_x$-allowed deformations that change the occupied irreps at $x$. For instance in $G_x = 41'$, four Wannier states in the representation $A\oplus B \oplus {}^1E{}^2E$ can be moved off $x$ because they form an induced representation of the lower symmetry position \cite{song2019real}. However, the RSIs $\delta_i = \{m(B) - m(A), m({}^1E{}^2E) - m(A)\}$ are invariant under this process. Here $m(\rho)$ is the multiplicity of the irrep $\rho$. In general, RSIs are invariant unless the gap is closed (changing the occupied states discontinuously) or the symmetries protecting them are broken. The RSIs of the non-crystalline PGs are computed in \App{app:tables} using the Smith Normal form \cite{song2019real}. We now formalize the Hofstadter SM invariants using RSIs. We assume that the $\phi=0$ ground state is gapped and has vanishing Chern number $C=0$ so that integer-valued \emph{local} RSIs are well-defined \cite{song2019real}. From the Streda formula \cite{Streda_1982, Dana_1985} $C \phi/2\pi = n  \mod 1$, $C=0$ means we only consider fixed integer fillings for all $\phi$. Here $n$ is the fractional number of electrons per unit cell.  If $n \notin \mathbb{N}$, then $C \neq 0$. 

\emph{Hofstadter SM.} Previously, we exemplified how $C_n$ enforces a gap closing due to irrep flow (equivalently, a change of RSIs) in the occupied states. Generally, with $M$ and $\mathcal{T}$ which relate the spectrum at $\pm\phi$, we obtain a finer classification by comparing the RSIs at $\phi =0,\Phi/2$ denoted $\delta^{\phi =0}, \delta^{\phi =\Phi/2}$.  A gap closing can be detected by an incompatibility of the RSIs $\delta^{\phi =0}$ and $\delta^{\phi =\Phi/2}$ when reduced to the $G^\phi_x$ subgroup. Perturbing away from $\phi = 0$ or $\phi = \Phi/2$, the PG is reduced to $G^\phi_x$ as the symmetries that reverse the flux are broken. The occupied states do not change under this infinitesimal perturbation (since $C=0$), and the RSIs of $G^\phi_x$ can be determined from $\delta^{\phi =0}$ and $\delta^{\phi =\Phi/2}$ by irrep reduction \cite{song2019real}. We denote the RSIs determined from the reduction of $G_x \to G_x^\phi$ and $G_x^{\Phi/2} \to G_x^\phi$ as $\delta^{\phi \to 0}_i$ and $\delta^{\phi \to \Phi/2}_i$ respectively ($i$ indexes the RSIs of $G^\phi_x$). The Hofstadter SM index is 
\bea
\label{eq:gapcompat}
\delta^{SM}_i = \delta^{\phi \to 0}_i - \delta^{\phi \to \Phi/2}_i \ . \\ 
\eea
We prove \Eq{eq:gapcompat} from the properties of the RSIs. If there is no gap closing between $0$ and $\Phi/2$ flux, then the RSIs of $G^\phi_x$ cannot change because the symmetries of $G_x^\phi$ exist at all flux. Hence if $\delta^{SM}_i \neq 0$, a gap closing must occur to change the occupied states. With $C_n$ symmetry alone, the gap closings from irrep flow are detected formally by Hofstadter SM index $\delta^{\phi =0}_i - \delta^{\phi =\Phi}_i$, which can be written solely in terms of $\delta^{\phi =0}$ since the irreps at $\Phi$ are determined by $U$ (\App{app:tables}). Moreover, we check exhaustively that \Eq{eq:gapcompat} also diagnoses the gap closings protected by irrep flow (even though they only compare RSIs at 0 and $\Phi/2$ flux) due to the degeneracies protected at $\phi = \Phi/2$. We list the Hofstadter SM indices in all PGs in \App{app:tables}, finding $\mathds{Z}$ indices with $C_n$ and $\mathds{Z}_2$ indices with $M\mathcal{T}$. 

We illustrate \Eq{eq:gapcompat} with $H^\phi_{4}$. \Tab{tab:RSIsnoSOC} contains the RSIs of $G_x = 41'$ and $G_x^{\Phi/2} = 4_\pi 1'$ (see \App{app:RSIexamples} for examples of the calculation). Their reduction to the RSIs of $G_x^\phi = 4$ follows from ${}^1E{}^2E \to {}^1E\oplus {}^2E$ near $\phi=0$ and $AB \to A \oplus B$ near $\phi = \Phi/2$ (see \Tab{tab:4irreps}). Using \Eq{eq:gapcompat}, we find a $\mathds{Z}^3$ Hofstadter SM index:
\bea
\label{eq:SM41'}
\delta^{SM}_i = \{ \delta^{\phi=0}_1, \delta^{\phi = 0}_2+\delta^{\phi = \frac{\Phi}{2}}_2, \delta^{\phi = 0}_2 - \delta^{\phi = \frac{\Phi}{2}}_1 + \delta^{\phi = \frac{\Phi}{2}}_2 \} \ .
\eea
We see that the SM phase is Peierls-indicated because $\delta^{\phi=0}_1 \neq 0$ enforces a gap closing, i.e. any zero-flux state with $\gamma_{1b} = \pi$ and $\delta^{\phi=0}_1 = m(B)-m(A) \neq 0$ at the 1b Wyckoff position is a Hofstadter SM. This is exactly the same gap closing diagnosed by irrep flow, since $A$ and $B$ irreps exchange between $\phi=0$ and $\phi = \Phi$. We generalize $H_4$ to a full lattice model in an obstructed atomic phase (see \App{app:flatbandham}) and show the protected crossings in \Fig{fig_OALcross}.

\emph{Hofstadter HOTI.} We now consider the Hofstadter HOTI phase defined by nontrivial flow of Wannier states through the bulk over $\phi \in (0, \Phi/2)$. On open boundary conditions, this flow is manifested as a pumping cycle of edge/corner states into the bulk. For the Wannier states to evolve continuously, we require that $\delta_i^{SM} = 0$ so that there is no enforced bulk gap closing (see \Eq{eq:gapcompat}). We develop topological invariants for these phases by detecting charge flow onto/off of a given Wyckoff position between $\phi = 0$ and $\Phi/2$. Explicitly, we compute the number of Wannier states at $\mbf{x}$ by counting their representations:
\bea
\label{eq:Nx}
N_{\mbf{x}} = \sum_{\rho \in G_\mbf{x}} m(\rho) \dim(\rho)\ .  \\
\eea
Of course, $N_{\mbf{x}}$ is not an adiabatic invariant because Wannier states can move onto and off of $\mbf{x}$ if they form an induced representation. However $N_{\mbf{x}}$ is adiabatically conserved modulo the dimension of the induced representation \cite{song2019real}. For instance, in PG $41'$, $N_{\mbf{x}} \mod 4$ is conserved (and can be written in terms of RSIs) because only multiples of 4 states can be moved while preserving $C_4$. Generally, we find that there exists $n_G \in \mathbb{N}$ such that
\bea
\label{eq:deltahoti}
\delta^{HOTI}_\mbf{x} = N_{\mbf{x}}^{\phi =0} - N_{\mbf{x}}^{\phi = \Phi/2} \mod n_G 
\eea
can be written in terms of RSIs. Recall that a nonzero RSI off an orbital position diagnoses corner states on open boundary conditions \cite{song2019real,2017PhRvB..96x5115B,song2019real,2020PhRvR...2d2038P,2019PhRvB..99x5151B}. Thus \Eq{eq:deltahoti}  diagnoses a Hofstadter HOTI phase because corner states are smoothly pumped onto/off of $\mbf{x}$ if $\delta^{HOTI}_\mbf{x} \neq 0$. In \App{app:tables}, we compute the compatibility conditions $\delta_i^{SM} = 0$ and then Hofstadter HOTI invariants for all PGs. 

We now give an example in magnetic PG $2'$ without SOC and $\gamma_\mbf{x} = \pi$, which can be obtained from the nearest-neighbor square lattice where $\Phi = 2\pi$. PG $2'$ has a single irrep $A$ where $D[C_2\mathcal{T}] = K$, which squares to $+1$. Orbitals can be moved offsite in $C_2\mathcal{T}$-symmetric pairs, so the RSI is $\delta = m(A) \mod 2$ and can be calculated from the nested Wilson loop \cite{2017PhRvB..96x5115B,2018arXiv181002373W} or the Stiefel-Whitney invariants \cite{2019ChPhB..28k7101A,2019PhRvB..99w5125A,2019PhRvX...9b1013A}. At $\Phi/2$ flux, PG $2'_\pi$ has a single 2D irrep $AA$ because $UC_2\mathcal{T}$ squares to $-1$. This PG has no RSI because the two states carrying $AA$ can always be removed offsite in opposite directions while respecting $UC_2\mathcal{T}$. Correspondingly, the Wilson loop is always trivial \cite{firstpaper}. In generic flux, $G^\phi = 1$ because $C_2\mathcal{T}$ is broken. Hence the gaps at $\phi = 0$ and $\phi = \Phi/2$ are trivially compatible: $\delta^{SM}_1 = 0$. The total charge $N_{\mbf{x}}$ is also simple:
\bea
\label{eq:hotic2t}
N_{\mbf{x}}^{\phi=0} \!= m(A) = \delta_1^{\phi=0}\, \text{mod }2, \, N_{\mbf{x}}^{\phi=\Phi/2}\!= 0 \text{ mod }2 .
\eea
 The Hofstadter HOTI invariant is simply $\delta^{HOTI}_\mbf{x} = \delta^{\phi = 0}_\mbf{x} \in \mathds{Z}_2$, and is Peierls-indicated. The HOTI index \Eq{eq:hotic2t} describes a pumping process where a single state at $\mbf{x}$ (which gives a corner state on open boundary conditions) is moved offsite as $\phi$ is increased and $C_2\mathcal{T}$ is broken. This Hofstadter HOTI phase is realized in models with $C_2\mathcal{T}$-protected fragile topology (see \App{app:qshexample}) \cite{firstpaper,2006Sci...314.1757B}. Indeed, the projective representation of $C_2\mathcal{T}$ has already been experimentally achieved in acoustics\cite{2022arXiv220713000M,2022arXiv220902349X}. 

\emph{Discussion}. The appearance of projective symmetries in strong flux enables zero-flus RSIs to constrain the Hofstadter spectrum. This work has completely classified the resulting 51 non-crystalline 2D PGs and demonstrated that the symmetries and topology of $H^{\phi=0}$, encoded in the RSIs, dictate universal features of its spectrum in flux. Our results give observable bulk signatures of obstructed atomic and fragile phases. Although we have focused on crystalline systems, acoustic materials offer alternative platforms where projective symmetries have already gathered interest \cite{2022arXiv220713000M,2021arXiv210711564X,2021arXiv210714579L,2021arXiv210413161X,PhysRevLett.127.076401}. Synthetic gauge fields in these platforms have been used to experimentally confirm irrep flow due \cite{2020Sci...367..797P,2021arXiv210502070L}. Additionally, we note that the Hofstadter topological indices derived here depend only on the local PG symmetries, and thus still apply to high-symmetry points in quasi-crystalline systems without translations \cite{PhysRevB.101.115413,2021arXiv210705635J,2021arXiv210308851H,2015PhRvB..91h5125T}. Lastly, the ever expanding family of moir\'e materials has already allowed access to the strong flux regime where signatures of the reentrant symmetries have already been proposed \cite{2022arXiv220408087K} and reentrant phases observed \cite{PhysRevLett.128.217701}. The single-particle projective symmetries mentioned here may also be approximately realized in continuum models, giving rise to otherwise impossible many-body phenomena in strong flux \cite{2021arXiv211111434H,2022arXiv220607717H,2021arXiv210804831S,2018arXiv181111786L}. 

\emph{Note Added}. A manuscript posted on the same day (\Ref{Cano}) also studies the topology of Hofstadter bands in magnetic flux. \Ref{Cano} employs a momentum space topological quantum chemistry approach at $\pi$ flux and obtains stable invariants in certain wallpaper groups. Instead, we work in real space and classify Hofstadter responses in flux for all (projective) point group symmetries. 

\emph{Acknowledgements}. B.A.B. and Z-D.S. were supported by the European Research Council (ERC) under the European Union's Horizon 2020 research and innovation programme (grant agreement No. 101020833),
the ONR Grant No. N00014-20-1-2303, 
the Schmidt Fund for Innovative Research, 
Simons Investigator Grant No. 404513, 
the Packard Foundation, 
the Gordon and Betty Moore Foundation through the EPiQS Initiative, Grant GBMF11070 and Grant No. GBMF8685 towards the Princeton theory program.
Further support was provided by the NSF-MRSEC Grant No. DMR-2011750, 
BSF Israel US foundation Grant No. 2018226, 
and the Princeton Global Network Funds.
 JHA is supported by a Hertz Fellowship and thanks the Marshall Aid Commemoration Commission for support during the earlier stages of this project.

%%%%%%%%%%%%%%%%%%%%%%%%%%%%%%%%%%
\bibliography{finalbib}
\bibliographystyle{aipnum4-1}
\bibliographystyle{unsrtnat}
%%%%%%%%%%%%%%%%%%%%%%%%%%%%%%%%%%
\newpage
\onecolumngrid
\appendix

\tableofcontents

\section{Review of the Hofstadter Hamiltonian}
\label{app:reviewhof}

%In units where the electric charge, $\hbar$, and the area of the unit cell are equal to one, $H^\phi$ is periodic under $\phi \to \phi + \Phi$ up to a unitary transformation. 

We begin by reviewing the Hofstadter Hamiltonian $H^\phi$ which is constructed from a general 2D tight-binding model using the Peierls substitution  \cite{firstpaper,PhysRevB.14.2239, 1933ZPhy...80..763P, PhysRev.84.814}. We study only single-particle physics in this work, so will use the single-particle bases $\ket{\mbf{R}, \al} = c^\dag_{\mbf{R}, \al} \ket{0}$ where $c^\dag_{\mbf{R},\al}$  creates an electron in orbital $\al = 1, \dots, n_{orb}$ at position $\mbf{R} + \mbf{r}_\al$, e.g. in the $\mbf{R}$th unit cell. Here $\mbf{R}$ is a point in the Bravais lattice which is spanned by the unit vectors $\mbf{a}_1, \mbf{a}_2$. It is convenient to normalize the area of the unit cell and require $\mbf{a}_1 \times \mbf{a}_2 = 1$, as well as choosing units where $\hbar$ and the electric charge $e$ are set to one. In addition, the cross product of 2D vectors is taken to be a scalar. In these units, $\pmb{\nabla} \times \mbf{A}(\mbf{r}) = \phi$ where $\mbf{A}(\mbf{r})$ is the vector potential of the magnetic field. As discussed in  \cite{firstpaper}, the Hofstadter Hamiltonian may be written in the Peierls substitution as 
\bea
\label{eq:hofham}
H^\phi &= \sum_{\mbf{R} \mbf{R}' \al \be} t_{\al \be}(\mbf{R} - \mbf{R}') \exp \lp i \varphi_{\mbf{R} + \mbf{r}_\al, \mbf{R}' + \mbf{r}_{\be}} \rp \ket{\mbf{R}, \al} \bra{\mbf{R}', \be}, \quad \varphi_{\mbf{R} + \mbf{r}_\al, \mbf{R}' + \mbf{r}_{\be}} = \int_{\mbf{R}' + \mbf{r}_\be}^{\mbf{R} + \mbf{r}_\al} \mbf{A}(\mbf{r}) \cdot d\mbf{r}  
\eea
where $t_{\al \be}(\mbf{R}-\mbf{R}')$ is the hopping matrix of the crystal at zero flux. The path of the line integral in \Eq{eq:hofham} must be specified for each hopping in the Hamiltonian. These so-called Peierls paths are determined by the overlap of the Wannier functions. Conventionally, the Peierls paths are chosen to be straight-line paths between the orbitals of the hoppings, but in general should be chosen to pass through the position of greatest orbital overlap in the case of extended Wannier functions (see \Ref{2018arXiv181111786L} for a detailed discussion). \Ref{firstpaper} proves that the spectrum of $H^\phi$ is invariant under the gauge of $\mbf{A}$, but we emphasize that the choice of Peierls paths is physical. In particular, the flux periodicity $\Phi$ of the Hofstadter Hamiltonian depends on the Peierls paths. \Ref{firstpaper} proved that $\Phi = 2\pi N, N \in \mathbb{N}$ where $N$ is equal to the lowest common denominator of the fractional areas of the unit cell enclosed by all possible Peierls paths. In particular, $N$ exists when all orbitals are located at rational positions and the Peierls paths are piecewise straight. More physically, $\frac{\phi}{N}$ is the smallest Aharonov-Bohm phase that a particle on the lattice can acquire when flux $\phi$ is inserted into each unit cell. For example, in a square lattice of $s$-orbitals with nearest-neighbor hoppings along the bonds, we find $N=1$ ($\Phi = 2\pi$) because all closed loops encircle an integer number of unit cells. It was demonstrated in \Ref{firstpaper} that $U H^{\phi} U^\dag = H^{\phi + \Phi}$ where (matching \Eq{eq:Umain} in the Main Text)
\bea
\label{eq:defU}
U &= \sum_{\mbf{R} \al} \exp \lp i \int_{\mbf{r}_0}^{\mbf{R} + \mbf{r}_\al} \mbf{\tilde{A}}(\mbf{r}) \cdot d\mbf{r} \rp \ket{\mbf{R}, \al} \bra{\mbf{R}, \al}, \quad \pmb{\nabla} \times \mbf{\tilde{A}} = \Phi, \\
\eea
and $\mbf{r}_0$ is an arbitrary but fixed position \emph{of an explicit orbital of} $H$ and the integral is taken along an arbitrary sequence of Peierls paths. 
Importantly, $U$ is independent of the sequence of Peierls paths as discussed in \Ref{firstpaper} because at $\phi = \Phi$, any deviation along Peierls paths adds a multiple of $2\pi$ to the integral. One can think of $U$ as a ``large" gauge transformation since $\mbf{\tilde{A}}$ carries nonzero flux but $U$ is single-valued. We will show that $U$ is responsible for creating projective representations of the symmetries. 

\Ref{firstpaper} focused on the momentum space description of the Hofstadter Hamiltonian, which only exists when $\phi$ is a rational multiple of $2\pi$, so that spatial periodicity is recovered in nonzero flux over an (enlarged) magnetic unit cell, thereby creating a conserved lattice momentum. In this work, we focus on the description of the Hofstadter Hamiltonian in position space. Our proofs do not require a momentum quantum number, and so the flux $\phi$ may take on any value in $\mathbb{R}$. For completeness, we recall expressions for the magnetic translation group operators $T_i(\phi)$ from \Ref{firstpaper}:
\bea
\label{eq:T}
T_i(\phi) &= \sum_{\mbf{R} \al} \exp \lp i \int_{\mbf{R} + \mbf{r}_\al}^{\mbf{R} + \mbf{r}_\al + \mbf{a}_i} \mbf{A} \cdot d\mbf{r} \, + \, i \chi_i(\mbf{R}+\mbf{r}_\al ) \rp \ket{\mbf{R} + \mbf{a}_i, \al} \bra{\mbf{R}, \al}, \qquad \chi_i(\mbf{r}) = \phi \, \mbf{a}_i \times \mbf{r} \\
\eea
where the path of integral is a straight line (not necessarily a Peierls path). They obey the algebra $T_1(\phi) T_2(\phi) = e^{i \phi} T_2(\phi) T_1(\phi)$, which is projective representation of the translation group. When $\phi = 2\pi \frac{p}{q}$ is rational, $[T_1(\phi), T_2^q(\phi)] = 0$, and there is a $1\times q$ magnetic unit cell.

\section{Symmetries in the Presence of a Magnetic Field}
\label{app:symflux}

In this Appendix, we provide the calculations used in the Main text which demonstrate that in the symmetric gauge, the zero-field rotational symmetries and anti-unitary mirrors remain symmetries of the Hamiltonian, and anti-unitary rotations and mirrors reverse the flux. 

\subsection{Symmetries In Flux}
\label{app:chicalc}

A tight-binding Hamiltonian may be written in terms of single particle kets as
\bea
\label{eq:Hinketbra}
H &= \sum_{\mbf{R} \mbf{R}' \al \be} t_{\al \be}(\mbf{R} - \mbf{R}') \ket{\mbf{R}, \al} \bra{\mbf{R}', \be} \ .
\eea
This Hamiltonian may possess non-trivial point group (PG) symmetries in addition to the translational symmetries of its space group. Specifically, a PG in two spatial dimensions is generated by rotations $C_n$, mirrors $M_i$, time reversal $\mathcal{T}$, and their products. Here $C_n \in SO(2), n = 2,3,4,6$ is an $n$-fold rotation, and $M_{i} \in O(2), i = 1,2$ is a reflection taking $x_i \to - x_i$. Time reversal $\mathcal{T}$ is an anti-unitary symmetry which commutes with $C_n$ and $M_i$. We take $C_n$ and $M_i$ to be defined about a fixed origin. 

Let $G$ denote the wallpaper group (also referred to as the plane group) containing all symmetries (unitary and anti-unitary) of the $H$. We denote the site-symmetry of a Wyckoff position $\mbf{x}$ by 
\bea
G_{\mbf{x}} = \{g \in G | g \mbf{x} = \mbf{x} \} 
\eea
where $g\mbf{x}$ is understood as the representation of $g$ in position space. Throughout this paper, we will mostly concern ourselves with the local structure of a PG where no translation operators are necessary (as opposed to the global structure of the full space group). This will simplify our results when a magnetic field is introduced. 

We write the representation of a unitary symmetry $g$ in terms of the single-particle states as
\bea
g &= \sum_{\mbf{R} \al \al'} D[g]_{\al' \al} \ket{g (\mbf{R} + \mbf{r}_\al) - \mbf{r}_{\al'},\al'} \bra{ \mbf{R}, \al}  \\
\eea
where the $n_{orb} \times n_{orb}$ matrix $D[g]$ forms the representation of $g$ on the orbitals and $D[g]_{\al' \al}  = 0$ unless $g \mbf{r}_\al - \mbf{r}_{\al'}$ is a lattice vector. For spinless/spinful particles, $D[C_n]^n= D[M_i]^2 = \pm 1$. We also consider anti-unitary symmetries containing time-reversal, $\mathcal{T}$, which obeys $\mathcal{T}^2 = \pm 1$ and commutes with all spatial symmetries.

Now we form the Hofstadter Hamiltonian by threading flux through the lattice in the Peierls substitution for \Eq{eq:Hinketbra}:
\bea
\label{eq:peierlssub}
t_{\al \be}(\mbf{R} - \mbf{R}')  \to t_{\al \be}(\mbf{R} - \mbf{R}') \exp \lp i \int_{\mbf{R}' + \mbf{r}_\be}^{\mbf{R} + \mbf{r}_\al} \mbf{A}(\mbf{r}) \cdot d\mbf{r} \rp
\eea
where $\mbf{A}(\mbf{r})$ is the vector potential satisfying $\pmb{\nabla} \times \mbf{A} = \phi$. (Recall that $\mbf{a}_1 \times \mbf{a}_2=1$.) We now study the PG symmetries in the presence of flux. We will find that rotations and anti-unitary mirrors are preserved when flux is inserted, but anti-unitary rotations and mirrors are broken. We show this explicitly in the symmetric gauge, and we note where our proofs can be extended to any gauge. 

We now seek to determine whether a symmetry $g$ may be extended to be symmetry of $H^\phi$ for $\phi \neq 0$. To do so, we assume that the Peierls paths which decorate the lattice are also symmetric under $g$, i.e if $\mathcal{C}$ is a Peierls path between $\mbf{r}$ and $\mbf{r}'$, then $g \mathcal{C}$ is also a Peierls path between $g\mbf{r}$ and $g\mbf{r}'$. We expect this assumption to be physically justified because the Peierls paths capture the chemistry of the local bonding, and hence should reflect their symmetries. Throughout this paper, we always choose Peierls paths that respect $G$. We make the ansatz in analogy to \Eq{eq:T}
\bea
\label{eq:symdef}
g^\phi &= \sum_{\mbf{R} \al \al'} D[g]_{\al' \al} \exp \lp i \chi^\phi_g( g(\mbf{R}+ \mbf{r}_\al))\rp \ket{g (\mbf{R} + \mbf{r}_\al) - \mbf{r}_{\al'},\al'} \bra{ \mbf{R}, \al}  \\
\eea
where $\chi_g^\phi(\mbf{r})$ is to be determined. We find an expression for $\chi_g^\phi$ by studying the action of $g^\phi$ on the Hofstadter Hamiltonian:
\bea
g^\phi H^\phi (g^\phi)^\dag &= \sum_{\mbf{R} \mbf{R}' \al \be} \sum_{\al' \be'} D[g]_{\al' \al} t_{\al \be}(\mbf{R} - \mbf{R}')  D^\dag[g]_{\be' \be}  \exp \lp i \int_{\mbf{R}' + \mbf{r}_\be}^{\mbf{R} + \mbf{r}_\al} \mbf{A}(\mbf{r}) \cdot d\mbf{r}+ i \chi_g^\phi(g ( \mbf{R} + \mbf{r}_\al)) - i \chi_g^\phi(g ( \mbf{R'} + \mbf{r}_\be)) \rp\\ & \qquad \times \ket{g(\mbf{R} + \mbf{r}_\al) - \mbf{r}_{\al'}, \al} \bra{g(\mbf{R}' + \mbf{r}_\be) - \mbf{r}_{\be'}, \be} \ . 
\eea
We relabel the lattice sites in the sum according to $\mbf{S} = g(\mbf{R} + \mbf{r}_\al) - \mbf{r}_{\al'}, \mbf{S}' = g(\mbf{R}' + \mbf{r}_\be) - \mbf{r}_{\be'}$ so that
\bea
\label{eq:gh}
g^\phi H^\phi (g^\phi)^\dag &= \sum_{\mbf{S} \mbf{S}' \al' \be'} \exp \lp i \int_{g^{-1}(\mbf{S}' + \mbf{r}_{\be'})}^{g^{-1}(\mbf{S} + \mbf{r}_{\al'})} \mbf{A}(\mbf{r}) \cdot d\mbf{r} + i \chi_g^\phi(\mbf{S} + \mbf{r}_{\al'}) -  i \chi_g^\phi(\mbf{S}' + \mbf{r}_{\be'}) \rp   \\ & \qquad \times \left[ \sum_{\al \be} D[g]_{\al' \al} t_{\al \be}(g^{-1}(\mbf{S} + \mbf{r}_{\al'}) - g^{-1}(\mbf{S}' + \mbf{r}_{\be'}))  D^\dag[g]_{\be' \be} \ket{\mbf{S}, \al} \bra{\mbf{S}', \be} \right] \ . 
\eea
We have assumed that $[g, H^{\phi = 0}] = 0$, i.e. that $g$ is a symmetry of the zero field Hamiltonian, so the bracketed portion of \Eq{eq:gh}, which is independent of $\phi$, is equal to $t_{\al' \be'}(\mbf{S} - \mbf{S}')$ because $[g^{\phi =0}, H^{\phi = 0}]$. Relabeling the sums, we find
\bea
\label{eq:ghg}
g^\phi H^\phi (g^\phi)^\dag &=  \sum_{\mbf{R} \mbf{R}' \al \be} t_{\al \be}(\mbf{R} - \mbf{R}') \exp \lp i \int_{g^{-1}(\mbf{R}' + \mbf{r}_{\be})}^{g^{-1}(\mbf{R} + \mbf{r}_{\al})} \mbf{A}(\mbf{r}) \cdot d\mbf{r} + i \chi_g^\phi(\mbf{R} + \mbf{r}_{\al}) -  i \chi_g^\phi(\mbf{R}' + \mbf{r}_{\be}) \rp \ket{\mbf{R}, \al} \bra{\mbf{R}', \be} \ .
\eea
We now study the integral in the exponent of \Eq{eq:ghg}. We change variables to $\mbf{r} = g^{-1}\mbf{s}$ (here $g$ is understood to act on the vector $\mbf{r}$) to find
\bea
\label{eq:intchange}
\int_{g^{-1}(\mbf{R}' + \mbf{r}_{\be})}^{g^{-1}(\mbf{R} + \mbf{r}_{\al})} \mbf{A}(\mbf{r}) \cdot d\mbf{r} &= \int_{\mbf{R}' + \mbf{r}_{\be}}^{\mbf{R} + \mbf{r}_{\al}} \mbf{A}(g^{-1}\mbf{s})\cdot g^{-1} d\mbf{s} \\
&= \int_{\mbf{R}' + \mbf{r}_{\be}}^{\mbf{R} + \mbf{r}_{\al}} g \mbf{A}(g^{-1}\mbf{r})\cdot d\mbf{r}  \\
\eea
where the final integral, upon invoking our assumption that the Peierls paths are $g$-symmetric, is to be evaluated along the Peierls path between $\mbf{R}' + \mbf{r}_{\be}$ and $\mbf{R} + \mbf{r}_{\al}$. Now we are ready to discuss the behavior of different kinds of symmetries (rotations and reflection, unitary and anti-unitary) in the presence of magnetic field. 

\subsubsection{Unitary Rotations}
\label{app:Cnchi}

Intuitively, because a transverse magnetic field does not break $C_n$ symmetry (although $\mbf{A} $ generically does), we expect $g^\phi H^\phi (g^\phi)^\dag = H^\phi$, where $g^{\phi=0} = C_n$. We consult \Eqs{eq:ghg}{eq:intchange} and find that for $[C^\phi_n, H^\phi] = 0$, we must require
\bea
\label{eq:chiint}
\int_{\mbf{R}' + \mbf{r}_{\be}}^{\mbf{R} + \mbf{r}_{\al}} C_n \mbf{A}(C_n^{-1}\mbf{r}) \cdot d\mbf{r} + \chi_{C_n}^\phi(\mbf{R} + \mbf{r}_{\al}) -  \chi_{C_n}^\phi(\mbf{R}' + \mbf{r}_{\be}) &= \int_{\mbf{R}' + \mbf{r}_\be}^{\mbf{R} + \mbf{r}_\al} \mbf{A}(\mbf{r}) \cdot d\mbf{r} \ .
\eea
So far this calculation has proceeded in a general gauge. We specialize now to the symmetric gauge $\mbf{A}(\mbf{r}) = \mathcal{A}(\mbf{r})$ where $\mathcal{A}(\mbf{r}) = \frac{\phi}{2} (-y,x)$. For convenience, we have chosen the origin of $\mathcal{A}(\mbf{r})$ and the rotation center of $C_n$ to be the same which is just a choice of gauge. We emphasize that the rotation center (Wyckoff position) \emph{does not} have to be the location of an orbital of the model. 

Noting that the integrals are along the same path in \Eq{eq:chiint}, we find that the requirement for $[C_n^\phi, H^\phi] = 0$ in the symmetric gauge is
\bea
\label{eq:chisteps}
 \chi_{C_n}^\phi(\mbf{R} + \mbf{r}_{\al}) -  \chi_{C_n}^\phi(\mbf{R}' + \mbf{r}_{\be}) &= \int_{\mbf{R}' + \mbf{r}_\be}^{\mbf{R} + \mbf{r}_\al} [ \mathcal{A}(\mbf{r}) - C_n \mathcal{A}(C_n^{-1}\mbf{r}) ] \cdot d\mbf{r}  &= 0  \\
 \eea
where the second equality follows because the symmetric gauge is symmetric: 
\bea
\null [C_n \mathcal{A}(C_n^{-1}\mbf{r}) ]_i = \frac{\phi}{2} [C_n]_{ii'} \eps_{i'j'} [C_n^{-1}]_{j' j} r_j = \frac{\phi}{2} \eps_{i'j'} [C_n]_{ii'} [C_n]_{jj'} r_j = \det C_n \frac{\phi}{2} \eps_{ij} r_j = [\mathcal{A}(\mbf{r})]_i  \\
\eea
where we used $C_n^T = C_n^{-1}$ and $\det C_n = +1$. We see that defining $\chi^\phi_{C_n}(\mbf{r} ) = 0$ is a solution to \Eq{eq:chiint}. Thus we find that in this gauge choice, $C_n$ remains a symmetry of $H^\phi$ without modification. For a general gauge, $C_n$ remains a symmetry but $\chi^\phi_{C_n}(\mbf{r} )$ must be nontrivial. To see this, we observe that in a general gauge $\mbf{A}$, \Eq{eq:chisteps} reads
\bea
\label{eq:chistepsgen}
 \chi_{C_n}^\phi(\mbf{R} + \mbf{r}_{\al}) -  \chi_{C_n}^\phi(\mbf{R}' + \mbf{r}_{\be}) &= \int_{\mbf{R}' + \mbf{r}_\be}^{\mbf{R} + \mbf{r}_\al} [ \mbf{A}(\mbf{r}) - C_n \mbf{A}(C_n^{-1}\mbf{r}) ] \cdot d\mbf{r} 
 \eea
which \emph{does} have a solution because $ \pmb{\nabla} \times [ \mbf{A}(\mbf{r}) - C_n \mbf{A}(C_n^{-1} \mbf{r}) ] = \phi - \phi  = 0$. Thus $\mbf{A}(\mbf{r})$ and $C_n \mbf{A}(C_n^{-1} \mbf{r})$ are related by a gauge transformation, so $\mbf{A}(\mbf{r}) =  C_n \mbf{A}(C_n^{-1} \mbf{r}) +  \pmb{\nabla} \la(\mbf{r})$. We now see that $\chi_{C_n}^\phi(\mbf{r}) = \la(\mbf{r})$ solves \Eq{eq:chistepsgen}. 

Although we have constructed an expression for $C_n^\phi$ in a general gauge at any flux, it is much more convenient to work in the symmetric gauge where the operator $C_n$ is independent of the flux. All conclusions we deduce about the spectrum in this gauge choice must be valid for any gauge choice because the spectrum is gauge invariant. 

\subsubsection{Unitary Mirrors}
\label{app:mirrorsymcalc}

Contrary to the unitary rotations, we will show now that unitary mirrors are broken by $\phi$ since they reverse the flux. We again work in the symmetric gauge centered at the origin. We will now show that 
\bea
\label{eq:tildegHg}
M_i^\phi H^\phi (M_i^{\phi})^\dag = H^{-\phi},  \\
\eea
so mirror symmetries are broken when $\phi \neq 0$. Working in the symmetric gauge, we use \Eq{eq:ghg} to evaluate the LHS of \Eq{eq:tildegHg} and find that $\chi_{M_i}$ must obey
\bea
\label{eq:chiintM}
\int_{\mbf{R}' + \mbf{r}_{\be}}^{\mbf{R} + \mbf{r}_{\al}} M_i \mathcal{A}(M_i^{-1} \mbf{r}) \cdot d\mbf{r} + \chi_{M_i}^\phi(\mbf{R} + \mbf{r}_{\al}) -  \chi_{M_i}^\phi(\mbf{R}' + \mbf{r}_{\be}) &= - \int_{\mbf{R}' + \mbf{r}_\be}^{\mbf{R} + \mbf{r}_\al} \mathcal{A}(\mbf{r}) \cdot d\mbf{r} 
\eea
with the minus sign on the RHS differing from \Eq{eq:chiint}. Using $M_{i} \mathcal{A}(M_{i}^{-1} \mbf{r})  = - \mathcal{A}(\mbf{r})$ which follows from
\bea
\null [M_i \mathcal{A}(M_i^{-1}\mbf{r}) ]_i = \frac{\phi}{2} [M_i]_{ii'} \eps_{i'j'} [M_i^{-1}]_{j' j} r_j = \frac{\phi}{2} \eps_{i'j'} [C_n]_{ii'} [C_n]_{jj'} r_j = \det C_n \frac{\phi}{2} \eps_{ij} r_j = [\mathcal{A}(\mbf{r})]_i  \\
\eea

 Hence we find that \Eq{eq:tildegHg} is satisfied for $\chi^\phi_{M_i}( \mbf{r}) = 0$. We see that in the symmetric gauge, $M_i$ is independent of $\phi$ and obeys $M_i H^\phi M_i^\dag = H^{-\phi}$.  This completes our definition of the unitary symmetries in flux.

\subsubsection{Time Reversal Symmetry}
\label{app:TRS}

We now discuss time reversal symmetry $\mathcal{T}$. In zero field, we define its position space representation as
\bea
\mathcal{T} &= \sum_{\mbf{R} \al} D[\mathcal{T}]_{\al \al'} \ket{\mbf{R}, \al'} \bra{\mbf{R} \al} K  \\
\eea
where $K$ is the complex conjugation operator and $D[\mathcal{T}]_{\al \al'}$ is nonzero only is $\mbf{r}_\al = \mbf{r}_{\al'}$. If $\mathcal{T}$ is a symmetry of $H^{\phi=0}$, we have that
\bea
D[\mathcal{T}]_{\al \al'} t^*_{\al' \be}(\mbf{R} - \mbf{R}') D^{-1}[\mathcal{T}]_{\be \be'} &=  t_{\al \be'}(\mbf{R} - \mbf{R}') \ . 
\eea
Using this expression, we act on the Hofstadter Hamiltonian and find
\bea
\mathcal{T} H^{\phi} \mathcal{T}^{-1} &= \sum_{\mbf{R} \mbf{R}' \al \be} D[\mathcal{T}]_{\al \al'} t^*_{\al' \be}D^{-1}[\mathcal{T}]_{\be \be'} (\mbf{R} - \mbf{R}') e^{-i \int_{\mbf{R}' + \mbf{r}_\be}^{\mbf{R} + \mbf{r}_\al} \mbf{A}(\mbf{r}) \cdot d\mbf{r}} \ket{\mbf{R}, \al} \bra{\mbf{R}', \be} \\
&= H^{-\phi} ,
\eea
so we see that $\mathcal{T}$ has a simple action on $H^\phi$ without any additional $\chi$ factors. We may then form the operators $C_n \mathcal{T}, M_i \mathcal{T}$ which obey
\bea
\label{eq:gTprop}
C_n \mathcal{T} H^{\phi} (C_n \mathcal{T})^{-1} &= H^{-\phi},  \\
M_i \mathcal{T} H^{\phi} (M_i \mathcal{T})^{-1} &= H^{\phi} \ . \\
\eea
We find that anti-unitary rotations $C_n \mathcal{T}$ reverse the sign of the flux and so are not symmetries at $\phi \neq 0$, whereas anti-unitary reflections $M_i \mathcal{T}$ reverse the sign twice, and so are preserved at all flux. We only consider point group symmetries in this work, and leave non-symmorphic symmetries for future work. 

\section{Hofstadter Realizations of Projective Point Groups}
\label{app:gammaprop}

In this Appendix, we show that the Hofstadter periodicity $H^{\phi + \Phi} = U H^\phi U^\dag$ allows for reentrant symmetries  $U C_n \mathcal{T},U M_i,U \mathcal{T}$ at $\phi = \Phi/2$. These symmetries can realize projective representations of the PGs characterized by nontrivial Schur multipliers. In this section, we derive expressions for the Schur multipliers $\gamma_\mbf{x}$ protected by $C_n$ symmetry at a Wyckoff position $\mbf{x}$. Next, we check that $\gamma_\mbf{x}$ is quantized to the values $n \gamma_\mbf{x} = 0 \mod 2\pi$ and we show that $\gamma_\mbf{x} = 0$ if $\mbf{x}$ is an atomic site. Finally, we determine the projective algebra of the mirror symmetries.

\subsection{Peierls Symmetry Indicators protected by $C_n$}
\label{app:gammacalc}
\label{app:atomiclimit}

Let $\mbf{x}$ be a high-symmetry Wyckoff position with site symmetry group $G_\mbf{x}$ such that $C_n \in G_{\mbf{x}}$, i.e $C_n$ is a rotation centered at $\mbf{x}$ so $C_n \mbf{x} = \mbf{x}$. We keep $\mbf{x}$ fixed throughout this section $\gamma_\mbf{x}$
 We define the Schur multiplier $\gamma_\mbf{x}$ appearing in the algebra of $C_n$ and $U$ by
\bea
\label{eq:appgammadef}
C_n U C_n^\dag &= e^{-i \gamma_{\mbf{x}}} U 
\eea
which shows the non-commutation of the rotation $C_n$ and $U$. \Eq{eq:appgammadef} holds in any gauge, but since $C_n$ depends on $\phi$ in a generic gauge choice, it is simplest to calculate $\gamma_{\mbf{x}}$ from \Eq{eq:appgammadef} in the symmetric gauge centered at the point $\mbf{x}$ as discussed in \App{app:chicalc}: $\mathcal{A}(\mbf{r}) = \frac{\phi}{2} \hat{z} \times (\mbf{r}-\mbf{x})$. It emphasize that $\gamma_{\mbf{x}}$ can be computed at each Wyckoff position in the unit cell with a $C_n$ symmetry (whether or not there is an orbital of the Hamiltonian there). We will show that $\gamma_{\mbf{x}}$ depends only on the Peierls paths of the model. In particular, $\gamma_{\mbf{x}}$ does not depend on the electron ground state or the positions of the Wannier functions. 

For generality's sake, we prove \Eq{eq:appgammadef} using the many-body expression for $U$ determined by \Eq{eq:defU} to be
\bea
U &= \exp \lp i  \sum_{\mbf{R} \al} c^\dag_{\mbf{R},\al}  c_{\mbf{R},\al} \int_{\mbf{r}_0}^{\mbf{R} + \mbf{r}_\al} \mathcal{A}^{\Phi}(\mbf{r}) \cdot d\mbf{r} \rp,   \quad \pmb{\nabla} \times \mathcal{A}^{\Phi} = \Phi  \\
\eea
and the integral is taken along Peierls paths, but is independent of the specific path because $\pmb{\nabla} \times \mathcal{A}^{\Phi} = \Phi$. First we recall that
\bea
C_n^\dag c^\dag_{\mbf{R},\al} C_n = \sum_\al D[C_n]_{\al' \al} c^\dag_{C_n(\mbf{R}+ \mbf{r}_{\al}) -  \mbf{r}_{\al'},\al'}
\eea
which is the definition of the symmetry $C_n$ on the electron operators, and  $D[C_n]_{\al' \al}$ is nonzero only if $C_n \mbf{r}_{\al} = \mbf{r}_{\al'} \mod \mbf{a}_i$. Now we can evaluate
\bea
\label{eq:UCU1}
C_n U C_n^{\dag} &= \exp \lp i  \sum_{\mbf{R} \al} C_n c^\dag_{\mbf{R},\al} c_{\mbf{R},\al} C_n^\dag \int_{\mbf{r}_0}^{\mbf{R} + \mbf{r}_\al} \mathcal{A}^{\Phi}(\mbf{r}) \cdot d\mbf{r} \rp \\
&= \exp \lp i  \sum_{\mbf{R} \al \al' \al''} c^\dag_{C_n (\mbf{R} + \mbf{r}_{\al}) - \mbf{r}_{\al'},\al'} \left[  D[C_n]_{\al' \al}  \int_{\mbf{r}_0}^{\mbf{R} + \mbf{r}_\al} \mathcal{A}^{\Phi}(\mbf{r}) \cdot d\mbf{r} \, D^\dag[C_n]_{\al \al''} \right] c_{C_n(\mbf{R} +\mbf{r}_{\al}) - \mbf{r}_{\al''},\al''}   \rp \ . \\
\eea
To simplify this, we shift $C_n (\mbf{R} + \mbf{r}_{\al}) - \mbf{r}_{\al'} \to \mbf{R}_{\al'}$. Additionally, $\mbf{r}_{\al'} = \mbf{r}_{\al''}$ when $D[C_n]_{\al' \al} D^\dag[C_n]_{\al \al''} \neq 0$. Then \Eq{eq:UCU1} can be rewritten as
\bea
C_n U C_n^{\dag} &= \exp \lp i  \sum_{\mbf{R}_{\al'}  \al' \al''} c^\dag_{\mbf{R}_{\al'},\al'} \left[ \sum_{\al} D[C_n]_{\al' \al}  \, D^\dag[C_n]_{\al \al''} \right] c_{\mbf{R}_{\al'},\al''}  \int_{\mbf{r}_0}^{C^{-1}_n(\mbf{R}_{\al'} + \mbf{r}_{\al'})} \mathcal{A}^{\Phi}(\mbf{r}) \cdot d\mbf{r} \rp \\
&= \exp \lp i  \sum_{\mbf{R}_{\al} \al} c^\dag_{\mbf{R}_{\al},\al} c_{\mbf{R}_{\al},\al}  \int_{\mbf{r}_0}^{C^{-1}_n(\mbf{R}_{\al} + \mbf{r}_{\al})} \mathcal{A}^{\Phi}(\mbf{r}) \cdot d\mbf{r} \rp \\
&= \exp \lp i  \sum_{\mbf{R} \al} c^\dag_{\mbf{R},\al} c_{\mbf{R},\al}  \int_{C_n \mbf{r}_0}^{\mbf{R} + \mbf{r}_{\al}} C_n \mathcal{A}^{\Phi}(C_n^{-1} \mbf{r}) \cdot d\mbf{r} \rp \\
\eea
where we have changed variables in the integral. In the symmetric gauge where $ C_n \mathcal{A}^{\Phi}(C_n^{-1} \mbf{r}) = \mathcal{A}^{\Phi}(\mbf{r})$, we find
\bea
C_n U C_n^{\dag} &= \exp \lp i  \sum_{\mbf{R} \al} c^\dag_{\mbf{R},\al} c_{\mbf{R},\al}  \int_{C_n \mbf{r}_0}^{\mbf{R} + \mbf{r}_{\al}} \mathcal{A}^{\Phi}(\mbf{r}) \cdot d\mbf{r} \rp \\
&= \exp \lp i  \sum_{\mbf{R} \al} c^\dag_{\mbf{R},\al} c_{\mbf{R},\al}  \left[ \int^{\mbf{r}_0}_{C_n \mbf{r}_0} \mathcal{A}^{\Phi}(\mbf{r}) \cdot d\mbf{r} + \int_{\mbf{r}_0}^{\mbf{R} + \mbf{r}_{\al}} \mathcal{A}^{\Phi}(\mbf{r}) \cdot d\mbf{r} \right] \rp \\
&= e^{-i \hat{N} \gamma_{\mbf{x}}} U \\
\eea
where $\hat{N}$ is the total particle operator, which we now set to 1 since our focus is on single-particle physics. The result is
\bea
\label{eq:gammaformula}
\gamma_{\mbf{x}} &= \int_{\mbf{r}_0}^{C_n\mbf{r}_0} \mathcal{A}^\Phi(\mbf{r}) \cdot d\mbf{r} \mod 2\pi \\
\eea
where we emphasize that $\mathcal{A}^\Phi(\mbf{r}) = \frac{\Phi}{2} \hat{z} \times (\mbf{r}-\mbf{x})$ and $C_n$ (with $C_n \mbf{x} = \mbf{x}$) are centered at $\mbf{x}$,  and $\mbf{r}_0 = \mbf{R} + \mbf{r}_\be$ is an arbitrary orbital position of the Hamiltonian, and the integral is taken along Peierls paths. \Eq{eq:gammaformula} appears to depend on $\mbf{r}_0$ and on the path of the integral, but we will prove momentarily that $\gamma_{\mbf{x}}$ is independent of both. We emphasize that $\gamma_{\mbf{x}}$ does depend on $\mbf{x}$, the Wyckoff position under consideration. We call $\gamma_{\mbf{x}}$ a Schur multiplier because it extends the usual PG representation to a projective representation, as will be discussed at length in \Sec{app:Luis}. 

\begin{figure}[h]
 \centering
\includegraphics[width=7cm]{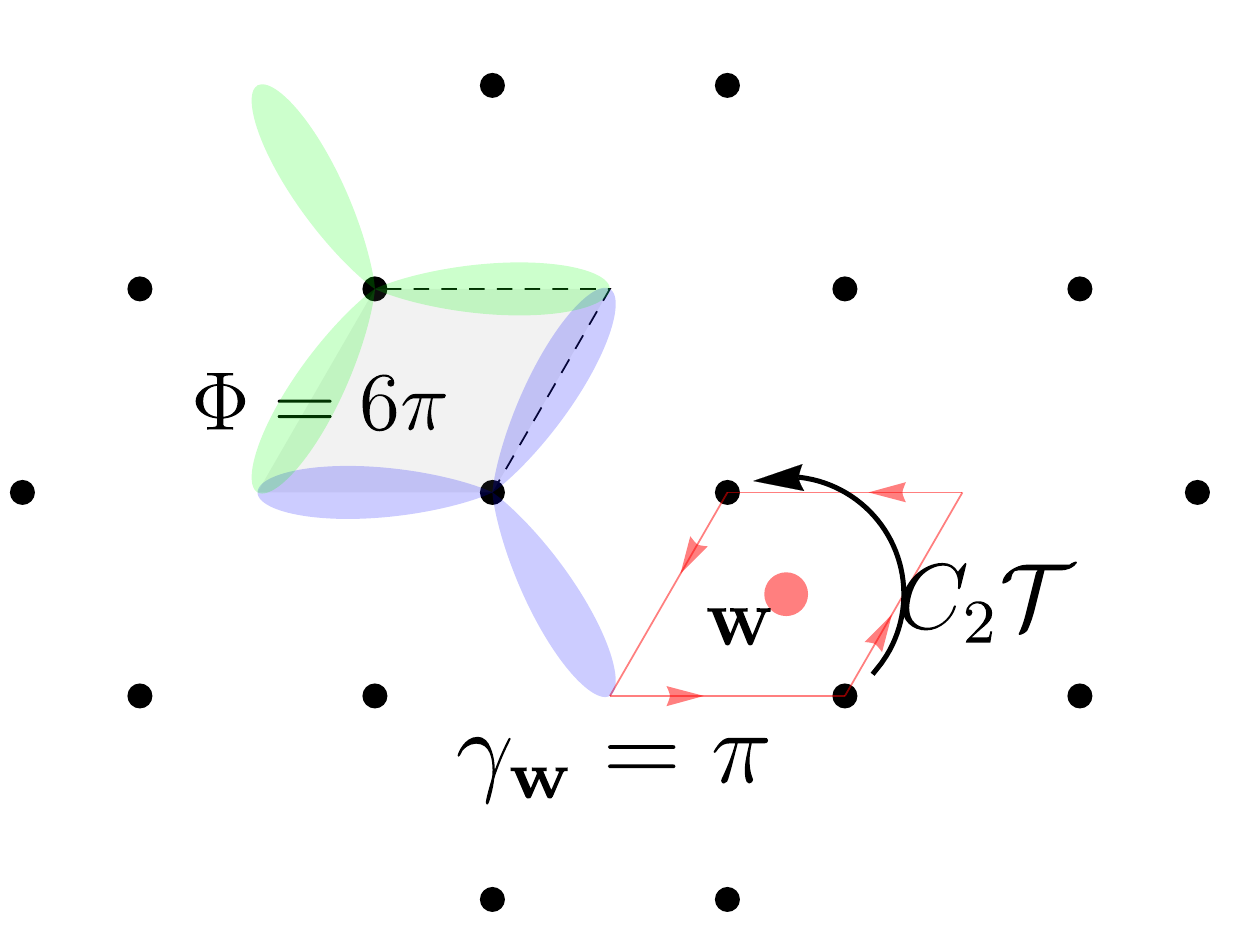} 
\caption{We show the orbitals and Peierls paths of a lattice model of twisted bilayer graphene. The Peierls paths yield a nontrivial Schur multiplier $\gamma_\mbf{w}$ for the $C_2 \mathcal{T}$ operator centered at the $\mbf{w}$ Wyckoff position.} 
\label{fig:TBGgamma}
\end{figure}

We now prove a few important properties of the Schur multipliers. First, $\gamma_\mbf{x}$ is independent of the sequence of Peierls paths chosen in the integral (\Eq{eq:gammaformula}). This follows immediately from the definition of $\pmb{\nabla} \times \mathcal{A}^\Phi = \Phi$: any closed loop on Peierls paths encloses $2\pi$ flux. So as discussed in \App{app:reviewhof}, \Eq{eq:gammaformula} can be computed along any sequence of Peierls paths without changing the value mod $2\pi$. Second, the Schur multiplier defined in \Eq{eq:gammaformula} is independent of $\mbf{r}_0$, the position of an arbitrary orbital of the Hamiltonian. Let $\mbf{r}_0'$ be a different arbitrary orbital of the Hamiltonian. Then
\bea
\int_{\mbf{r}_0'}^{C_n\mbf{r}_0'} \mathcal{A}^\Phi(\mbf{r}) \cdot d\mbf{r} \mod 2\pi &= \lp \int_{\mbf{r}_0'}^{\mbf{r}_0} +  \int_{\mbf{r}_0}^{C_n\mbf{r}_0} +  \int_{C_n \mbf{r}_0}^{C_n\mbf{r}_0'} \rp \mathcal{A}^\Phi(\mbf{r}) \cdot d\mbf{r}  \mod 2\pi \\
&= \lp \int_{\mbf{r}_0'}^{\mbf{r}_0} + \int_{C_n \mbf{r}_0}^{C_n\mbf{r}_0'} \rp \mathcal{A}^\Phi(\mbf{r}) \cdot d\mbf{r} + \gamma_{\mbf{x}} \mod 2\pi \\
&= \lp \int_{\mbf{r}_0'}^{\mbf{r}_0} - \int^{C_n \mbf{r}_0}_{C_n\mbf{r}_0'} \rp \mathcal{A}^\Phi(\mbf{r}) \cdot d\mbf{r} + \gamma_{\mbf{x}} \mod 2\pi \\
&= \gamma_{\mbf{x}} \mod 2\pi \\
\eea
where in the last line we have made use of the identity
\bea
 \int^{C_n \mbf{r}_0}_{C_n\mbf{r}_0'}  \mathcal{A}^\Phi(\mbf{r}) \cdot d\mbf{r}  &=  \int^{ \mbf{r}_0}_{\mbf{r}_0'}  \mathcal{A}^\Phi(C_n \mbf{r}) \cdot d(C_n \mbf{r}) = \int^{ \mbf{r}_0}_{\mbf{r}_0'}  C_n^{-1} \mathcal{A}^\Phi(C_n \mbf{r}) \cdot d\mbf{r}  =  \int^{\mbf{r}_0}_{\mbf{r}_0'}  \mathcal{A}^\Phi(\mbf{r}) \cdot d\mbf{r}  
\eea
making use of the symmetric gauge. Thus $\gamma_\mbf{x}$ is independent of the choice of $\mbf{r}_0$. The formula in \Eq{eq:gammaformula} does not appear to be gauge-invariant since the path of integration is not closed. We now provide a gauge-invariant formula for the Schur multiplier protected by $C_n$. Let $\mathcal{C}$ denote an arbitrary sequence of Peierls  paths from $\mbf{r}_0$ to $C_n \mbf{r}_0$. Using \Eq{eq:usefulidentity}, we find
\bea
\label{eq:gammagaugeinv}
\gamma_{\mbf{x}}  &= \frac{1}{n}\sum_{j=1}^n \int_{C_n^j \mathcal{C}}\mathcal{A}^\Phi(\mbf{r}) \cdot d\mbf{r} = \frac{1}{n} \oint_{\mathcal{C}_\mbf{x}}\mathcal{A}^\Phi(\mbf{r}) \cdot d\mbf{r} =  \frac{1}{n} \oint_{\mathcal{C}_\mbf{x}}\mbf{A}^\Phi(\mbf{r}) \cdot d\mbf{r} \mod 2\pi \\
\eea
where we have defined $\mathcal{C}_\mbf{x} = \sum_{j=1}^n C_n^j \mathcal{C}$ which is a closed, $C_n$-symmetric path (along Peierls paths) which encircles $\mbf{x}$ (since $C_n$ rotates $\mbf{r}_0$ around $\mbf{x}$). In the last equality, we replaced the symmetric gauge with a generic gauge $\mbf{A}$ because the path is closed. \Eq{eq:gammagaugeinv} has a very simple interpretation: the Schur multiplier is the fractional Aharanov-Bohm phase acquired by the electron in $1/n$th of a rotation. We now observe that $\gamma_{\mbf{x}}$ is quantized. Recall that any closed loop along Peierls path at $\phi = \Phi$ encloses a multiple of $2\pi$ flux. Hence from \Eq{eq:gammagaugeinv}, we see that  $\gamma_{\mbf{x}} \in \frac{2\pi}{n} \mathds{Z}_n$. Lastly, we observe that $\gamma_{\mbf{x}} = 0$ whenever $\mbf{x} = \mbf{R} + \mbf{r}_\be$ is an explicit orbital of the Hamiltonian, e.g. when $\mbf{x}$ is an atomic position. This is because we could choose $ \mbf{r}_0 = \mbf{x}$, so the path of integration in \Eq{eq:gammagaugeinv} is a single point. Using \Eq{eq:gammagaugeinv}, it is easy to compute $\gamma_\mbf{x}$ at any high-symmetry Wyckoff position $\mbf{x}$ with a $C_n$ symmetry. To give another example (see \Fig{fig:symmetrybehavior} of the Main Text), we consider the tight-binding model of twisted bilayer graphene in \Ref{PhysRevLett.123.036401}. The lattice is shown in \Fig{fig:TBGgamma} with the orbitals (black dots) at the corners of the unit cell. The Wannier functions \cite{2018PhRvX...8c1087K,2018PhRvX...8c1088K} of the continuum Bistritzer-MacDonald model \cite{2011PNAS..10812233B} are sketched in blue and green, and the Peierls paths (dashed) are chosen to follow their profile \cite{2018arXiv181111786L}. The smallest closed loop (grey) encircles a third of a unit cell, so $\Phi = 6\pi$. The rotation symmetries of the model at $C_2\mathcal{T}$ and $C_3$. The $C_2\mathcal{T}$ symmetry has a Schur multiplier determined by the $C_2$ operator:
\bea
C_2\mathcal{T} U (C_2\mathcal{T})^{-1} &= C_2\mathcal{T} U \mathcal{T} C_2^\dag = C_2U^\dag C_2^\dag = (C_2U C_2^\dag)^\dag  =  e^{i \gamma_\mbf{x}} U
\eea
where we used that $\mathcal{T} U \mathcal{T} = U^\dag$ \cite{firstpaper}. We will compute $\gamma_\mbf{x}$ at the three Wyckoff positions of the unit cell: 1a (the center of the unit cell), 2b (the corners of the hexagon), and 3c (the Kagome positions in the middle of the hexagon edges). We will compute $\gamma_{1a}$ corresponding to the $C_6 \mathcal{T}$ symmetry centered at the origin of the unit cell. Because the Peierls paths in \Fig{fig:TBGgamma} go through the 1a position perpendicular to $\mathcal{A}(\mbf{r})$ in the symmetric gauge, the integral \Eq{eq:gammagaugeinv} gives $\gamma_{1a} = 0$. At the 2b position (denoted by black dots), there is a $C_3$ symmetry. However, $\gamma_{2b} = 0$ because there is an orbital of the model at $2b$. As a check, one could also take a hexagonal path with the area of the unit cell centered at the $2b$ position. Then $\gamma_{2b} = \frac{1}{3} \Phi \mod 2\pi = 0 $. Lastly at the $3c$ position (the red dot in \Fig{eq:gammagaugeinv}), there is a $C_{2z}\mathcal{T}$ symmetry. We can compute $\gamma_{3c}$ from the $C_2$-symmetric path shown in red, which encloses $\frac{\Phi}{3} = 2\pi$ flux. Thus there is a nontrivial Schur multiplier with $\gamma_{\mbf{w}} = \frac{1}{2} \frac{\Phi}{3} = \pi \mod 2\pi$. 

\subsection{$U$-$M_i$ Algebra}
\label{eq:UMalg}

We now study the algebra of $M_i$ with $U$. We will see that although $U$ and $M_i$ do not commute in general, we will show that their algebra is not gauge invariant and can be ``lifted" to a non-projective representation, unlike the $C_n, U$ algebra in \App{app:gammacalc}. We will show explicitly that this is because mirrors reverse the flux and map $U$ to $U^\dag$. 

We proceed with a direct calculation. Following identically the steps of \Eq{eq:UCU1}, we arrive at
\bea
\label{eq:UMU1}
M_i U M_i^{\dag}  &= \exp \lp i  \sum_{\mbf{R} \al} c^\dag_{\mbf{R},\al} c_{\mbf{R},\al}  \int_{M_i \mbf{r}_0}^{\mbf{R} + \mbf{r}_{\al}} M_i \mathcal{A}^{\Phi}(M_i^{-1} \mbf{r}) \cdot d\mbf{r} \rp \ . \\
\eea
At this point, our calculation differs because mirror reverses the flux, e.g. $M_i \mathcal{A}^{\Phi}(M_i^{-1} \mbf{r}) = - \mathcal{A}^{\Phi}(\mbf{r})$. Hence we find
\bea
\label{eq:UMU2}
M_i U M_i^{\dag} &= \exp \lp - i  \sum_{\mbf{R} \al} c^\dag_{\mbf{R},\al} c_{\mbf{R},\al}  \int_{M_i \mbf{r}_0}^{\mbf{R} + \mbf{r}_{\al}} \mathcal{A}^{\Phi}(\mbf{r}) \cdot d\mbf{r} \rp \\
&= \exp \lp -i  \sum_{\mbf{R} \al} c^\dag_{\mbf{R},\al} c_{\mbf{R},\al}  \left[ \int^{\mbf{r}_0}_{M_i \mbf{r}_0} \mathcal{A}^{\Phi}(\mbf{r}) \cdot d\mbf{r} + \int_{\mbf{r}_0}^{\mbf{R} + \mbf{r}_{\al}} \mathcal{A}^{\Phi}(\mbf{r}) \cdot d\mbf{r} \right] \rp \\
&= e^{-i \xi_{\mbf{x},i}} U^\dag \\
\eea
where we have restricted to single-particle states and defined
\bea
 \label{eq:mirrorgamma}
\xi_{\mbf{x},i} &= \int_{\mbf{r}_0}^{M_i \mbf{r}_0} \mathcal{A}^\Phi(\mbf{r}) \cdot d\mbf{r} \mod 2\pi \ . \\
\eea
The major difference between \Eq{eq:mirrorgamma} and \Eq{eq:appgammadef} is that $M_i$ related $U$ to $U^\dag$. Note that $U$ is a unitary operator and can be rescaled by an overall phase. Taking $U \to e^{i \xi_{\mbf{x},i}/2} U$ removes the $e^{-i \xi_{\mbf{x},i}}$ phase in \Eq{eq:UMU2}. This rescaling has a physical meaning. Recall that $M_i$ reverses the flux so that $UM_i$ is a mirror symmetry at $\Phi/2$ flux. Then we check
\bea
(e^{i \xi_{\mbf{x},i}/2} UM_i)^2 &= e^{i \xi_{\mbf{x},i}} UM_i UM_i = e^{i \xi_{\mbf{x},i}} U M_i U M_i^{\dag} M_i^2 = M_i^2 \\
\eea
so in fact the rescaled operator $e^{i \xi_{\mbf{x},i}/2} UM_i$ has the same eigenvalues as $M_i$, and is the appropriate symmetry. In \App{app:Luis}, we will construct a general projective PG which may contain mirrors $e^{i \xi_{\mbf{x},i}/2} UM_i$ and rotation $C_n$. In this case, the PG will have a projective representation because of the algebra between $U$ and $C_i$. 

For completeness, we mention that the translation operators $T_i \in G$ also obey projective algebras with $U$ and can furnish projective \emph{space group} representations. Expressions for the projective phases in terms of $\Phi$ and the Peierls paths were worked out in App. D.2.b. of \Ref{firstpaper}. Because the Hofstadter topological indices we consider in this work are local to a given Wyckoff position, we do not need to consider the translation operators. 

\section{Construction of the Projective Groups}
\label{app:Luis}

In this Appendix, we derive the irreps of the projective point groups (PGs) which are realized in the Hofstadter Hamiltonian using a central extension method where the projective phases $e^{i \gamma_\mbf{x}}$ are consider as additional commuting elements in the group. We include tables for all 51 of the non-crystalline projective irreps with and without SOC and their Real Space invarians (RSIs). 

\subsection{Setup}

In order to make the discussion self-contained, we briefly review the symmetries of the Hamiltonian with and without flux. Throughout this discussion, we consider a single Wyckoff position and consider only the symmetries in its site-symmetry group. 

At $\phi = 0$, all possible PG symmetries are generated by $C_n, M$ and $\mathcal{T}$ denoting the $n$-fold rotation, mirror, and time-reversal (which is anti-unitary) respectively. These symmetries form the 31 magnetic point groups in 2D. (The conventional terminology ``magnetic" refers to the anti-unitary symmetries, not the magnetic flux.) As proved in \App{app:symflux}, $M$ and $\mathcal{T}$ reverse the magnetic field:
\bea
\mathcal{T}^{-1} H^\phi \mathcal{T} &= H^{-\phi}, \quad M_i^{\dag} H^\phi M_i &= H^{-\phi} \ . 
\eea
and $C_n$ and $M \mathcal{T}$ are preserved as symmetries at all flux. In the Peierls substitution, \App{app:reviewhof} demonstrated that $H^{\phi+\Phi} = U H^\phi U^\dag$ where $U$ is a unitary operator which is diagonal in the position basis. Now we consider the Hamiltonian at $\phi = \Phi/2$. Because $C_n$ and $M_i \mathcal{T}$ are symmetries at all flux, they are part of the symmetry group. In addition, there are reentrant symmetries: $U\mathcal{T}, UC_n \mathcal{T}, UM$. For instance, we check
\bea
(U \mathcal{T})^{-1} H^{\Phi/2} (U \mathcal{T}) &= \mathcal{T}^{-1} U^\dag H^{\Phi/2} U \mathcal{T}  = \mathcal{T}^{-1} H^{-\Phi/2} \mathcal{T} =  H^{\Phi/2} \\
\eea
Thus we find that the site-symmetry group (for a fixed Wyckoff position) at $\phi = \Phi/2$ is generated by
\bea
C_n, M \mathcal{T}, U\mathcal{T}, UC_n \mathcal{T}, UM
\eea
where the final three operators are the new, non-crystalline symmetries. We must now determine their group structure (dropping the spatial notation since we work at a fixed Wyckoff position), which we do using the relations derived in \App{app:gammaprop}:
\bea
\label{eq:PGpres}
U C_n = e^{i \gamma} C_n U, \quad UM = e^{i \xi} M U^\dag, \quad U\mathcal{T} = \mathcal{T} U^\dag
\eea
where $\gamma \in \frac{2\pi}{n} \mathds{Z}_n$ is the projective phase characterizing the representation. The zero-flux symmetries $C_n,M, \mathcal{T}$ obey their usual group structure, which is 
\bea
\null [\mathcal{T}, C_n] = [\mathcal{T}, M] = 0, \quad MC_nM^\dag = C_n^{\dag}
\eea
which holds with and without SOC.  Additionally, $C_n^n = M^2 = \mathcal{T}^2 = +1 \ (-1)$ without (with) SOC. The phase $\xi$ is gauge-\emph{dependent} as proved in \Eq{eq:UMalg}. To simply the group algebra, we take $U \to e^{i \xi/2} U$, which removes the $\xi$ factor from \Eq{eq:PGpres}. (To check that this rescaling does not interfere with the $U,\mathcal{T}$ algebra, recall that $\mathcal{T}$ is anti-unitary and can be rescaled by an arbitrary complex phase). Going forward in our discussion of the projective point groups at $\phi = \Phi/2$, we assume that $U$ is scaled such that $\xi = 0$. 

To name the projective PGs systematically, we use a notation matching the conventional PGs where a $C_n$ symmetry is notated with a $n$, mirror with $m$, and $\mathcal{T}$ with a prime ${}'$. For instance, the PG $2'm'm$ denotes the group generated by $C_2\mathcal{T}$ and $M$, which contains the anti-unitary mirror $M C_2\mathcal{T}$. In analogy, we denote the projective $C_n$ group relation by $n_\gamma$, e,g,  $2_\pi'm'm$ is generated by $UC_2\mathcal{T}$ and $UM$ where $\gamma = \pi$. Throughout, $\gamma$ refers to the projective phase of the $U$ and $C_n$ or $C_n\mathcal{T}$ algebra with the largest $n$ in the point group. For example, in $4_{\pi/2}mm$,  $C_4 U = e^{i \pi/2} C_4 U$. 

We now discuss the multiplication tables of the projective groups. Let us work in a convention where all symmetries with $U$ are ordered so that $U$ is the left-most operator. As an example, in $2_\pi1' = \{\pm1, \pm C_2, \pm UC_2\mathcal{T}\}$, we would write $C_2 \cdot U\mathcal{T} = - U C_2 \mathcal{T}$. This will allow us to systematically discuss the projective phases of operators.

Let $G$ be a conventional PG and $G_\gamma$ be the possibly projective PG formed from its symmetries at $\phi = \Phi/2$. The multiplication table of $G_\gamma$ can be written as a projective representation $D_\gamma[g], \ g \in G$ via
\bea
D_\gamma[C_n] &= C_n, \quad  D_\gamma[M \mathcal{T}] = M \mathcal{T} \\
D_\gamma[\mathcal{T}] &= U\mathcal{T}, \quad D_\gamma[C_n \mathcal{T}] = U C_n \mathcal{T}, \quad D_\gamma[M] = U M  \\
\eea
where we have employed the aforementioned rescalings. The projective group multiplication is given by
\bea
\label{eq:grpupmult}
D_\gamma[\mathcal{O}_1] D_\gamma[\mathcal{O}_2] = e^{i \theta_\gamma(\mathcal{O}_1, \mathcal{O}_2)} D_\gamma[\mathcal{O}_1 \mathcal{O}_2]  \ .
\eea
A definition of the Schur multiplier (also called the 2-cocycle) $\theta_\gamma(\mathcal{O}_1 \mathcal{O}_2)$ for each pair of operators defines the projective representation. Let us take $2_\gamma 1'$ as an example, with $\gamma= \pi$ the nontrivial projective representation:
\bea
D_\gamma[C_2] D_\gamma[\mathcal{T}] &= C_2 U\mathcal{T} =  e^{i \theta_\gamma(C_2, \mathcal{T})} D_\gamma[C_2 \mathcal{T}] =  e^{i \theta_\gamma(C_2, \mathcal{T})} U C_2 \mathcal{T} \\
\eea
so we find that $C_2 U\mathcal{T} = e^{i \theta_\gamma(C_2, \mathcal{T})} U C_2 \mathcal{T}$ which fixes $e^{i \theta_\gamma(C_2, \mathcal{T})} = - 1$.

We now calculate the Schur multiplier for the projective PGs defined by the Hofstadter symmetries in \Eq{eq:PGpres}. Only 4 kinds of elements appear in the groups: rotations, mirrors, anti-unitary rotations, and anti-unitary mirrors. The projective phase of different group elements may all be determined from \Eq{eq:PGpres}. We now define projective representations $D_\gamma[g]$ for all $g \in G$ with the following conventions
\bea
D_\gamma[C^m_n] &= C^m_n, \quad D_\gamma[C^m_n \mathcal{T}] = U C^m_n \mathcal{T}, \quad D_\gamma[M C_n^m] = e^{\frac{i}{2} m \gamma} U M C_n^m, \quad D_\gamma[M C_n^m \mathcal{T}] =  e^{-\frac{i}{2}m \gamma} M C_n^m \mathcal{T} \ .
\eea
We have chosen the phases in $D_\gamma[M C_n^m]$ so that
\bea 
D_\gamma[M C_n^m] D_\gamma[M C_n^m] &= e^{ im \gamma} U M C_n^m U M C_n^m = e^{ i m \gamma} e^{- i m \gamma}M C_n^m M C_n^m = M^2 \\
\eea
so that the mirror operators all square to $M^2 = \pm1$ with/without SOC. The choice of prefactor on $D[M C_n^m \mathcal{T}]$ is for convenience --- the overall factor of an anti-unitary symmetry is a matter of convention. We also choose a convention where the power of $C_n$ is taken to be positive by convention, i.e. a reflection about the line $y=-x$ is given by $M C^3_4$ where $M$ reflects about the $y$-axis. In the following formulae, it is important to use $C^3_4$ and not $C^{-1}_4$. To stress this, we define $[a]_n = a \mod n \in\{0,\dots, n-1\}$. Using these conventions, it is straightforward to compute $\theta$ from its definition in \Eq{eq:grpupmult} (we give examples afterwards):
\bea
\label{eq:projectiverelations}
\theta_\gamma(C_n^k, C_n^\ell) &= 0 \\
\theta_\gamma(C_n^k, C_n^\ell \mathcal{T}) &= - \gamma k \\
\theta_\gamma(C_n^k, M C_n^\ell) &= \frac{\gamma}{2} (\ell -2k - [\ell - k]_n) \\
\theta_\gamma(C_n^k, M C_n^\ell \mathcal{T}) &= \frac{\gamma}{2} (-\ell + [\ell - k]_n) \\
\\
\theta_\gamma(M C_n^k, C_n^\ell) &= \frac{\gamma}{2} (k - [k +\ell]_n) \\
\theta_\gamma(M C_n^k, C_n^\ell \mathcal{T}) &= \frac{\gamma}{2} (-k + [\ell+k]_n) \\
\theta_\gamma(M C_n^k, M C_n^\ell) &= \frac{\gamma}{2} (\ell - k) \\
\theta_\gamma(M C_n^k, M C_n^\ell \mathcal{T}) &= \frac{\gamma}{2} (k -\ell) \\
\\
\theta_\gamma(C_n^k \mathcal{T}, C_n^\ell) &= 0 \\
\theta_\gamma(C_n^k\mathcal{T}, C_n^\ell \mathcal{T}) &= \gamma k \\
\theta_\gamma(C_n^k \mathcal{T}, M C_n^\ell) &= \frac{\gamma}{2} (2k - \ell + [\ell - k]_n) \\
\theta_\gamma(C_n^k \mathcal{T}, M C_n^\ell \mathcal{T}) &= \frac{\gamma}{2} (\ell - [\ell - k]_n) \\
\\
\theta_\gamma(M C_n^k \mathcal{T}, C_n^\ell) &= \frac{\gamma}{2}( -k + [k+l]_n) \\ 
\theta_\gamma(M C_n^k \mathcal{T}, C_n^\ell \mathcal{T}) &= \frac{\gamma}{2}(k- [k+l]_n) \\
\theta_\gamma(M C_n^k \mathcal{T}, M C_n^\ell) &= \frac{\gamma}{2} (k - \ell)\\
\theta_\gamma(M C_n^k \mathcal{T}, M C_n^\ell \mathcal{T}) &= \frac{\gamma}{2} (\ell - k)\\
\eea

Using these formulae, the projective phase tables are fully determined. To be consistent about the sign, one must take $\ell, k>0$ in these expressions. 

We derive the first four lines in \Eq{eq:projectiverelations} to illustrate the method. The $C_n$ case is trivial:
\bea
D_\gamma[C^k_n] D_\gamma[C^\ell_n] &= e^{i \theta_\gamma(C^k_n, C^\ell_n)} D_\gamma[C^k_n C^\ell_n]  \\
C^k_n C^\ell_n &= e^{i \theta_\gamma(C^k_n, C^\ell_n)} D_\gamma[C^{k+\ell}_n] \\
C^{k+\ell}_n &= e^{i \theta_\gamma(C^k_n, C^\ell_n)} C^{k+\ell}_n \\
\eea
which shows $\theta_\gamma(C^k_n, C^\ell_n) = 0$. The first nontrivial case comes from $C^\ell_n\mathcal{T}$:
\bea
D_\gamma[C^k_n] D_\gamma[C^\ell_n\mathcal{T}] &= e^{i \theta_\gamma(C^k_n, C^\ell_n \mathcal{T})} D_\gamma[C^k_n C^\ell_n\mathcal{T}]  \\
C^k_n U C^\ell_n\mathcal{T} &= e^{i \theta_\gamma(C^k_n, C^\ell_n \mathcal{T})} D_\gamma[C^{k+\ell}_n\mathcal{T}]  \\
e^{- i k \gamma} U C^k_n C^\ell_n\mathcal{T} &= e^{i \theta_\gamma(C^k_n, C^\ell_n \mathcal{T})} U C^{k+\ell}_n\mathcal{T}  \\
\eea
from which we conclude $\theta_\gamma(C^k_n, C^\ell_n \mathcal{T}) = -\gamma k$. We now encounter the cases with mirror symmetry. We find
\bea
D_\gamma[C^k_n] D_\gamma[M C^\ell_n] &= e^{i \theta_\gamma(C^k_n, M C^\ell_n)} D_\gamma[C^k_n M C^\ell_n]  \\
C^k_n e^{\frac{i}{2} \ell \gamma} U M C^\ell_n &= e^{i \theta_\gamma(C^k_n, M C^\ell_n)} D_\gamma[M C^{\ell-k}_n]  \\
e^{\frac{i}{2} \ell \gamma - i k \gamma } U C^k_n M C^\ell_n &= e^{i \theta_\gamma(C^k_n, M C^\ell_n)} e^{\frac{i}{2} [\ell-k]_n \gamma } U M C^{\ell-k}_n \\
e^{\frac{i}{2} \ell \gamma - i k \gamma } U M C^{k-\ell}_n &= e^{i \theta_\gamma(C^k_n, M C^\ell_n)} e^{\frac{i}{2} [\ell-k]_n \gamma } U M C^{\ell-k}_n \\
\eea
from which we see that $\theta_\gamma(C^k_n, M C^\ell_n) = \frac{1}{2} \ell \gamma - k \gamma - \frac{1}{2} [\ell-k]_n \gamma = \frac{\gamma}{2} (\ell  - 2k - [\ell-k]_n)$. Lastly, we come to the anti-unitary symmetry:
\bea
D_\gamma[C^k_n] D_\gamma[M C^\ell_n \mathcal{T}] &= e^{i \theta_\gamma(C^k_n, M C^\ell_n \mathcal{T})} D_\gamma[C^k_n M C^\ell_n \mathcal{T}]  \\
e^{-\frac{i}{2}\ell \gamma} M C_n^{\ell-k} \mathcal{T}  &= e^{i \theta_\gamma(C^k_n, M C^\ell_n \mathcal{T})}  e^{-\frac{i}{2}[k-\ell] \gamma}  M C^{k-\ell}_n \mathcal{T}  \\
\eea
from which we derive $ \theta_\gamma(C^k_n, M C^\ell_n \mathcal{T}) = \frac{\gamma}{2}(-\ell +[k-\ell]_n)$. In this example we did not need to use the anti-unitarity of $\mathcal{T}$ because all complex phases were to the left of $\mathcal{T}$, but in general it is important to remember that $\mathcal{T}$ complex conjugates. 

\subsection{Calculation of the irreducible projective representations}
\label{sec:irrepconstruction}
The calculation of the projective representations of the 51 magnetic projective and, for completeness, the well-studied 31 ordinary (non-projective) magnetic plane PGs has been divided into two steps. First we have calculated the representations of the unitary 21 groups. In a second step, we have calculated the irreducible (co)representations of the remaining 61 anti-unitary groups, starting from the irreps of their unitary maximal subgroup. In the next two sections we describe the algorithm used in the calculation.

\subsubsection{Projective representations of the unitary groups}
\label{unitary}
For the calculation of the irreps of the projective groups without anti-unitary operators we use a central extension technique, where a larger ordinary (non-projective) group, called the \emph{central extension}, is defined and whose ordinary (vector) representations are calculated in a first step. The irreps of the central extension include the projective irreps of the original (projective) group and a set of spurious irreps that must be removed from the final set. There are spurious irreps because we have dramatically enlarged the group by considering e.g. $g= e^{i \gamma}$ as its own element of the group. To remove the spurious irreps, we require that $D[e^{i \gamma}] = e^{i \gamma}$ in a non-spurious irrep and we reject all irreps where this is not satisfied. In principle, this approach is identical to the approach used to construct the spinful ``double" groups where the central element ${}^d1$ is introduced as the $2\pi$ rotation in spin space. The resulting central extension has irreps where $D[{}^d1] = +1$ which are kept for the spinless groups and where $D[{}^d1] = -1$ which are kept for the spinful groups. We first provide the general construction and give an explicit example in \Eq{Luisexample}. We also give a few examples of a  physical but non-rigorous construction of the irreps in \App{app:RSIexamples}. 

We start by constructing the central extension of a given projective group. Let $G_{\gamma}$ be a projective finite group whose multiplication law is given by Eq. (\ref{eq:grpupmult}) (in the rest of this section we will ommit the subindex $\gamma$ in the notation of the group for simplicity). According to the relations (\ref{eq:projectiverelations}), in a PG that contains a $n$-fold rotational axis and $\gamma=2\pi m/n$, all the phases have rational values $\theta_{\gamma}(\mathcal{O}_1,\mathcal{O}_2)=2\pi j/N$ with $j=0,\ldots,N-1$, and the integer $N$ counts the number of projective phases that multiply each original group element. $N$ is given by the expression
\begin{equation}
\label{Nfac}
N=\frac{f\cdot n}{GCD(f\cdot n,[m]_n)}
\end{equation}
being $f=2$ for PGs that contain a mirror plane $M$ and $f=1$ for those PGs that do not contain a mirror plane. We check \Eq{Nfac} exhaustively from the values of $\theta_{\gamma}(\mathcal{O}_1,\mathcal{O}_2)$ in \Eq{eq:projectiverelations}. For instance in $2_\pi mm$ without SOC, the elements of the larger group obtained from the central extension are $1, -1, C_2, -C_2, UM_x, -UM_x, UM_y, -UM_y$ since $e^{i \gamma} = -1$ is considered as its own element (as opposed to a projective phase). Here $N=2$ because each original element $g$ yields $+g, -g$ in the central extension. 

$GCD(i,j)$ is the greatest common divisor of $i$ and $j$. Then we can define the \emph{central extension} $G_C$ as the set of elements $(\mathcal{O}_k,\frac{2\pi j}{N})$ with $\mathcal{O}_k\in G$ and $j=0,\ldots,N-1$. Under the operation,
\begin{equation}
\label{multcentral}
\left(\mathcal{O}_1,\frac{2\pi j_1}{N}\right)\cdot\left(\mathcal{O}_2,\frac{2\pi j_2}{N}\right)=\left(\mathcal{O}_1\mathcal{O}_2,\frac{2\pi j_1}{N}\pm\frac{2\pi j_2}{N}+\theta_{\gamma}(\mathcal{O}_1,\mathcal{O}_2) \mod {2\pi}\right)
\end{equation}
$G_C$ is an ordinary finite group defined by the group multiplication \Eq{multcentral}. On the right side, the sign that multiplies $\frac{2\pi j_2}{N}$ is $+1(-1)$ when $\mathcal{O}_1$ is a unitary (anti-unitary) operation.  Note that for $N=1$ the groups $G$ and $G_C$ are isomorphic and $G$ is thus an ordinary (non-projective) group. Note also that, except for $N=1$, the elements $\{(\mathcal{O}_k,0)\}$ with $\mathcal{O}_k\in G$ do not form a subgroup of $G_C$. We use the central extension technique because we can now use existing algorithms to determine the irreps of the centrally extended group. 

Because $G_C$ is a well-defined finite group, we can calculate its irreps following the standard \emph{induction} procedure used to calculate the irreps of the point and space groups (see for instance Ref. \cite{repres} or \Ref{doublerepres} for a complete introduction). Because the technique is standard, we provide review of the procedure here and an example in \App{Luisexample}. We refer the reader to \Ref{repres} for a complete treatment. 

In this section we will assume that $G$ is a double plane PG whose elements act on both orbital and spin space. The resulting set of irreps can be divided into two subsets: one subset of irreps are the relevant ones in systems without SOC (single-valued irreps) and a second subset corresponds to systems with SOC (double-valued irreps). Therefore, the consideration of $G$ as a double group allows the calculation of both types of irreps at once, and the single irreps are in 1:1 correspondence with the irreps of the corresponding single group.

In crystallographic point and space groups, and also in the central extension $G_C$ defined above, it is always possible to find a group-(normal) subgroup induction chain,
\begin{equation}
\label{chain}
G_C=H_0\rhd H_1\rhd \ldots\rhd H_{m-1}\rhd  H_m\rhd \ldots\rhd H_n 
\end{equation}
where the index $|H_{m-1}|/|H_m|$ is 2 or 3 for all $m$, being the first group of the chain $H_n$ an Abelian group. The irreps of $G_C$ can be obtained from the irreps of $H_n$ by applying $n-1$ times the induction procedure moving up along the group-subgroup chain (\ref{chain}). In this way, the irreps of $H_{m-1}$ are obtained from the irreps of its normal subgroup $H_m$ until we reach the top of the chain $G_C$.

In our case, the most obvious choice for $H_n$ as the starting point of the induction chain is the subgroup of $G_C$ formed by the set of operations, $\left(E,\frac{2\pi j}{N}\right)$ with $j=0,\ldots,N-1$. Since the subgroup $$\left(E,\frac{2\pi j}{N}\right)$$ is Abelian, the irreps of $H_n$ are 1-dimensional and, by choosing as generator of $H_n$ the element $\left(E,\frac{2\pi}{N}\right)$, the matrices of the $N$ different irreps for this operation are ($\rho^k$ labels the different irreps),
\begin{equation}
\label{irr:abel}
D_{\gamma}[\rho^k]\left(E,\frac{2\pi}{N}\right)=e^{i2\pi\frac{k}{N}}\hspace{1cm}k=0,\ldots,N-1
\end{equation}
The matrices of all the elements of $H_n$ are obtained as powers of these $1\times1$ matrices. In principle, all the irreps of the central extension $G_C$ can be calculated following the process that will be detailed in the next paragraphs. However, as stated above, the central extension $G_C$ enlarges the group $G$ by a factor $N$, i.e., $|G_C|=N|G|$ and thus it introduces spurious irreps. We define the \emph{physical} irreps $D_{\gamma}[\rho^k](g)$, $g\in G_C$ as those which satisfy $D_{\gamma}[\rho^k](E,\frac{2\pi j}{N})=e^{i2\pi\frac{j}{N}}$. This is satisfied only by one of the irreps in (\ref{irr:abel}) with $k=j$. Essentially, we require $D_\gamma[\rho](E,\gamma k) = e^{i \gamma k}$ and reject irreps which do not obey this relation. For instance, all finite groups contain the trivial irrep for which $D_\gamma[\rho^{\textrm{id}}](R,\frac{2\pi j}{N}) = 1$ for all $R$ and $j$. This irrep is unphysical when $\gamma \neq 0$ because it would map $D_\gamma[\rho^{\textrm{id}}](E, \gamma)$ into 1. Therefore, we only keep a single irrep in (\ref{irr:abel}) for the next steps of the calculation.

 In the induction procedure, in a group-subgroup pair $H\rhd M$, we first decompose the supergroup $H$ into coset representatives of the subgroup $M$,
\begin{equation}
\label{eq:cosets}
H=Mh_1 \cup  M h_2 \cup  \ldots \cup M h_n
\end{equation}
where the first representative $h_1\in M$ is usually taken as the identity, and for $i\neq1$, $h_i\in\mathcal{H}$ but $h_i\notin M$. As it has been mentioned  above, with a good election of the group-subgroup chain, in crystallographic PGs, and also in the central extension $G_C$ defined above, $n=2$ or $n=3$. When $n=3$ we make the election $h_3=h_2^2$.

In a next step, the irreps of $M$ are distributed into orbits under the supergroup $H$. Let be $\rho$ an irrep of $M$ and $D[\rho](m)$ the matrix of this irrep of the symmetry operation $m\in M$. As $M$ is a normal subgroup of $H$, the symmetry operations obtained through conjugation of $m\in M$ by the coset representatives $h_i$ in Eq. (\ref{eq:cosets}) belong to $M$, i.e., $h_i^{-1}mh_i\in M$. Therefore, the set of matrices $D[\rho](h_i^{-1}mh_i)$ form an irrep of $M$. It is said that this irrep belongs to the orbit of $\rho$ in $H$ and that all the irreps in the same orbit are mutually conjugated in $H$. Given an irrep $\rho$, two different kinds of orbits can result:
\begin{itemize}
	\item The irrep $\rho$ is self-conjugated in $H$, i.e., the set of matrices $D[\rho](h_i^{-1}mh_i)$ correspond to the same irrep as the set of matrices $D[\rho](m)$. In our case, as the index $|H|/|M|$ takes only the values 2 or 3, the irrep $\rho$ of $M$ induces a reducible representation $\rho\uparrow H$ in $H$ which is reduced into 2 (when $n=2$) or 3 (when $n=3$) irreps in $H$ of the same dimension of $\rho$.
	\item The orbit contains 2 (for $n=2$) or 3 (for $n=3$) different irreps. These 2 or 3 irreps $\rho^i$ of $M$ \emph{combine} to give a single irrep $\rho^1\uparrow H=\rho^2\uparrow H(=\rho^3\uparrow H)$ of $H$ whose dimension is 2 (for $n=2$) or 3 (for $n=3$) times the dimension of $\rho$.
\end{itemize}
Now we proceed to calculate the matrices of the induced irreps $\rho\uparrow H$ from the matrices of the irreps $\rho$ of $M$ in the previous two cases following Ref. \cite{repres}:
\begin{itemize}
	\item When the irrep $\rho$ of $M$ is self-conjugated, it induces thus $n=2,3$ irreps in $H$ of the same dimension of $\rho$. We denote these induced irreps as $\rho_h^i$ with $i=1,2$ or $i=1,2,3$. In all the induced irreps (2 or 3), the matrices of the symmetry operations $h\in M,H$ can be taken as $D[\rho](m)$, i.e., the same matrices of these operations in $\rho$, $D[\rho_h^i](m)=D[\rho](m)$ for all $i$. The matrices of the coset representative $h_2$ in \ref{eq:cosets} for $n=2$ are,
	\begin{equation}
	\label{coset2}
	D[\rho_h^1](h_2)=-D[\rho_h^2](h_2)=U
	\end{equation}
where the matrix $U$ fulfills the conditions,
	\begin{eqnarray}
	\label{coset2a}
	D[\rho](h_2^{-1}mh_2)&=&U^{-1}D[\rho](m)U\\
	\label{coset2b}
	U^2&=&D[\rho](h_2^2) \ .
	\end{eqnarray}
Note that, in the last relation, $h_2^2\in M$.
	
The matrices of the coset representative $h_2$ in \ref{eq:cosets} for $n=3$ are,
		\begin{equation}
		D[\rho_h^1](h_2)=\epsilon D[\rho_h^2](h_2)=\epsilon^2 D[\rho_h^3](h_2)=U\\
		\end{equation}
		with $\epsilon=e^{2\pi i/3}$ and the matrix $U$ fulfills these conditions,
	\begin{eqnarray}
	D[\rho](h_2^{-1}mh_2)&=&U^{-1}D[\rho](m)U\\
	U^3&=&D[\rho](h_2^3)
	\end{eqnarray}
	\item When two (for $n=2$) or three (for $n=3$) irreps $\rho^i$ are included in the same orbit though conjugation by $h_2$ (and $h_2^2$ for $n=3$), they induce a single irrep $\rho_h$. The matrices of the symmetry operations that belong to $M$ are the direct sum of the matrices of the irreps of $M$ in the same orbit, i.e., $D[\rho_h](m)=D[\rho^1](m)\oplus D[\rho^2](m)$ for $n=2$ and $D[\rho_h](m)=D[\rho^1](m)\oplus D[\rho^2](m)\oplus D[\rho^3](m)$ for $n=3$. The matrices of $h_2$ are,
	\begin{equation}
	\label{inducedn2}
	D[\rho_h](h_2)=\left(
	\begin{array}{cc}
	0&D[\rho^1](h_2^2)\\
	\mathbb{1}&0
	\end{array}\right)
	\end{equation}
	for $n=2$ and,
	\begin{equation}
	D[\rho_h](h_2)=\left(
	\begin{array}{ccc}
	0&0&D[\rho^1](h_2^3)\\
	\mathbb{1}&0&0\\
	0&\mathbb{1}&0
	\end{array}\right)
	\end{equation}
	for $n=3$, being $\mathbb{1}$ the identity matrix of the same dimension as $D[\rho^1]$.
\end{itemize}

We systematically apply the induction process described here to every group-subgroup pair in the chain \ref{chain}, thereby calculating the projective irreps of the central extension of the 21 unitary projective PGs, starting from the unique physically relevant irrep of $H_n$ at the beginning of the chain. We stress that we have always chosen the second group of the chain to be 
\begin{equation}
H_{n-1}=H_n(E,0)+H_n\,(^dE,0)
\end{equation}
being $^dE$ the symmetry operation of the double group that, according to the notation of \cite{doublerepres}, represents the identity in the orbital space and the inversion in the spin space (geometrically, it can be interpreted as a $2\pi$ rotation, that inverts the spin of the fermions). The single relevant irrep of $H_n$ induces two 1-dimensional irreps into $H_{n-1}$: one irrep for which $D[\rho^s](\,^dE,0)=1$ (single-valued irrep or no-SOC irrep in our context) and another one for which $D[\rho^d](\,^dE,0)=-1$ (double-valued irrep or SOC irrep). In the next steps of the induction chain single (double)-valued irreps induce always single (double)-valued irreps. 

Once the irreps of the central extension $G_C$ have been calculated, in the final step, and according to the following isomorphism between the subset $\{(R,0)\}$ of $G_C$ and $G$,
\begin{equation}
\label{isomor}
(R,0)\rightarrow R,
\end{equation}
for every projective irrep we assign the matrix of the symmetry operation $(R,0)$ to $R\in G$. The isomorphism \Eq{isomor} simply takes an element $e^{i \varphi} g \in G_C$ to $g \in G$ for all central elements $e^{i \varphi}, \varphi = \frac{2\pi k}{N}, k = 0,\dots, N-1$. 

\subsubsection{Projective representations of the anti-unitary groups}
\label{antiunitary}
We have followed the procedure described in ref. \cite{bradley}, based on Wigner's original works \cite{wigner1932,wigner1959}, in the calculation of the irreducible (co)representations of the 61 non-unitary (magnetic) plane groups $G$. For every group, we first construct the central extension $G_C$ of $G$ as described in the previous section and determine its unitary (maximal) subgroup $G_C^U$, being the group-subgroup index $|G_C|/|G_C^U|=2$. Following the procedure explained in the previous section, we calculate first the  physically relevant irreps of $G_C^U$.

\begin{figure}[h]
	\centering
	\includegraphics[scale=0.7]{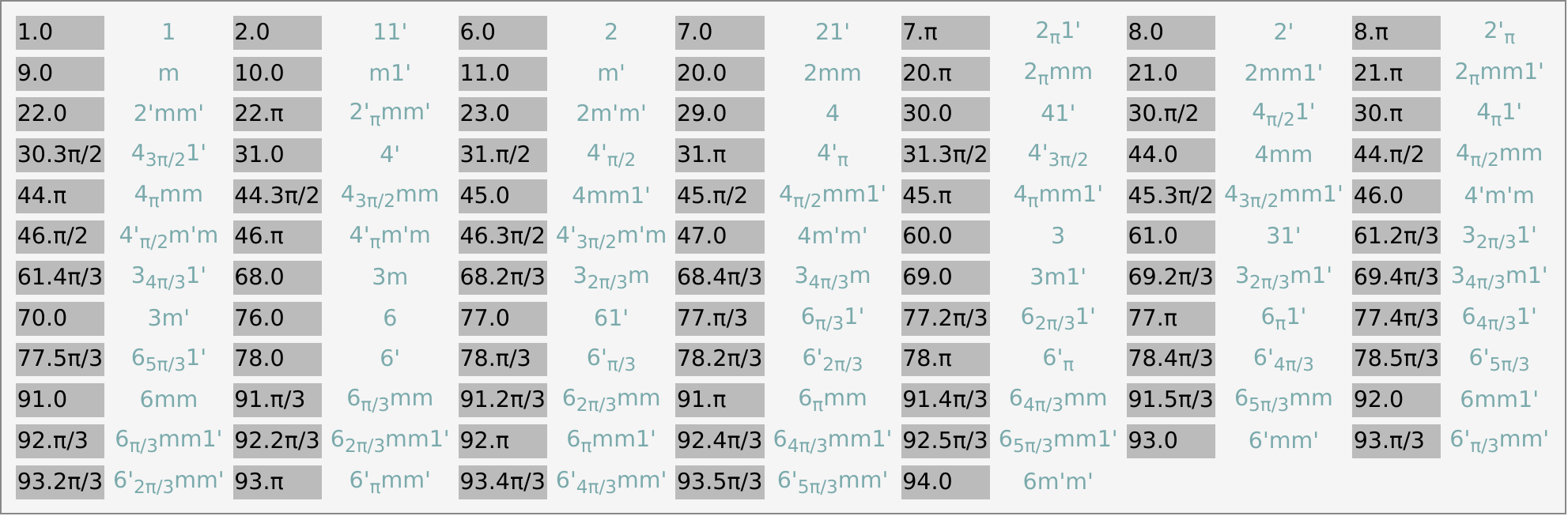} 
	\caption{Screenshot of the main input of the \href{http://www.cryst.ehu.es/cryst/projectiverepres}{ProjectiveRep PG} tool in the Bilbao Crystallographic Server.} 
	\label{fig:maininput}
\end{figure}

\begin{figure}[h]
	\centering
	\includegraphics[scale=0.6]{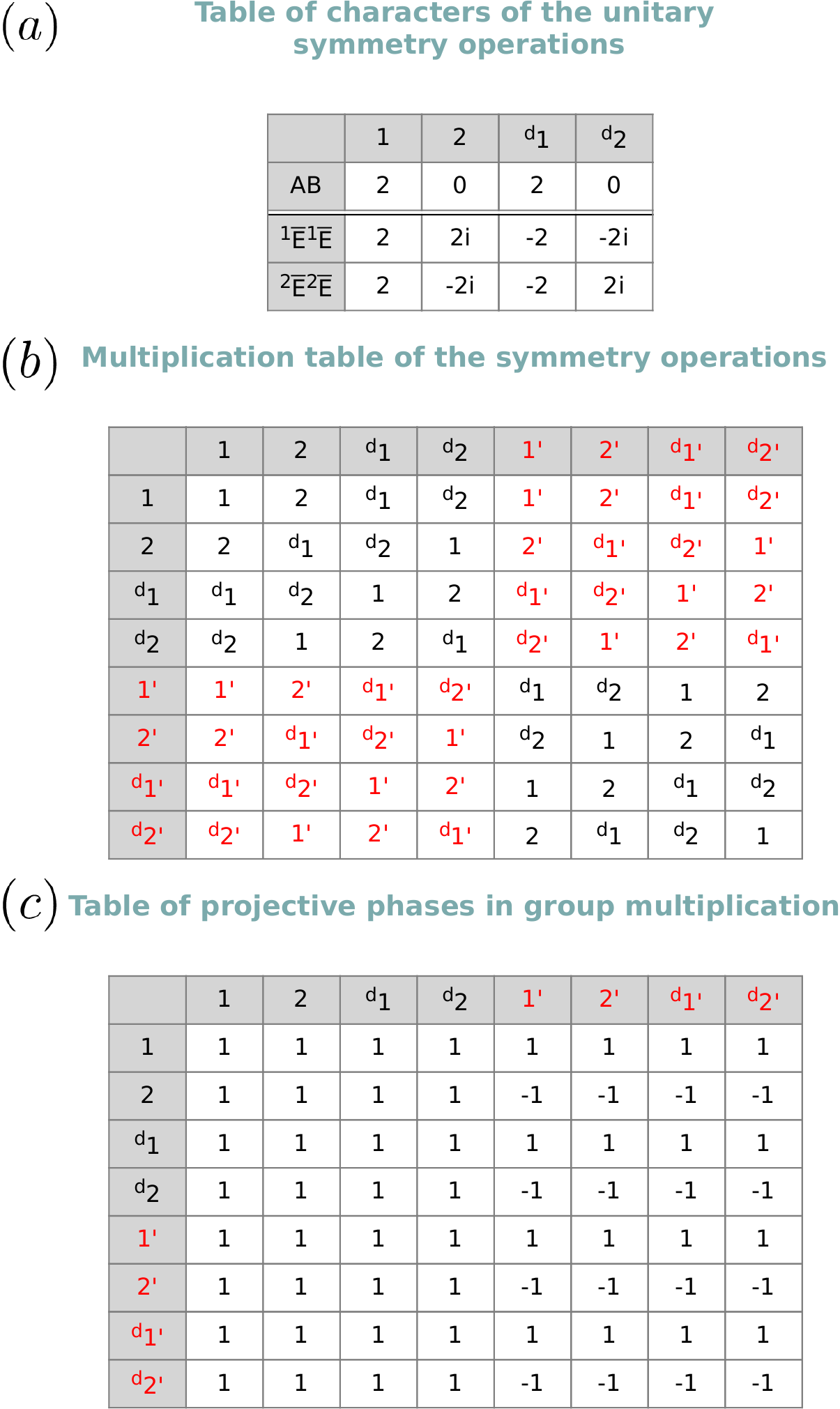} 
	\caption{Partial output of the tool \href{http://www.cryst.ehu.es/cryst/projectiverepres}{ProjectiveRep PG} when the projective plane PG $2_{\pi}'$ has been chosen in the main menu reproduced in \Fig{fig:maininput}. (a) Characters of the irreps for the unitary operations. The first column shows the notation of the irrep. (b) Cayley multiplication table of the group chosen. (c) Projective phases given by $\theta_\gamma(\mathcal{O}_1,\mathcal{O}_2)$ in eqs. (\ref{eq:projectiverelations}).} 
	\label{fig:rest}
\end{figure}

\begin{figure}[H]
	\centering
	\includegraphics[width=7cm]{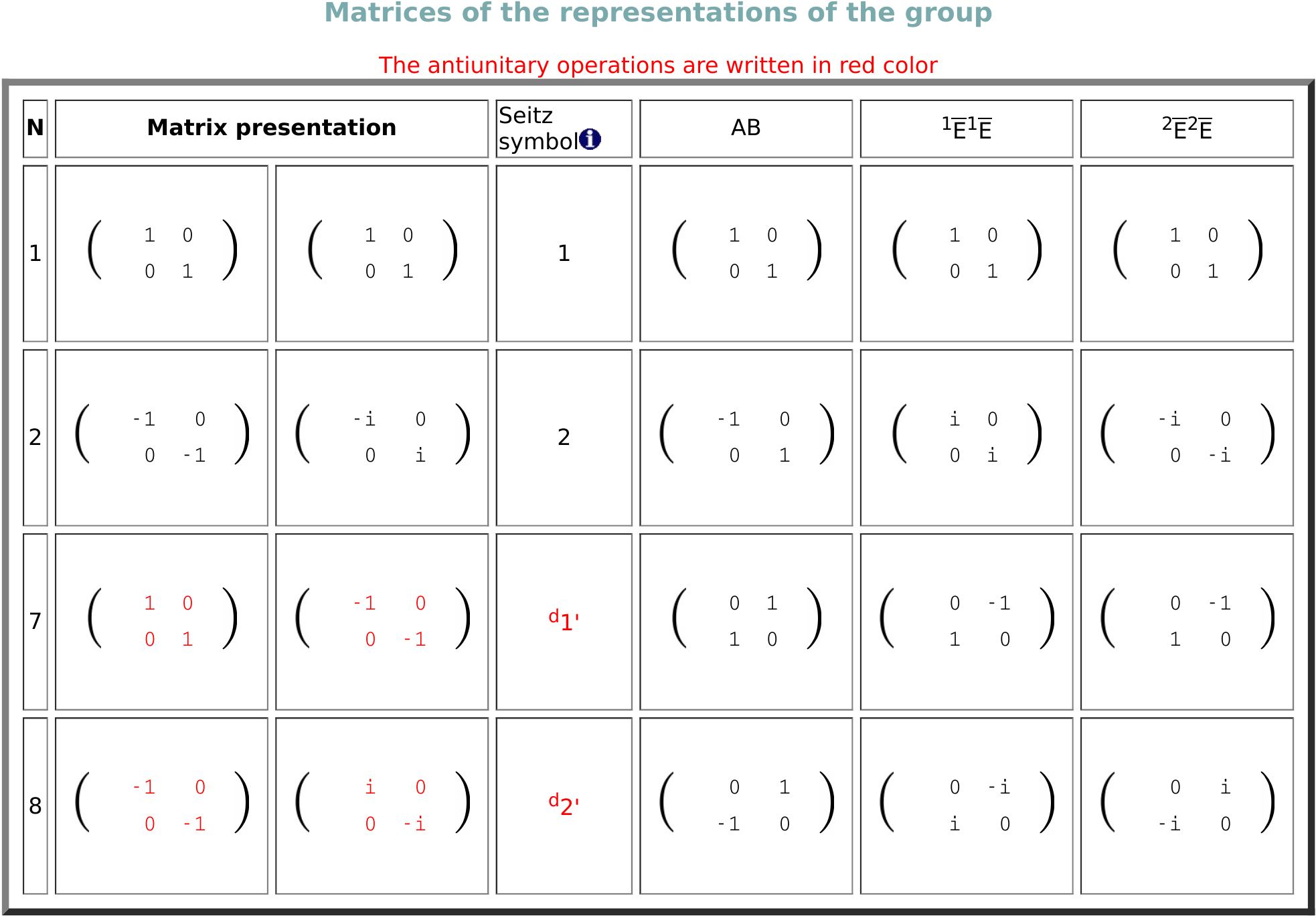} 
	\caption{Matrices of the representations for the projective plane PG $2_{\pi}1'$ given by the tool \href{http://www.cryst.ehu.es/cryst/projectiverepres}{ProjectiveRep PG}. The figure includes only the data for a selected subset of symmetry operations. The first and second columns give the matrices of the PG symmetry operation in the orbital and spin space, respectively, e.g. $\mathbb{1}, C_2, \mathcal{T}, C_2\mathcal{T}$ in row order. The third column gives the Seitz notation of the symmetry operation, where the\, "$d$" superscript was introduced in ref.  \cite{doublerepres} to distinguish the two operations with the same orbital matrix but that differ in the spin space. The rest of columns give the matrices of the representations.} 
	\label{matrices}
\end{figure}

In the next step, we choose an anti-unitary operation $g_A=(R',0)$ as a representative, where $R'$ represents an anti-unitary element of the point group $G$. If the time-reversal symmetry $\mathcal{T}$ belongs to $G$, we choose this operation as representative, $g_A=(\mathcal{T},0)$. Otherwise, we choose a two-fold axis $g_A=(C_2\mathcal{T},0)$ or a mirror plane $g_A=(M\mathcal{T},0)$. Following ref. \cite{bradley}, taking a given irrep $\rho$ of $G_C^U$ with matrices $D[\rho](g_U)$ of the symmetry operations $g_U\in G_C^U$, we calculate the following set of matrices for all $g_U\in G_C^U$, 
\begin{equation}
\label{conjugate}
D^{*}[\rho](g_A^{-1}g_Ug_A)
\end{equation}
As $g_A^{-1}g_Ug_A\in G_C^U$ because $g_A$ is anti-unitary, the set of matrices (\ref{conjugate}) are well defined, and they also form an irreducible representation $\rho'$ of $G_C^U$. Depending on the relation between the sets of matrices $\{D[\rho](g_U)\}$ and $\{D^{*}[\rho](g_A^{-1}g_Ug_A)\}$, we obtain three possible results, denoted as cases (a), (b) and (c) in ref. \cite{bradley}:
\begin{enumerate}
	\item[(a)] The sets of matrices $\{D[\rho](g_U)\}$ and $\{D^{*}[\rho](g_A^{-1}g_Ug_A)\}$ are equivalent. There exist a unitary matrix $\mathcal{U}$ such that 
	\begin{equation}
	\label{equivalence}
	D[\rho](g_U)=\mathcal{U}D^{*}[\rho](g_A^{-1}g_Ug_A)\mathcal{U}^{-1}\hspace{1cm}\forall g_U\in G_C^U
	\end{equation}
	and $\mathcal{U}\mathcal{U}^{*}=D[\rho](g_A^2)$. In this case, the representations are equivalent, $\rho\equiv\rho'$, the resulting (co)representation has the same dimension of $\rho$, and it is usually denoted by the same label $\rho$. The matrices $D'[\rho]$ of the (co)representation for the unitary elements are the same as the matrices of the original irrep $\rho$ of the unitary subgroup $G_C^U$, i.e., $D'[\rho](g_U)=D[\rho](g_U)$, and the matrix for the anti-unitary operation $g_A$ is $D'[\rho](g_A)=\mathcal{U}$. The matrices of the rest of anti-unitary elements are obtained by group multiplication.
	\item[(b)]  The sets of matrices $\{D[\rho](g_U)\}$ and $\{D^{*}[\rho](g_A^{-1}g_Ug_A)\}$ also fulfill eq. (\ref{equivalence}), but $\mathcal{U}\mathcal{U}^{*}=-D[\rho](g_A^2)$. In this case, the (co)representation $\rho'$ is the direct sum of two copies of the original irrep, $\rho'=\rho\oplus\rho$ and it is usually denoted as $\rho\rho$. The matrices of the unitary elements $g_U$ are the direct sum of $D[\rho](g_U)$ and $D^{*}[\rho](g_A^{-1}g_Ug_A)$, i.e., $D'[\rho\rho](g_U)=D[\rho](g_U)\oplus D^{*}[\rho](g_A^{-1}g_Ug_A)$, and the matrix of the anti-unitary element $g_A$ is,
	\begin{equation}
	\label{antimatrix}
	D'[\rho\rho](g_A)=\left(\begin{array}{cc}
	0&D[\rho](g_A^2)\\
	\mathbb{1}&0
	\end{array}\right)
	\end{equation}
	 The matrices of the rest of anti-unitary elements are obtained by group multiplication.
	\item[(c)] The sets of matrices $\{D[\rho](g_U)\}$ and $\{D^{*}[\rho](g_A^{-1}g_Ug_A)\}$ do not fulfill eq. (\ref{equivalence}). They correspond to different irreps of $G_C^U$, $\rho$ and $\sigma$, with $\rho\not\equiv\sigma$. The (co)representation is the direct sum of the two non-equivalent irreps and it is usually denoted as $\rho\sigma$. The matrices of the unitary operations are, as in case (b), the direct sum $D'[\rho\sigma](g_U)=D[\rho](g_U)\oplus D^{*}[\rho](g_A^{-1}g_Ug_A)$ and the matrix of the anti-unitary operation $g_A$ is also given by eq. (\ref{antimatrix}). The matrices of the rest of anti-unitary elements are obtained by group multiplication.
\end{enumerate}
The tables of irreducible representations of the 31 non-projective and the 51 projective groups have been implemented in the ~\href{http://www.cryst.ehu.es/cryst/projectiverepres}{ProjectiveRep PG}~tool~(\url{http://www.cryst.ehu.es/cryst/projectiverepres} in the Bilbao Crystallographic Server \cite{aroyo2006bilbao}.
\Fig{fig:maininput} presents the main menu of the program. Clicking on the symbol of the group, the program gives an output that contains four tables. We reproduce in \Fig{fig:rest} and \Fig{matrices} the output for the projective plane PG $2_{\pi}'$.
 \Fig{fig:rest}a gives the characters of the irreps for the unitary operations of the group, \Fig{fig:rest}b includes the Cayley multiplication table of the group and \Fig{fig:rest}c the projective phases (\ref{eq:projectiverelations}). Finally, \Fig{matrices} gives the matrices of the representations for the unitary and anti-unitary (in red) symmetry operations.

\subsection{Example: Calculation of the irreducible projective representations of the point group $2_{\pi}mm1'$}
\label{Luisexample}
As an example of the application of the procedure explained in section \ref{app:Luis} we calculate the projective representations of the anti-unitary point group $G=2_{\pi}mm1'$. According to the general procedure we first determine the irreps of the central extension $G_C^U$ of the maximal unitary subgroup $G^U=2_{\pi}mm$ and, using these irreps, we calculate the co-representations of the central extension $G_C$ of $G$. Finally, applying the isomorphism (\ref{isomor}), we get the projective co-representations of the anti-unitary group $G=2_{\pi}mm1'$.

The first step of the procedure explained in section \ref{antiunitary} is the definition of the central extension of the projective group $2_{\pi}mm$. The group contains a 2-fold axis ($n=2$) and  $\gamma=\pi$. Therefore  $m=1$ in the general definition of $\gamma=2\pi n/m$. As the group contains a mirror plane, $f=2$. These values give $N=4$ in equation (\ref{Nfac}). The elements of the central extension are thus,
\begin{equation}
(E,\pi j/2),(\,^dE,\pi j/2),(2_z,\pi j/2),(\,^d2_z,\pi j/2),(m_x,\pi j/2),(\,^dm_x,\pi j/2),(m_y,\pi j/2),(\,^dm_y,\pi j/2)
\end{equation}
with $j=0,1,2,3$. The point group operations $2_z$ and $2_z^d$ represent a 2-fold rotation arount the $z$ axis in the orbital space but differ in their action on the spin space, and $m_x(m_y)$ and $\,^dm_x(\,^dm_y)$ are the two operations that represent a mirror plane in the orbital space perpendicular to $x(y)$ but differ in their action on the spin space. 

Following the procedure of section \ref{unitary}, we set the group-subgroup chain \ref{chain} as,
\begin{eqnarray}
\label{example:chain}
H_3&=&(E,0),(E,\pi/2),(E,\pi),(E,3\pi/2)\nonumber\\
H_2&=&H_3(E,0)\cup H_3(\,^dE,0)\\
H_1&=&H_2(E,0)\cup H_2(2_z,0)\nonumber\\
G_C^U=H_0&=&H_1(E,0)\cup H_1(m_x,0)\nonumber
\end{eqnarray}
$H_3$ is an abelian cyclic group with 4 elements whose generator can be chosen as $(E,\pi/2)$. The matrices of the 4 irreps for the generator chosen are,
\begin{equation}
D_{\pi}[\rho^k](E,\pi/2)=e^{i\pi k/2},\hspace{0.5cm}k=0,1,2,3
\end{equation}
However, only the irrep $k=1$ represents a physical projective representation (since we require the element $D[(E, \pi/2)] \sim D[e^{ i \frac{\pi}{2}}] = e^{i \pi/2}$), with the other 3 being spurious irreps (see the text surrounding equation (\ref{irr:abel})). This is the irrep $A$ of the identity (non-projective) single point group 1. Next, we calculate the irreps of the next group of the group-subgroup chain (\ref{example:chain}), $H_2$. Each member of the group $H_1$ conjugates into itself under $(\,^dE,0)$, i.e.,
\begin{equation}
(\,^dE,0)^{-1}(E,\pi j/2)(\,^dE,0)=(E,\pi j/2)
\end{equation}
Then, the single irrep $A$ of $H_1$ is self-conjugated and its orbit contains only the (unique) irrep $A$. It induces two irreps of dimension 1 into $H_2$. The matrices of the symmetry operations $(E,\pi j/2)$ that belong to $H_3$ are exactly the same in both irreps,
\begin{equation}
D[A\uparrow H_2](E,\pi j/2)=e^{i\pi j/2}
\end{equation}
and the matrices of the coset representative $(\,^dE,0)$ in the two induced irreps are (see equations (\ref{coset2}), (\ref{coset2a}) and (\ref{coset2b})) $D[A\uparrow H_2](\,^dE,0)=\pm1$. The irrep with the $+$ sign corresponds to a single-valued irrep (relevant in our context when SOC is not considered) and the irrep with the $-$ sign is double-valued (relevant in our context when SOC is considered). These two irreps are the irreps of the ordinary (non-projective) double point group 1 and are denoted as $A$ and $\overline{A}$, respectively.

Climbing up the chain (\ref{example:chain}) we now calculate the irreps of $H_1$. Again, all the symmetry elements of $H_2$ are self-conjugated under the coset representative $(2_z,0)$ and the orbit of the irrep $A$ of $H_2$ contains only the irrep $A$. Likewise the orbit of $\overline{A}$ contains only $\overline{A}$. Each irrep induces thus two irreps of dimension 1. For the single irrep, the matrices of the coset representative $(2_z,0)$ are $D[A\uparrow H_1](2_z,0)=\pm1$. For the double irrep,  the matrices of the coset representative $(2_z,0)$ are $D[\overline{A}\uparrow H_1](2_z,0)=\pm i$. The four irreps are thus the irreps of the ordinary (non-projective) point group 2 and are denoted as $A,B$ (single-valued irreps) and $\,^1\overline{E},\,^2\overline{E}$ (double-valued irreps). Table (\ref{tab:centralexth1}) gives the matrices $D[\rho^i]((R,\pi j/2))$ of the elements of the central extension $H_1$ of the four the physically relevant (non-spurious) irreps $\rho^i$.

\begin{table}[h!]
	    \centering
	    \begin{tabular}{r|rrrr}
	    	$h$&$(E,\pi j/2)$&$(\,^dE,\pi j/2)$&$(2_z,\pi j/2)$&$(\,^d2_z,\pi j/2)$\\
	    	\hline
		$A$&$e^{i\pi j/2}$&$e^{i\pi j/2}$&$e^{i\pi j/2}$&$e^{i\pi j/2}$\\
		$B$&$e^{i\pi j/2}$&$e^{i\pi j/2}$&$-e^{i\pi j/2}$&$-e^{i\pi j/2}$\\
		$^1\overline{E}$&$e^{i\pi j/2}$&$-e^{i\pi j/2}$&$ie^{i\pi j/2}$&$-ie^{i\pi j/2}$\\
		$^2\overline{E}$&$e^{i\pi j/2}$&$-e^{i\pi j/2}$&$-ie^{i\pi j/2}$&$ie^{i\pi j/2}$		
	    \end{tabular} 
	    \caption{Physically relevant (non-spurious) irreducible representations of the central extension $H_1$ (see group-subgroup chain (\ref{example:chain})) of the double point group $2$. 
	    	\label{tab:centralexth1}
	    } % title of Table	    
\end{table}
Up to this point the results of the calculation of the irreducible representations of the central extensions $H_3$, $H_2$ and $H_1$ does not differ from the standard calculation of the irreps following the standard way. These intermediate groups $H_3$, $H_2$ and $H_1$ are the central extensions of ordinary (non-projective) groups and we recover the known result once the isomorphism has been applied to assign the matrices of the irreps to the $R$ point group operations of the ordinary groups (see eq. (\ref{isomor})). However, the last group $G_C^U=H_0$ in of the group-subgroup chain (\ref{example:chain}) contains mirror planes  $UM$ which are the Hofstadter symmetries that render the group projective. To see this explicitly, the conjugation of the group elements $h\in H_1$ under the coset representative $(m_x,0)$ (note that $(m_x,0)^{-1}=(\,^dm_x,0)$) gives the following relations,
\begin{eqnarray}
\label{conj1}
(m_x,0)^{-1}(E,\pi j/2)(m_x,0)&=&(E,\pi j/2)\\
\label{conj2}
(m_x,0)^{-1}(\,^dE,\pi j/2)(m_x,0)&=&(\,^dE,\pi j/2)\\
\label{conj3}
(m_x,0)^{-1}(2_z,\pi j/2)(m_x,0)&=&(\,^d2_z,\pi (j-2)/2\textrm{ mod }2\pi)\\
\label{conj4}
(m_x,0)^{-1}(\,^d2_z,\pi j/2)(m_x,0)&=&(2_z,\pi (j-2)/2\textrm{ mod }2\pi)
\end{eqnarray}
The last two expressions we explain in detail since they define the nontrivial projective phases of $2_\pi mm$. For instance, the conjugation (\ref{conj3}) can be split into two steps In the first step, we multiply the first two elements $(m_x,0)^{-1}=(\,^dm_x,0)$ and $(2_z,\pi j/2)$ according to the multiplication law of the central extension (eq. (\ref{multcentral})):
\begin{equation}
\label{multelem1}(
(\,^dm_x,0)\cdot(2_z,\pi j/2)=(\,^dm_x2_z,0+\pi j/2+\theta_{\pi}(\,^dm_x,2_z)\textrm{ mod 2}),\hspace{0.5cm}j=0,1,2,3
\end{equation}
But, using the general relations (\ref{eq:projectiverelations}),
\begin{equation}
\theta_{\gamma}(MC_n^k,C_n^{\ell})=\frac{\gamma}{2}(k-[k+\ell]_n)
\end{equation}
and substituting $\gamma=\pi$, $n=2$, $k=0$ and $\ell=1$, the phase is $\theta_{\pi}(\,^dm_x,2_z))=-\pi/2$. Therefore the relation (\ref{multelem1}) is,
\begin{equation}
(\,^dm_x,0)\cdot(2_z,\pi j/2)=(\,^dm_x2_z,\pi (j-1)/2\textrm{ mod 2}),\hspace{0.5cm}j=0,1,2,3
\end{equation}
Finally,
\begin{equation}
\label{phase2}
(\,^dm_x2_z,\pi (j-1)/2\textrm{ mod 2})\cdot,(m_x,0)=(\,^d2_z,\pi (j-1)/2+0+\theta_{\gamma}(\,^dm_x2_z,m_x))\hspace{0.5cm}j=0,1,2,3
\end{equation}
To calculate the phase $\theta_{\gamma}$, we make use of the general relation (\ref{eq:projectiverelations}),
\begin{equation}
\theta_{\gamma}(MC_n^k,MC_n^{\ell})=\frac{\gamma}{2}(\ell-k)
\end{equation}
with $n=2$, $k=1$ and $\ell=0$. The resulting phase is $\theta_{\gamma}(\,^dm_x2_z,m_x))=-\pi/2$. Introducing this value into eq. (\ref{phase2}) we get the result shown in eq. \ref{conj3}.

The relations (\ref{conj1})-(\ref{conj4}) indicate that, under conjugation of the elements $h\in H_2$ by the coset $g=(m_x,0)$, the first two columns in table (\ref{tab:centralexth1}) remain unchanged but the last two columns swap and change the sign. Then,
\begin{eqnarray}
D[A](g^{-1}hg)=D[B](h)\\
D[B](g^{-1}hg)=D[A](h)\\
D[\,^1\overline{E}](g^{-1}hg)=D[\,^1\overline{E}](h)\\
D[\,^2\overline{E}](g^{-1}hg)=D[\,^2\overline{E}](h)
\end{eqnarray}
On the one hand, the two single valued irreps $A$ and $B$ belong to the same orbit, but both $\,^1\overline{E}$ and $\,^2\overline{E}$ are self-conjugated. The irreps $A$ and $B$ \emph{combine} to induce a 2-dimensional single-valued irrep, which we denote as $AB$. The  2-dimensional matrices of the elements $h\in H_2$ are diagonal in the induced $AB$ irrep, and the two elements in the diagonal are just the two values in table (\ref{tab:centralexth1}) for the irreps $A$ and $B$. The matrix of the coset representative $(m_x,0)$ in $AB$ can be obtained using the relation (\ref{inducedn2}). On the other hand, each double-valued irrep induces two 1-dimensional irreps: those irreps induced from $\,^1\overline{E}$ are denoted as $\,^1\overline{E}_1$ and $\,^2\overline{E}_1$ and those irreps induced from $\,^2\overline{E}$ as $\,^1\overline{E}_2$ and $\,^2\overline{E}_2$. They are obtained following the same procedure used in the calculation of the previous irreps in $H_2$ and $H_1$. Table (\ref{tab:centralexth2}) gives the traces of the irreps of the central extension $G_C^U=H_0$ of the unitary point group $2_{\pi}mm$. The projective representations of this group $2_{\pi}mm$ can be obtained if we assign to the point group operation $R$ the matrix of the irrep of the element $(R,j=0)$ in the table (\ref{tab:centralexth2}) (see eq. (\ref{isomor})), which gives the irreps of each operator $g \in G$ within the central extension $e^{i \varphi} g \in G_C$. 

\begin{table}[h!]
	\centering
	\begin{tabular}{r|rrrrrrrr}
		$h$&$(E,\pi j/2)$&$(\,^dE,\pi j/2)$&$(2_z,\pi j/2)$&$(2_z^d,\pi j/2)$&$(m_x,\pi j/2)$&$(\,^dm_x,\pi j/2)$&$(m_y,\pi j/2)$&$(\,^dm_y,\pi j/2)$\\
		\hline
		$AB$&$2e^{i\pi j/2}$&$-2e^{i\pi j/2}$&0&0&0&0&0&0\\
		$^1\overline{E}_1$&$e^{i\pi j/2}$&$-e^{i\pi j/2}$&$ie^{i\pi j/2}$&$-ie^{i\pi j/2}$&$ie^{i\pi j/2}$&$-ie^{i\pi j/2}$&$ie^{i\pi j/2}$&$-ie^{i\pi j/2}$\\
		$^2\overline{E}_1$&$e^{i\pi j/2}$&$-e^{i\pi j/2}$&$ie^{i\pi j/2}$&$-ie^{i\pi j/2}$&$-ie^{i\pi j/2}$&$ie^{i\pi j/2}$&$-ie^{i\pi j/2}$&$ie^{i\pi j/2}$\\
		$^1\overline{E}_2$&$e^{i\pi j/2}$&$-e^{i\pi j/2}$&$-ie^{i\pi j/2}$&$ie^{i\pi j/2}$&$ie^{i\pi j/2}$&$-ie^{i\pi j/2}$&$-ie^{i\pi j/2}$&$ie^{i\pi j/2}$\\	
		$^2\overline{E}_2$&$e^{i\pi j/2}$&$-e^{i\pi j/2}$&$-ie^{i\pi j/2}$&$ie^{i\pi j/2}$&$-ie^{i\pi j/2}$&$ie^{i\pi j/2}$&$ie^{i\pi j/2}$&$-ie^{i\pi j/2}$
	\end{tabular} 
	\caption{Physically relevant (non-spurious) irreducible representations of the central extension $H_0=2_{\pi}mm$ (see group-subgroup chain (\ref{example:chain}). $j=0,1,2,3$). 
		\label{tab:centralexth2}
	} % title of Table	    
\end{table}

Finally, we apply the procedure detailed in section \ref{antiunitary} to calculate the co-representations of the central extension of the magnetic projective group $G=2_{\pi}mm1'$ from the irreps of the central extension of its unitary subgroup $G_C^U=2_{\pi}mm$ given in table (\ref{tab:centralexth2}).

First we choose as representative anti-unitary element of the central extension $G_C$ the element $g_A=(\mathcal{T},0)$, where $\mathcal{T}$ is the time-reversal symmetry. Making use of the relations (\ref{eq:projectiverelations}), we calculate the conjugation of the elements of the central extension $H_1$ under $g_A$. The conjugation gives the following result,

\begin{eqnarray}
\label{conj12}
(\mathcal{T},0)^{-1}(E,\pi j/2)(\mathcal{T},0)&=&(E,-\pi j/2\textrm{ mod }2\pi)\\
\label{conj22}
(\mathcal{T},0)^{-1}(2_z,\pi j/2)(\mathcal{T},0)&=&(2_z,-\pi (j+2)/2\textrm{ mod }2\pi)\\
\label{conj32}
(\mathcal{T},0)^{-1}(m_x,\pi j/2)(\mathcal{T},0)&=&(m_x,-\pi j/2\textrm{ mod }2\pi)\\
\label{conj42}
(\mathcal{T},0)^{-1}(m_y,\pi j/2)(\mathcal{T},0)&=&(m_y,-\pi j/2\textrm{ mod }2\pi)
\end{eqnarray}
and equivalent relations for the elements whose first component is the symmetry operation $\,^dE$, $\,^d2_z$, $\,^dm_x$ or $\,^dm_y$. The factor $(-1)$ that multiplies $j$ on the right hand side is due to the anti-unitary character of $\mathcal{T}$ (see the general multiplication law (\ref{multcentral}) of the central extension). Using these relations and the characters of the irreps of the unitary subgroup $G^U$ given in table (\ref{tab:centralexth2}), we can construct the table of $D[\rho]^{*}(g_A^{-1}hg_A)$ magnitudes (see eq. (\ref{conjugate}) for each irrep $\rho$ of the unitary subgroup and each symmetry operation $h\in G_C^U$. Table (\ref{tab:centralexth3}) shows the results of the calculation.

\begin{table}[h!]
	\centering
	\begin{tabular}{r|rrrrrrrr}
		$h$&$(E,\pi j/2)$&$(\,^dE,\pi j/2)$&$(2_z,\pi j/2)$&$(2_z^d,\pi j/2)$&$(m_x,\pi j/2)$&$(\,^dm_x,\pi j/2)$&$(m_y,\pi j/2)$&$(\,^dm_y,\pi j/2)$\\		
		\hline
		$D[AB]^{*}(g_A^{-1}hg_A)$&$2e^{i\pi j/2}$&$2e^{i\pi j/2}$&0&0&0&0&0&0\\
		$D[^1\overline{E}_1]^{*}(g_A^{-1}hg_A)$&$e^{i\pi j/2}$&$-e^{i\pi j/2}$&$ie^{i\pi j/2}$&$-ie^{i\pi j/2}$&$-ie^{i\pi j/2}$&$ie^{i\pi j/2}$&$-ie^{i\pi j/2}$&$ie^{i\pi j/2}$\\
		$D[^2\overline{E}_1]^{*}(g_A^{-1}hg_A)$&$e^{i\pi j/2}$&$-e^{i\pi j/2}$&$ie^{i\pi j/2}$&$-ie^{i\pi j/2}$&$ie^{i\pi j/2}$&$-ie^{i\pi j/2}$&$ie^{i\pi j/2}$&$-ie^{i\pi j/2}$\\
		$D[^1\overline{E}_2]^{*}(g_A^{-1}hg_A)$&$e^{i\pi j/2}$&$-e^{i\pi j/2}$&$-ie^{i\pi j/2}$&$ie^{i\pi j/2}$&$-ie^{i\pi j/2}$&$ie^{i\pi j/2}$&$ie^{i\pi j/2}$&$-ie^{i\pi j/2}$\\	
		$D[^2\overline{E}_2]^{*}(g_A^{-1}hg_A)$&$e^{i\pi j/2}$&$-e^{i\pi j/2}$&$-ie^{i\pi j/2}$&$ie^{i\pi j/2}$&$ie^{i\pi j/2}$&$-ie^{i\pi j/2}$&$-ie^{i\pi j/2}$&$ie^{i\pi j/2}$
	\end{tabular} 
	\caption{Tables of characters $D[\rho]^{*}(g_A^{-1}hg_A)$ after the conjugation of the irreps $D[\rho](h)$ of table (\ref{tab:centralexth2}) under the anti-unitary symmetry operation $g_A=(\mathcal{T},0)$). 
		\label{tab:centralexth3}
	} % title of Table	    
\end{table}
Comparing the traces of tables (\ref{tab:centralexth2}) and (\ref{tab:centralexth3}), we can check that the single-valued irrep $AB$ if self-conjugated (therefore the irrep $AB$ induces two copies of a type (a) irreducible co-representation labeled also as $AB$), the double-valued irreps $^1\overline{E}_1$ and $^2\overline{E}_1$ are mutually conjugated and they induce a type (c) irreducible co-representation labeled as $^1\overline{E}_1\,^2\overline{E}_1$ and $^1\overline{E}_2$ and $^2\overline{E}_2$ are also mutually conjugated and induce a type (c) irreducible co-representation labeled as $^1\overline{E}_2\,^2\overline{E}_2$.

The matrices of the unitary elements for the single-valued co-representation $AB$ are exactly the same as the matrices of these elements in the irrep $AB$ of the unitary subgroup and the matrix of the representative anti-unitary element $g_A=(\mathcal{T},0)$ is $D[AB](\mathcal{T},0)=\mathcal{U}$, where $\mathcal{U}$ is the matrix that fulfills eq. (\ref{equivalence}).

The matrices of the unitary elements $h$ of the type (c) irreducible co-representation $^1\overline{E}_1\,^2\overline{E}_1$ are the direct sum of the matrices of $h$ in the two irreps  $^1\overline{E}_1$ and $\,^2\overline{E}_1$ of the unitary subgroup. The matrix of the anti-unitary element $g_A=(\mathcal{T},0)$ can be calculated using eq. (\ref{antimatrix}).The matrices of the other type (c) co-representation $^1\overline{E}_2$ and $^2\overline{E}_2$ are obtained in exactly the same way.

Table (\ref{tab:centralexth4}) shows the characters of the co-representations (only for unitary operations) of the central extension of the projective point group $2_{\pi}mm1'$ and table (\ref{tab:centralexth5}) the final set of co-representations of the projective point group $2_{\pi}mm1'$, once the isomorphism (\ref{isomor}) has been assumed.

\begin{table}[h!]
	\centering
	\begin{tabular}{r|rrrrrrrr}
		$h$&$(E,\pi j/2)$&$(\,^dE,\pi j/2)$&$(2_z,\pi j/2)$&$(2_z^d,\pi j/2)$&$(m_x,\pi j/2)$&$(\,^dm_x,\pi j/2)$&$(m_y,\pi j/2)$&$(\,^dm_y,\pi j/2)$\\		
		\hline
		$AB$&$2e^{i\pi j/2}$&$2e^{i\pi j/2}$&0&0&0&0&0&0\\
		$^1\overline{E}_1\,^2\overline{E}_1$&$2e^{i\pi j/2}$&$-2e^{i\pi j/2}$&$2ie^{i\pi j/2}$&$-2ie^{i\pi j/2}$&0&0&0&0\\
		$^1\overline{E}_2\,^2\overline{E}_2$&$2e^{i\pi j/2}$&$-2e^{i\pi j/2}$&$-2ie^{i\pi j/2}$&$2ie^{i\pi j/2}$&0&0&0&0
	\end{tabular} 
	\caption{Table of characters of the irreducible co-representations of the central extension $G_C$ of the projective point group $2_{\pi}mm1'$. The table shows only the characters of the unitary elements.
		\label{tab:centralexth4}
	} % title of Table	    
\end{table}

\begin{table}[h!]
	\centering
	\begin{tabular}{r|rrrrrrrr}
		&$E$&$\,^dE$&$2_z$&$2_z^d$&$m_x$&$\,^dm_x$&$m_y$&$\,^dm_y$\\		
		\hline
		$AB$&$2$&$-2$&0&0&0&0&0&0\\
		$^1\overline{E}_1\,^2\overline{E}_1$&$2$&$-2$&$2i$&$-2i$&0&0&0&0\\
		$^1\overline{E}_2\,^2\overline{E}_2$&$2$&$-2$&$-2i$&$2i$&0&0&0&0
	\end{tabular} 
	\caption{Table of characters of the irreducible co-representations of the projective point group $2_{\pi}mm1'$. The table shows only the characters of the unitary elements.
		\label{tab:centralexth5}
	} % title of Table	    
\end{table}

\section{Tables of RSIs and Hofstadter Topological Invariants}
\label{app:tables}

In this section, tabulate the RSIs of all 31 magnetic PGs and the 51 projective PGs using the irreps obtained in \App{sec:irrepconstruction}. We then write the Hofstadter topological invariants in terms of these RSIs, giving indices which diagnose protected gap closings (Hofstadter SMs) or protected boundary state pumping (Hofstadter HOTIs). All projective groups and their irreps can be found at \href{www.cryst.ehu.es/cryst/projectiverepres}{Bilbao Crystallographic Server}.

\subsection{Real Space Invariants}
\label{app:RSIexamples}

Because RSIs play a primary role in our classification, we briefly review their construction. First defined in \Ref{song2019real}, RSIs are local, adiabatic invariants determined by the Wannier functions of the ground state. Each high symmetry Wyckoff position has a set of RSIs determined by its PG, and the RSIs cannot be changed by any perturbation unless the gap is closed or a symmetry is broken. RSIs are protected by PG symmetries alone (\emph{not} translations), making them well suited to describing phases in flux. Throughout, we focus on a single Wyckoff position $\mbf{x}$ with PG $G_x$. For brevity of notation, we will sometimes drop the $\mbf{x}$ or $x$ subscript, but one should recall that the projective representation, RSIs, and Hofstadter indices are defined for every $\mbf{x}$ individually.  \Ref{song2019real} gave a group-theoretic procedure to construct the RSIs. We give two physically motivated examples of this procedure which illustrate the rigorous theoretic construction of the irreps and RSIs of the new projective groups. 

First we choose PG $4_\gamma1'$ without SOC which is a point group that can be realized at $\Phi/2$ flux. We will derive the RSIs for all the possible $\gamma = 0 , \pm \pi/2, \pi$. At generic points away from $\mbf{x}$, the PG $4_\gamma1'$ to its subgroup $1'$. A state $\ket{\psi}$ off $\mbf{x}$ transforms in an irrep of $1'$ which has only the trivial irrep $A$ carried by the state $\ket{\psi}$: $U\mathcal{T} \ket{\psi} = \ket{\psi}$. We now calculate the induced representation $A \uparrow 4_\gamma1'$. Because $\ket{\psi}$ is off $\mbf{x}$, the states $C_4^j\ket{\psi}, j = 0,\dots3,$ are necessarily distinct and carry a representation of $4_\gamma 1'$. Using \Eq{eq:gammadef}, we observe that on the $C_4^j\ket{\psi}$ basis, the symmetries $g \in G_{\mbf{x}}$ transform in the representation
\bea
\label{eq:RSIinduct}
D[C_4] &= 
\bpm
 0 & 0 & 0 & 1 \\
 1 & 0 & 0 & 0 \\
 0 & 1 & 0 & 0 \\
 0 & 0 & 1 & 0 \\
\epm \!, D[U\mathcal{T}] = 
\bpm
 1 & 0 & 0 & 0 \\
 0 & e^{i \gamma} & 0 & 0 \\
 0 & 0 & e^{2i \gamma} & 0 \\
0 & 0 & 0 & e^{3i \gamma} \\
\epm K \\
\eea
where $D[g]$ is the representation of $g\in 4_\gamma 1'$ induced by $\ket{\psi}$ and $K$ is complex conjugation. $D[C_4]$ and $D[U\mathcal{T}]$ can be simultaneously block diagonalized into irreps of $4_\gamma 1'$ using the $D[C_4]$ eigenbasis. Explicitly, this eigenbasis is obtained from the unitary operator
\bea
V^\dag D[C_4] V &= \left(
\begin{array}{cccc}
 1 & 0 & 0 & 0 \\
 0 & -1 & 0 & 0 \\
 0 & 0 & i & 0 \\
 0 & 0 & 0 & -i \\
\end{array}
\right), \qquad V = \left(
\begin{array}{cccc}
 \frac{1}{2} & -\frac{1}{2} & -\frac{i}{2} & \frac{i}{2} \\
 \frac{1}{2} & \frac{1}{2} & -\frac{1}{2} & -\frac{1}{2} \\
 \frac{1}{2} & -\frac{1}{2} & \frac{i}{2} & -\frac{i}{2} \\
 \frac{1}{2} & \frac{1}{2} & \frac{1}{2} & \frac{1}{2} \\
\end{array}
\right) \ .
\eea
The $1,-1,i,-i$ eigenvalues are conventionally referred to as the $A,B,{}^2E, {}^1E$ irreps of PG $4$. In this basis where $D[C_4]$ is diagonal, it is easy to read off the irreps from the structure of $V^\dag D[U\mathcal{T}]V$. We compute
\bea
\gamma &= 0: \qquad V^\dag D[U\mathcal{T}] V = \left(
\begin{array}{cccc}
 1 & 0 & 0 & 0 \\
 0 & 1 & 0 & 0 \\
 0 & 0 & 0 & 1 \\
 0 & 0 & 1 & 0 \\
\end{array}
\right) \\ 
\gamma &= \pi/2: \qquad V^\dag D[U\mathcal{T}] V  = \left(
\begin{array}{cccc}
 0 & 0 & 0 & -i \\
 0 & 0 & -i & 0 \\
 0 & -i & 0 & 0 \\
 -i & 0 & 0 & 0 \\
\end{array}
\right) \\
\gamma &= \pi: \qquad V^\dag D[U\mathcal{T}] V  = \left(
\begin{array}{cccc}
 0 & -1 & 0 & 0 \\
 -1 & 0 & 0 & 0 \\
 0 & 0 & -1 & 0 \\
 0 & 0 & 0 & -1 \\
\end{array}
\right)\\
\gamma &= -\pi/2: \qquad V^\dag D[U\mathcal{T}] V  = \left(
\begin{array}{cccc}
 0 & 0 & i & 0 \\
 0 & 0 & 0 & i \\
 i & 0 & 0 & 0 \\
 0 & i & 0 & 0 \\
\end{array}
\right)\ . \\
\eea
We consider each individually. At $\gamma = 0$, we see that the matrices split into two $1\times1$ blocks (the 1D irreps $A$, $B$) and a 2D irrep where $U\mathcal{T}$ is off-diagonal in the ${}^1E, {}^2E$ irreps. We call this two irrep ${}^1E{}^2E$. These irreps are the familiar irreps of $41'$. Thus we have computed the induction
\bea
A \uparrow 41' = A \oplus B \oplus {}^1E{}^2E
\eea
under which the follows invariants $\delta_1 = m(B) - m(A), \delta_2 = m({}^1E{}^2E) - m(A)$ are invariant. These invariants are the RSIs of $41'$. Technically this construction does not prove $\delta_1, \delta_2$ are the complete set of invariants, but a more formal procedure using the Smith Normal form confirms this is the case \cite{song2019real}. Since our aim here is to give an intuitive construction, we refer to \Ref{song2019real} in the remaining examples to show completeness. 

Now we proceed to the new projective groups. For $\gamma = \pi/2$, we see that there are two $2\times 2$ blocks of $U\mathcal{T}$ connecting the $A, {}^1E$ and $B,{}^2E$. These are the two 2D irreps $A{}^1E$ and $B {}^2E$. Thus we have
\bea
A \uparrow 4_{\pi/2}1' = A\, {}^2E \oplus B\,{}^1E
\eea
under which there is a single invariant RSI $\delta_1 = m(B{}^2E) - m(A {}^2E)$ for $4_{\pi/2}1'$. The case of $\gamma = - \pi/2$ is related by complex conjugation and gives $\delta_1 = m(B{}^1E) - m(A {}^1E)$ as the RSI of $4_{-\pi/2}1'$. 

Lastly, for $\gamma = \pi$, following the same steps we find
\bea
A \uparrow 4_{\pi}1' = AB \oplus {}^1E \oplus {}^2E
\eea
leading to the two RSIs $\delta_1 = -m({}^2E) + m({}^1E), \delta_2 = -m({}^2E) + m(AB)$. 

Our second example is in group $2_\pi'$ which is generated by $UC_2\mathcal{T}$ satisfying $(UC_2\mathcal{T})^2 = UC_2U^\dag C_2\mathcal{T}^2= e^{i \gamma} C_2^2\mathcal{T}^2 = e^{i \gamma} = -1$ with and without SOC because $(C_2\mathcal{T})^2 = (\pm 1)^2 = +1$ where $\pm$ is with/without SOC. This group has a single 2D irrep which we denote $AA$. The irrep must be $2D$ because of Kramers' theorem, since $UC_2\mathcal{T}$ is an anti-unitary operator squaring to $-1$. We now determine the RSIs of $2_\pi'$. If $\ket{\psi}$ is a state off the Wyckoff position, then it induces the representation
\bea
\label{eq:duc2tex}
D[UC_2\mathcal{T}] = \bpm 0 & 1 \\ -1 & 0 \epm K
\eea
on the basis of states $\ket{\psi}, UC_2\mathcal{T} \ket{\psi}$. \Eq{eq:duc2tex} is the irrep $AA$ of $2_\pi'$, so we see that $A \uparrow 2_\pi' = AA$. Thus the number of irreps $AA$ can always be adiabatically increased or decreased by $1$, and $2_\pi'$ has no nontrivial RSI. 

Our last example is for the group $2_\pi 1'$ with SOC generated by $C_2^2 = -1, (U\mathcal{T})^2 = -1$ and with $C_2^\dag U C_2 = e^{i \gamma} U = - U$. First let us heuristically construct the irreps of this group. We pick an eigenstate of $C_2$ such that $C_2 \ket{\pm i} = \pm i \ket{\pm i}$. Then we observe $U\mathcal{T} \ket{\pm i}$ is necessarily a distinct state by Kramers theorem, since $(U\mathcal{T})^2 = -1$. The $C_2$ eigenvalue of this state is
\bea
C_2 U\mathcal{T} \ket{\pm i}  =  - U\mathcal{T} C_2 \ket{\pm i} = - U\mathcal{T} (\pm i) \ket{\pm i} = - (\mp i) U\mathcal{T}  \ket{\pm i} = \pm i \, U\mathcal{T}  \ket{\pm i} \ .
\eea
Thus we have found two $2D$ irreps denoted ${}^1\overline{E}{}^1\overline{E}$ and ${}^2\overline{E}{}^2\overline{E}$. 

Now we determine the RSIs of $2_\pi1'$. A point away from the Wyckoff position $\mbf{x}$ has the point group $1'$ with SOC which has a 2D irrep $\overline{EE}$ (a Kramers pair) spanned by the states $\ket{\psi}, U\mathcal{T} \ket{\psi}$. Acting with $C_2$, we obtain the distinct states $C_2\ket{\psi}, C_2 U\mathcal{T} \ket{\psi}$. On the ordered basis $\ket{\psi}, U\mathcal{T} \ket{\psi}, C_2\ket{\psi}, C_2 U\mathcal{T} \ket{\psi}$, we find that the representation induced by these states is
\bea
D[C_2] &= \bpm 
0 & 0 & -1 & 0 \\
0 & 0 & 0 & -1 \\
1 & 0 & 0 & 0 \\
0 & 1 & 0 & 0 \\
\epm,  
\eea
which has eigenvalues $i,i,-i,-i$. As such, we find $\overline{EE} \uparrow 2_\pi1' = {}^1\overline{E}{}^1\overline{E} \oplus {}^2\overline{E}{}^2\overline{E}$. Hence we find a single RSI $\delta_1 = -m({}^1\overline{E}{}^1\overline{E}) + m({}^2\overline{E}{}^2\overline{E})$. 

All projective groups and their irreps can be found at \href{www.cryst.ehu.es/cryst/projectiverepres}{Bilbao Crystallographic Server}.

\subsection{No SOC}

 \Tab{tab:RSIsnoSOC} is a complete list of all RSIs in the conventional 31 2D magnetic PGs and the 51 new projective PGs.  \Tab{tab:RSIsnoSOC} is calculated using the group theoretical method of \Ref{song2019real} (or equivalently the physical description in \Eq{eq:RSIinduct}). We find that all RSIs groups are products of $\mathds{Z}$ and/or $\mathds{Z}_2$. All projective groups and their irreps can be found at \href{www.cryst.ehu.es/cryst/projectiverepres}{Bilbao Crystallographic Server}.

\begin{center}
	\begin{scriptsize}
	\label{tab:RSIsnoSOC}
\begin{longtable}{l|lllll}
		\caption{Real Space Invariants (RSIs) without SOC in all 82 PGs}\\ 
	\hline
	1&&\\
\hline
$11'$&&\\
\hline
$2$&$\delta_1=+m(A)-m(B)$&\\
\hline
$21'$&$\delta_1=+m(A)-m(B)$&\\
\hline
$2_{\pi}1'$&&\\
\hline
$2'$&$\delta_1=+m(A)\hspace{0.5cm}\textrm{mod 2}$&\\
\hline
$2_{\pi}'$&&\\
\hline
$m$&$\delta_1=-m(A')+m(A'')$&\\
\hline
$m1'$&$\delta_1=-m(A')+m(A'')$&\\
\hline
$m'$&$\delta_1=+m(A)\hspace{0.5cm}\textrm{mod 2}$&\\
\hline
$2mm$&$\delta_1=+m(A_{1})+m(A_{2})-m(B_{2})-m(B_{1})$&\\
\hline
$2_{\pi}mm$&&\\
\hline
$2mm1'$&$\delta_1=+m(A_{1})+m(A_{2})-m(B_{2})-m(B_{1})$&\\
\hline
$2_{\pi}mm1'$&&\\
\hline
$2'mm'$&$\delta_1=-m(A')+m(A'')\hspace{0.5cm}\textrm{mod 2}$&\\
\hline
$2_{\pi}'m'm$&&\\
\hline
$2m'm'$&$\delta_1=+m(A)-m(B)$&\\
\hline
$4$&$\delta_1=-m(A)+m(B)$&$\delta_2=-m(A)+m(\,^{2}E)$\\&$\delta_3=-m(A)+m(\,^{1}E)$&\\
\hline
$41'$&$\delta_1=-m(A)+m(B)$&$\delta_2=-m(A)+m(\,^{2}E^{1}E)$\\
\hline
$4_{\pi/2}1'$&$\delta_1=+m(\,^{2}EB)-m(\,^{1}EA)$&\\
\hline
$4_{\pi}1'$&$\delta_1=-m(\,^{2}E)+m(\,^{1}E)$&$\delta_2=-m(\,^{2}E)+m(AB)$\\
\hline
$4_{3\pi/2}1'$&$\delta_1=-m(\,^{2}EA)+m(\,^{1}EB)$&\\
\hline
$4'$&$\delta_1=+m(A)-2m(BB)$&\\
\hline
$4_{\pi/2}'$&$\delta_1=+m(AB)\hspace{0.5cm}\textrm{mod 2}$&\\
\hline
$4_{\pi}'$&$\delta_1=-m(B)+2m(AA)$&\\
\hline
$4_{3\pi/2}'$&$\delta_1=+m(AB)\hspace{0.5cm}\textrm{mod 2}$&\\
\hline
$4mm$&$\delta_1=-m(A_{1})-m(A_{2})+m(B_{1})+m(B_{2})$&$\delta_2=-m(A_{1})-m(A_{2})+m(E)$\\
\hline
$4_{\pi/2}mm$&$\delta_1=+m(\,^{2}EB)-m(\,^{1}EA)$&\\
\hline
$4_{\pi}mm$&$\delta_1=-m(\,^{2}E')-m(\,^{2}E'')+m(\,^{1}E')+m(\,^{1}E'')$&$\delta_2=-m(\,^{2}E')-m(\,^{2}E'')+m(AB)$\\
\hline
$4_{3\pi/2}mm$&$\delta_1=-m(\,^{2}EA)+m(\,^{1}EB)$&\\
\hline
$4mm1'$&$\delta_1=-m(A_{1})-m(A_{2})+m(B_{1})+m(B_{2})$&$\delta_2=-m(A_{1})-m(A_{2})+m(E)$\\
\hline
$4_{\pi/2}mm1'$&$\delta_1=+m(\,^{2}EB)-m(\,^{1}EA)$&\\
\hline
$4_{\pi}mm1'$&$\delta_1=-m(\,^{2}E')-m(\,^{2}E'')+m(\,^{1}E')+m(\,^{1}E'')$&$\delta_2=-m(\,^{2}E')-m(\,^{2}E'')+m(AB)$\\
\hline
$4_{3\pi/2}mm1'$&$\delta_1=-m(\,^{2}EA)+m(\,^{1}EB)$&\\
\hline
$4'm'm$&$\delta_1=+m(A_{1})+m(A_{2})-2m(B_{2}B_{1})$&\\
\hline
$4_{\pi/2}'m'm$&$\delta_1=+m(AB)\hspace{0.5cm}\textrm{mod 2}$&\\
\hline
$4_{\pi}'m'm$&$\delta_1=-m(B')-m(B'')+2m(A'A'')$&\\
\hline
$4_{3\pi/2}'m'm$&$\delta_1=+m(AB)\hspace{0.5cm}\textrm{mod 2}$&\\
\hline
$4m'm'$&$\delta_1=-m(A)+m(B)$&$\delta_2=-m(A)+m(\,^{2}E)$\\&$\delta_3=-m(A)+m(\,^{1}E)$&\\
\hline
$3$&$\delta_1=-m(A)+m(\,^{2}E)$&$\delta_2=-m(A)+m(\,^{1}E)$\\
\hline
$31'$&$\delta_1=-m(A)+m(\,^{2}E^{1}E)$&\\
\hline
$3_{2\pi/3}1'$&$\delta_1=+m(\,^{2}E)-m(\,^{1}EA)$&\\
\hline
$3_{4\pi/3}1'$&$\delta_1=+m(\,^{1}E)-m(\,^{2}EA)$&\\
\hline
$3m$&$\delta_1=-m(A_{1})-m(A_{2})+m(E)$&\\
\hline
$3_{2\pi/3}m$&$\delta_1=+m(\,^{2}E'')+m(\,^{2}E')-m(\,^{1}EA)$&\\
\hline
$3_{4\pi/3}m$&$\delta_1=+m(\,^{1}E')+m(\,^{1}E'')-m(\,^{2}EA)$&\\
\hline
$3m1'$&$\delta_1=-m(A_{1})-m(A_{2})+m(E)$&\\
\hline
$3_{2\pi/3}m1'$&$\delta_1=+m(\,^{2}E'')+m(\,^{2}E')-m(\,^{1}EA)$&\\
\hline
$3_{4\pi/3}m1'$&$\delta_1=+m(\,^{1}E'')+m(\,^{1}E')-m(\,^{2}EA)$&\\
\hline
$3m'$&$\delta_1=-m(A)+m(\,^{2}E)$&$\delta_2=-m(A)+m(\,^{1}E)$\\
\hline
$6$&$\delta_1=-m(A)+m(B)$&$\delta_2=-m(A)+m(\,^{2}E_{1})$\\&$\delta_3=-m(A)+m(\,^{2}E_{2})$&$\delta_4=-m(A)+m(\,^{1}E_{1})$\\&$\delta_5=-m(A)+m(\,^{1}E_{2})$&\\
\hline
$61'$&$\delta_1=-m(A)+m(B)$&$\delta_2=-m(A)+m(\,^{2}E_{1}^{1}E_{1})$\\&$\delta_3=-m(A)+m(\,^{2}E_{2}^{1}E_{2})$&\\
\hline
$6_{\pi/3}1'$&$\delta_1=-m(\,^{2}E_{1}^{2}E_{2})+m(\,^{1}E_{2}A)$&$\delta_2=-m(\,^{2}E_{1}^{2}E_{2})+m(\,^{1}E_{1}B)$\\
\hline
$6_{2\pi/3}1'$&$\delta_1=-m(\,^{1}E_{1})+m(\,^{1}E_{2})$&$\delta_2=-m(\,^{1}E_{1})+m(\,^{2}E_{1}A)$\\&$\delta_3=-m(\,^{1}E_{1})+m(\,^{2}E_{2}B)$&\\
\hline
$6_{\pi}1'$&$\delta_1=-m(\,^{1}E_{1}^{2}E_{2})+m(\,^{1}E_{2}^{2}E_{1})$&$\delta_2=-m(\,^{1}E_{1}^{2}E_{2})+m(AB)$\\
\hline
$6_{4\pi/3}1'$&$\delta_1=-m(\,^{2}E_{1})+m(\,^{2}E_{2})$&$\delta_2=-m(\,^{2}E_{1})+m(\,^{1}E_{1}A)$\\&$\delta_3=-m(\,^{2}E_{1})+m(\,^{1}E_{2}B)$&\\
\hline
$6_{5\pi/3}1'$&$\delta_1=-m(\,^{1}E_{1}^{1}E_{2})+m(\,^{2}E_{2}A)$&$\delta_2=-m(\,^{1}E_{1}^{1}E_{2})+m(\,^{2}E_{1}B)$\\
\hline
$6'$&$\delta_1=-m(A)+m(\,^{2}E^{1}E)$&$\delta_2=+m(A)\hspace{0.5cm}\textrm{mod 2}$\\
\hline
$6_{\pi/3}'$&$\delta_1=+2m(\,^{2}E^{2}E)-m(\,^{1}EA)$&\\
\hline
$6_{2\pi/3}'$&$\delta_1=+m(\,^{1}E)-m(\,^{2}EA)$&$\delta_2=+m(\,^{1}E)\hspace{0.5cm}\textrm{mod 2}$\\
\hline
$6_{\pi}'$&$\delta_1=+m(\,^{1}E^{2}E)-2m(AA)$&\\
\hline
$6_{4\pi/3}'$&$\delta_1=+m(\,^{2}E)-m(\,^{1}EA)$&$\delta_2=+m(\,^{2}E)\hspace{0.5cm}\textrm{mod 2}$\\
\hline
$6_{5\pi/3}'$&$\delta_1=+2m(\,^{1}E^{1}E)-m(\,^{2}EA)$&\\
\hline
$6mm$&$\delta_1=-m(A_{1})-m(A_{2})+m(B_{1})+m(B_{2})$&$\delta_2=-m(A_{1})-m(A_{2})+m(E_{2})$\\&$\delta_3=-m(A_{1})-m(A_{2})+m(E_{1})$&\\
\hline
$6_{\pi/3}mm$&$\delta_1=-m(\,^{2}E_{1}^{2}E_{2})+m(\,^{1}E_{2}A)$&$\delta_2=-m(\,^{2}E_{1}^{2}E_{2})+m(\,^{1}E_{1}B)$\\
\hline
$6_{2\pi/3}mm$&$\delta_1=-m(\,^{1}E_{1}')-m(\,^{1}E_{1}'')+m(\,^{1}E_{2}')+m(\,^{1}E_{2}'')$&$\delta_2=-m(\,^{1}E_{1}')-m(\,^{1}E_{1}'')+m(\,^{2}E_{1}A)$\\&$\delta_3=-m(\,^{1}E_{1}')-m(\,^{1}E_{1}'')+m(\,^{2}E_{2}B)$&\\
\hline
$6_{\pi}mm$&$\delta_1=-m(\,^{1}E_{1}^{2}E_{2})+m(\,^{1}E_{2}^{2}E_{1})$&$\delta_2=-m(\,^{1}E_{1}^{2}E_{2})+m(AB)$\\
\hline
$6_{4\pi/3}mm$&$\delta_1=-m(\,^{2}E_{1}')-m(\,^{2}E_{1}'')+m(\,^{2}E_{2}')+m(\,^{2}E_{2}'')$&$\delta_2=-m(\,^{2}E_{1}')-m(\,^{2}E_{1}'')+m(\,^{1}E_{1}A)$\\&$\delta_3=-m(\,^{2}E_{1}')-m(\,^{2}E_{1}'')+m(\,^{1}E_{2}B)$&\\
\hline
$6_{5\pi/3}mm$&$\delta_1=-m(\,^{1}E_{1}^{1}E_{2})+m(\,^{2}E_{2}A)$&$\delta_2=-m(\,^{1}E_{1}^{1}E_{2})+m(\,^{2}E_{1}B)$\\
\hline
$6mm1'$&$\delta_1=-m(A_{1})-m(A_{2})+m(B_{1})+m(B_{2})$&$\delta_2=-m(A_{1})-m(A_{2})+m(E_{2})$\\&$\delta_3=-m(A_{1})-m(A_{2})+m(E_{1})$&\\
\hline
$6_{\pi/3}mm1'$&$\delta_1=-m(\,^{2}E_{1}^{2}E_{2})+m(\,^{1}E_{2}A)$&$\delta_2=-m(\,^{2}E_{1}^{2}E_{2})+m(\,^{1}E_{1}B)$\\
\hline
$6_{2\pi/3}mm1'$&$\delta_1=-m(\,^{1}E_{1}')-m(\,^{1}E_{1}'')+m(\,^{1}E_{2}')+m(\,^{1}E_{2}'')$&$\delta_2=-m(\,^{1}E_{1}')-m(\,^{1}E_{1}'')+m(\,^{2}E_{1}A)$\\&$\delta_3=-m(\,^{1}E_{1}')-m(\,^{1}E_{1}'')+m(\,^{2}E_{2}B)$&\\
\hline
$6_{\pi}mm1'$&$\delta_1=-m(\,^{1}E_{1}^{2}E_{2})+m(\,^{1}E_{2}^{2}E_{1})$&$\delta_2=-m(\,^{1}E_{1}^{2}E_{2})+m(AB)$\\
\hline
$6_{4\pi/3}mm1'$&$\delta_1=-m(\,^{2}E_{1}')-m(\,^{2}E_{1}'')+m(\,^{2}E_{2}')+m(\,^{2}E_{2}'')$&$\delta_2=-m(\,^{2}E_{1}')-m(\,^{2}E_{1}'')+m(\,^{1}E_{1}A)$\\&$\delta_3=-m(\,^{2}E_{1}')-m(\,^{2}E_{1}'')+m(\,^{1}E_{2}B)$&\\
\hline
$6_{5\pi/3}mm1'$&$\delta_1=-m(\,^{1}E_{1}^{1}E_{2})+m(\,^{2}E_{2}A)$&$\delta_2=-m(\,^{1}E_{1}^{1}E_{2})+m(\,^{2}E_{1}B)$\\
\hline
$6'mm'$&$\delta_1=-m(A_{1})-m(A_{2})+m(E)$&$\delta_2=-m(A_{1})+m(A_{2})\hspace{0.5cm}\textrm{mod 2}$\\
\hline
$6_{\pi/3}'mm'$&$\delta_1=+2m(\,^{2}E'^{2}E'')-m(\,^{1}EA_{1})$&\\
\hline
$6_{2\pi/3}'mm'$&$\delta_1=+m(\,^{1}E')+m(\,^{1}E'')-m(\,^{2}EA_{1})$&$\delta_2=-m(\,^{1}E')+m(\,^{1}E'')\hspace{0.5cm}\textrm{mod 2}$\\
\hline
$6_{\pi}'mm'$&$\delta_1=+m(E)-2m(A_{1}A_{2})$&\\
\hline
$6_{4\pi/3}'mm'$&$\delta_1=+m(\,^{1}E')+m(\,^{1}E'')-m(\,^{2}EA_{1})$&$\delta_2=-m(\,^{1}E')+m(\,^{1}E'')\hspace{0.5cm}\textrm{mod 2}$\\
\hline
$6_{5\pi/3}'mm'$&$\delta_1=+2m(\,^{2}E'^{2}E'')-m(\,^{1}EA_{1})$&\\
\hline
$6m'm'$&$\delta_1=-m(A)+m(B)$&$\delta_2=-m(A)+m(\,^{2}E_{1})$\\&$\delta_3=-m(A)+m(\,^{2}E_{2})$&$\delta_4=-m(A)+m(\,^{1}E_{1})$\\&$\delta_5=-m(A)+m(\,^{1}E_{2})$&\\
\hline

\end{longtable}
\end{scriptsize}
\end{center}

We now enumerate all symmetry-protected Hofstadter topological invariants in \Tab{tab:HofinvnoSOC} which diagnose Hofstadter SM phases and Hofstadter HOTI phases. We organize the table by listing the crystalline PG at $G^{\phi = 0}$, the (projective) PG at the half-period flux $G^{\phi = \Phi/2}$, and the reduced PG in generic flux $G^{\phi}$. We also calculate the particle number (or total charge) $N^\phi$ in terms of the RSIs at $\phi = 0$ and $\phi = \Phi/2$. 
The Hofstadter SM invariants are deduced by requiring compatibility of the $\phi =0$ and $\phi = \Phi$ RSIs in groups with only $C_n$ or $C_n,M\mathcal{T}$ where there are no projective point groups at $\Phi/2$ flux since there are no reentrant symmetries ($C_n$ and $M\mathcal{T}$ are symmetries at all flux). In groups with $M, \mathcal{T},$ or $C_n\mathcal{T}$, the Hofstadter SM invariants are deduced by requiring compatibility of the $\phi =0$ and $\phi = \Phi/2$ RSIs when they are reduced in flux to the RSIs of $G^\phi$. In this way, we obtain a finer classification due to the projective representation at $\Phi/2$ created by $UM, U\mathcal{T}$, or $UC_n\mathcal{T}$. If any of the SM invariants are nonzero, a bulk gap closing is enforced. The Hofstadter HOTI indices are given by $N^{\phi=0}-N^{\phi=\Phi/2} \mod n_G$ where $n_G \in \mathbb{N}$ corresponds to the smallest number of particles which can be symmetrically removed from the Wyckoff position. 

We provide a few examples of the calculations in \Tab{tab:HofinvnoSOC}. First we consider PG $2$ at $\phi = 0$. The irreps are $A,B$ corresponding to the $C_2$ eigenvalues $\pm1$, and the RSI of PG $2$ defined is $\delta^{\phi=0}_1 = m(B) - m(A)$ where $m(\rho)$ is the multiplicity of the irrep $\rho$. Because $C_2$ is a symmetry at all flux, there are no reentrant symmetries and we write the Hofstadter SM indices between $\phi=0$ and $\phi = \Phi$ determined by irrep flow (\Eq{eq:irrepflow} of the Main Text). If $C_2^\dag U C_2 = e^{i \gamma} U$ with $\gamma \neq 0$, then an irrep of $C_2$ eigenvalue $\la$ at $\phi =0$ flows to an irrep of eigenvalue $e^{-i \gamma} \la \neq \la$ at $\phi = \Phi$. Let us take the nontrivial case of $\gamma = \pi$ so that $A$ and $B$ irreps interchange between $0$ and $\Phi$ flux. Thus $\delta^{\phi = \Phi}_1 = - \delta^{\phi = 0}_1$ and we compute
\bea
\delta^{SM}_1 = \delta_1^{\phi=0} - \delta_1^{\phi=\Phi} = 2 \delta_1^{\phi=0}
\eea
which is nonzero whenever $\delta_1^{\phi=0} \neq 0$, corresponding to a mismatch in the number of $A$ and $B$ irreps at $\phi =0$. Thus $\delta^{SM}_1$ is Peierls-indicated. In general, any nonzero RSI at $\phi=0$ in the groups $2,3,4,6, 2m'm', 3m', 4m'm',6m'm'$ leads to a gap closing protected by irrep flow if $\gamma \neq 0$. This is because the RSIs in these groups are all zero iff the numbers of each irrep are equal \cite{song2019real}. Thus a nonzero $\gamma$ will cause nontrivial irrep flow iff there is a nonzero RSI at $\phi=0$. This is also true with SOC since the groups $2,3,4,6, 2m'm', 3m', 4m'm',6m'm'$ with SOC are identical to those without SOC up to an overall phase factor. 

As a second example, we pick PG $G^{\phi=0} = 2'$ with $\gamma = \pi$ so that $G^{\phi=\Phi/2} = 2_\pi'$. At $\phi =0$, there is a single RSI of $2'$ which is $\delta_1^{\phi=0} = m(A) \mod 2$ since states can only move away from the $C_2\mathcal{T}$ center in pairs (the only irrep of $2'$ is $A$). At $\phi = \Phi/2$ where $(UC_2\mathcal{T})^2 = -1$, the only irrep is $AA$ which has a $UC_2\mathcal{T}$-related pair of states, so there is no nontrivial RSI since $AA = A \uparrow 2'_\pi$ is automatically an inducted representation. There is no SM invariant in this case because there is no symmetry at generic $\phi$ to protect a gap closing. However, there is a HOTI index. The number of states mod 2 at $\phi= 0$ is $m(A) \mod 2 = \delta_1^{\phi=0} \mod 2$ and the number of states mod 2 at $\phi= \Phi/2$ is $2m(AA) \mod 2 = 0 \mod 2$. Thus the HOTI index is $\delta_1^{HOTI} = \delta_1^{\phi=0} \mod 2$ which is Peierls-indicated. There is a simple physical picture behind $\delta_1^{HOTI}$. If $\delta_1^{\phi=0} \neq 0$, then there must be an odd number of states at $\phi = 0$. But at $\phi = \Phi/2$, the projective group $2_\pi'$ ensures an even number of states, so there is necessarily nonzero pumping. 

Lastly we consider PG $G^{\phi =0 } = 4'$ with $\gamma = \pi/2$ such that $G^{\phi = \Phi/2} = 4_{\pi/2}'$. We will see that these symmetries protect a Hofstadter SM (protected by $C_2 = (C_4\mathcal{T})^2$ which is a symmetry at all flux) and Hofstadter HOTI index (protected by the reentrant symmetry $UC_4\mathcal{T}$). Let us first understand the SM index. Since $C_4^\dag U C_4 = e^{i \pi/2} U$ with $\gamma = \pi/2$ (even though $C_4$ is a not a symmetry, it is still an operator with a nontrivial commutation relation with $U$), we see that $C_2 = (C_4\mathcal{T})^2$ symmetry obeys $C_2^\dag U C_2 = e^{i \pi} U$ which causes irrep flow from $\phi =0$ to $\phi = \Phi$. This irrep flow is diagnosed also from the $\phi =0$ and $\phi = \Phi/2$ groups from the formal $\delta_1^{SM}$ invariant 
\bea
\delta_1^{SM} = \delta^{\phi \to 0}_1 - \delta^{\phi \to \Phi/2}_1 \ . 
\eea
Let us show this explicitly. At $\phi = 0$, the RSI of $4'$ is $\delta^{\phi=0}_1 = m(A) - 2m(BB)$ where $A$ and $B$ indicate the $C_2$ eigenvalues of the states. Turning on infinitesimal flux breaks $4' \to 2$ whose RSI is $\delta^{\phi\to0}_1 = m(A) - m(B)$. The irrep $A$ of $4'$ reduces to the $A$ irrep of $2$, and the $BB$ irrep of $4'$ reduces to the $B \oplus B$ representation of $2$. Thus we have $\delta^{\phi\to0}_1 = \delta^{\phi=0}_1$. At $\phi = \Phi/2$, the $G^{\phi=\Phi/2} = 4_{\pi/2}'$ has a single irrep $AB$ and the RSI $\delta_1^{\phi = \Phi/2} = m(AB) \mod 2$. Since $AB \to A \oplus B$, reducing $4_{\pi/2}' \to 2$ by adding infinitesimal flux always yields an equal number of $A$ and $B$ irreps, so $\delta^{\phi \to \Phi/2}_1 = 0$. Thus $\delta_1^{SM} = \delta^{\phi=0}_1$, and thus is Peierls-indicated as predicted by irrep flow. Secondly, let us understand the HOTI index. The number of states at $\phi = 0$ is adiabatically defined mod 4 because $C_4\mathcal{T}$ ensures only quartets of states can be moved onto/off of the Wyckoff position. Indeed, we see that $N^{\phi=0} = m(A) + 2m(BB) \mod 4 = m(A) - 2m(BB) \mod 4 = \delta^{\phi=0}_1 \mod 4$ which is an adiabatic invariant. Similarly at $\phi = \Phi/2$, $N^{\phi=\Phi/2} = 2m(AB) \mod 4 = 2 \delta^{\phi=\Phi/2} \mod 4$. Thus the HOTI index detecting a difference in the number of states after pumping flux is
\bea
\delta^{HOTI} = N^{\phi=0} - N^{\phi=\Phi/2} \mod 4 = \delta^{\phi=0}_1  - 2 \delta^{\phi=\Phi/2}  \mod 4 \ . 
\eea
However, we also must enforce the compatibility condition $\delta_1^{SM} = 0$ so that there is no gap closing and the Wannier states evolve continuously. Since $\delta_1^{SM} = \delta^{\phi=0}_1$, we see that the HOTI index reduces to $\delta^{HOTI} = 2 \delta^{\phi=\Phi/2}  \mod 4$ which is $\mathds{Z}_2$ classified. It is not Peierls-indicated because it depends on the RSI $\delta^{\phi=\Phi/2}$ at $\phi = \Phi/2$. As an example, if $ \delta^{\phi=0}_1 = 0$ and $\delta^{\phi=\Phi/2} = 1 $, then $\delta^{HOTI} = 2 \mod 4$ states are continuously pumped onto the Wyckoff position as $\phi$ is increased to $\Phi/2$.

We find that the SM invariants are $\mathds{Z}$-valued with rotation symmetries and $\mathds{Z}_2$-valued with $M\mathcal{T}$. In some cases, an SM index depends only on the $\phi = 0$ RSIs and $\gamma$ which determines the projective group at $\phi = \Phi/2$ and hence is Peierls-indicated. The HOTI invariants are only $\mathds{Z}_2$-valued and in some cases (like with $2_\pi'$) are Peierls-indicated. The value of the HOTI index is the number of charges transferred, e.g. in PG $4'$ with $\gamma = \pi$, a nonzero the HOTI index $2 \delta^{\Phi/2}_1 \mod 4$ means that $2 \mod 4$ electrons are pumped off the Wyckoff position as $\phi$ is increased to $\Phi/2$. All projective groups and their irreps can be found at \href{www.cryst.ehu.es/cryst/projectiverepres}{Bilbao Crystallographic Server}.

	\begin{scriptsize}
		\label{tab:HofinvnoSOC}
		\begin{longtable}{|l|l|l|l|l|l|l|}
		\caption{Hofstadter topological invariants without SOC. For brevity, we abbreviate $\delta_i^{\phi=0} \to \delta_i^0$ and $\delta_i^{\phi = \Phi/2}\to \delta_i^{\Phi/2}$, etc. }\\ 
			\hline
			$G^{\phi=0}$&$G^{\phi=\Phi/2}$&$G^{\phi}$&$N^{\phi=0}$&$N^{\phi=\Phi/2}$&SM &HOTI\\
			\hline
			$1$&$1$&1&&&&\\
			\hline
			$11'$&$11'$&1&&&&\\
			\hline
			$m$&$m$&1&$\dff$ mod 2&$\dFf$ mod 2&&$\dff-\dFf$ mod 2\\
			\hline
			$m1'$&$m1'$&$m'$&$\dff$ mod 2&$\dFf$ mod 2&$\dff-\dFf$ mod 2&\\
			\hline
			$m'$&$m'$&$m'$&$\dff$ mod 2&$\dFf$ mod 2&$\dff-\dFf$ mod 2&\\
			\hline
			$2$&$2$&2&$\dff$ mod 2&$\dFf$ mod 2& $\delta_1^0- \delta_1^{\Phi}$& \\
			\hline
			$2'$&$2'$&1&$\dff$ mod 2&$\dFf$ mod 2&&$\dff-\dFf$ mod 2\\
			\hline			
			$2'$&$2_{\pi}'$&1&$\dff$ mod 2&0 mod 2&&$\dff$ mod 2\\
			\hline
			$21'$&$21'$&2&$\dff$ mod 2&$\dFf$ mod 2&$\dff-\dFf$&\\
			\hline
			$21'$&$2_{\pi}1'$&2&$\dff$ mod 2&0 mod 2&$\dff$ &\\
			\hline
			$2mm$&$2mm$&2&$\dff$ mod 2&$\dFf$ mod 2&$\dff-\dFf$&\\
			\hline
			$2mm$&$2_{\pi}mm$&2&$\dff$ mod 2&0 mod 2&$\dff$&\\
			\hline
			$2mm1'$&$2mm1'$&$2m'm'$&$\dff$ mod 2&$\dFf$ mod 2&$\dff-\dFf$&\\
			\hline
			$2mm1'$&$2_{\pi}mm1'$&$2m'm'$&$\dff$ mod 2&0 mod 2&$\dff$&\\
			\hline
			$2'mm'$&$2'mm'$&$m'$&$\dff$ mod 2&$\dFf$ mod 2&$\dff-\dFf$&\\
			\hline
			$2'mm'$&$2_{\pi}'mm'$&$m'$&$\dff$ mod 2&0 mod 2&$\dff$&\\
			\hline
			$2m'm'$&$2m'm'$&$2m'm'$&$\dff$ mod 2&$\dFf$ mod 2&$\delta_1^0- \delta_1^{\Phi}$& \\
			\hline
			$4$&$4$&4&$\dff+\dfs+\dft$ mod 4&$\dFf+\dFs+\dFt$ mod 4&$\delta_1^0- \delta_1^{\Phi},\delta_2^0- \delta_2^{\Phi},\delta_3^0- \delta_3^{\Phi}$&\\
			\hline
			$41'$&$41'$&4&$\dff+2\dfs$ mod 4&$\dFf+2\dFs$ mod 4&$\dff-\dFf$,$\dfs-\dFs$&\\
			\hline
			$41'$&$4_{\pi/2}1'$&4&$\dff+2\dfs$ mod 4&$2\dFf$ mod4&$\dff$, $\dfs$, $\dFf$&\\
			\hline
			$41'$&$4_{\pi}1'$&4&$\dff+2\dfs$ mod 4&$\dFf+2\dFs$ mod 4&$\dff$, $\dFf$, $\dfs+\dFs$&\\
			\hline
			$41'$&$4_{3\pi/2}1'$&4&$\dff+2\dfs$ mod 4&$2\dFf$ mod4&$\dff$, $\dfs$, $\dFf$&\\
			\hline
			$4'$&$4'$&2&$\dff$ mod 4&$\dFf$ mod 4&$\dff-\dFf$&\\
			\hline
			$4'$&$4_{\pi/2}'$&2&$\dff$ mod 4&$2\dFf$ mod 4&$\dff$&$2\dFf$ mod 4\\
			\hline
			$4'$&$4_{\pi}'$&2&$\dff$ mod 4&$-\dFf$ mod 4&$\dff-\dFf$&$2\dFf$ mod 4\\
			\hline
			$4'$&$4_{3\pi/2}'$&2&$\dff$ mod 4&$2\dFf$ mod 4&$\dff$&$2\dFf$ mod 4\\
			\hline
			$4mm$&$4mm$&4&$\dff+2\dfs$ mod 4&$\dFf+2\dFs$ mod 4&$\dff-\dFf$, $\dfs-\dFs$&\\
			\hline
			$4mm$&$4_{\pi/2}mm$&4&$\dff+2\dfs$ mod 4&$2\dFf$ mod 4&$\dff$, $\dfs$, $\dFf$&\\
			\hline
			$4mm$&$4_{\pi}mm$&4&$\dff+2\dfs$ mod 4&$\dFf+2\dFs$ mod 4&$\dff$, $\dFf$, $\dfs+\dFs$&\\
			\hline
			$4mm$&$4_{3\pi/2}mm$&4&$\dff+2\dfs$ mod 4&$2\dFf$ mod 4&$\dff$, $\dfs$, $\dFf$&\\
			\hline
			$4mm1'$&$4mm1'$&$4m'm'$&$\dff+2\dfs$ mod 4&$\dFf+2\dFs$ mod 4&$\dff-\dFf$, $\dfs-\dFs$&\\
			\hline
			$4mm1'$&$4_{\pi/2}mm1'$&$4m'm'$&$\dff+2\dfs$ mod 4&$2\dFf$&$\dff$, $\dfs$, $\dFf$&\\
			\hline
			$4mm1'$&$4_{\pi}mm1'$&$4m'm'$&$\dff+2\dfs$ mod 4&$\dFf+2\dFs$ mod 4&$\dff$, $\dFf$, $\dfs+\dFs$&\\
			\hline
			$4mm1'$&$4_{3\pi/2}mm1'$&$4m'm'$&$\dff+2\dfs$ mod 4&$2\dFf$ mod 4&$\dff$, $\dfs$, $\dFf$&\\
			\hline
			$4'm'm$&$4'm'm$&$2m'm'$&$\dff$ mod 4&$\dFf$ mod 4&$\dff-\dFf$&\\
			\hline
			$4'm'm$&$4_{\pi/2}'m'm$&$2m'm'$&$\dff$ mod 4&$2\dFf$ mod 4&$\dff$&$2\dFf$ mod 4\\
			\hline
			$4'm'm$&$4_{\pi}'m'm$&$2m'm'$&$\dff$ mod 4&$-\dFf$ mod 4&$\dff-\dFf$&$2\dFf$ mod 4\\
			\hline
			$4'm'm$&$4_{3\pi/2}'m'm$&$2m'm'$&$\dff$ mod 4&$2\dFf$ mod 4&$\dff$&$2\dFf$ mod 4\\
			\hline
			$4m'm'$&$4m'm'$&4m'm'&$\dff+\dfs+\dft$ mod 4&$\dFf+\dFs+\dFt$ mod 4& $\delta_1^0- \delta_1^{\Phi},\delta_2^0- \delta_2^{\Phi},\delta_3^0- \delta_3^{\Phi}$ &\\
			\hline
			$3$&$3$&3&$\dff+\dfs$ mod 3&$\dFf+\dFs$ mod 3&$\delta_1^0- \delta_1^{\Phi},\delta_2^0- \delta_2^{\Phi}$&\\
			\hline
			$31'$&$31'$&3&$2\dff$ mod 3&$2\dFf$ mod 3&$\dff-\dFf$&\\
			\hline
			$31'$&$3_{2\pi/3}1'$&3&$2\dff$ mod 3&$\dFf$ mod 3&$\dff$, $\dFf$&\\
			\hline
			$31'$&$3_{4\pi/3}1'$&3&$2\dff$ mod 3&$\dFf$ mod 3&$\dff$, $\dFf$&\\
			\hline
			$3m$&$3m$&3&$2\dff$ mod 3&$2\dFf$ mod 3&$\dff-\dFf$&\\
			\hline
			$3m$&$3_{2\pi/3}m$&3&$2\dff$ mod 3&$\dFf$ mod 3&$\dff$, $\dFf$&\\
			\hline
			$3m$&$3_{4\pi/3}m$&3&$2\dff$ mod 3&$\dFf$ mod 3&$\dff$, $\dFf$&\\
			\hline
			$3m1'$&$3m1'$&$3m'$&$2\dff$ mod 3&$2\dFf$ mod 3&$\dff-\dFf$&\\
			\hline
			$3m1'$&$3_{2\pi/3}m1'$&$3m'$&$2\dff$ mod 3&$\dFf$ mod 3&$\dff$, $\dFf$&\\
			\hline
			$3m1'$&$3_{4\pi/3}m1'$&$3m'$&$2\dff$ mod 3&$\dFf$ mod 3&$\dff$, $\dFf$&\\
			\hline
			$3m'$&$3m'$&$3m'$&$\dff+\dfs$ mod 3&$\dFf+\dFs$ mod 3&$\delta_1^0- \delta_1^{\Phi},\delta_2^0- \delta_2^{\Phi}$&\\
			\hline
			$6$&$6$&6&$\dff+\dfs+\dft+$ &$\dFf+\dFs+\dFt+$ &$\delta^0_1-\delta^{\Phi}_1,\delta^0_2-\delta^{\Phi}_2,\delta^0_3-\delta^{\Phi}_3,$&\\
			&&&$\delta^0_4+\delta^0_5$ mod 6& $\delta^{\Phi/2}_4+\delta^{\Phi/2}_5$ mod 6 &$\delta^0_4-\delta^{\Phi}_4,\delta^0_5-\delta^{\Phi}_5$&\\
			\hline
			$61'$&$61'$&6&$\dff+2\dfs+2\dft$ mod 6&$\dFf+2\dFs+2\dFt$ mod 6&$\delta_1^0-\delta_1^{\Phi/2},\delta_2^0-\delta_2^{\Phi/2},\delta_3^0-\delta_3^{\Phi/2}$&\\
			\hline
			$61'$&$6_{\pi/3}1'$&6&$\dff+2\dfs+2\dft$ mod 6&$2\dFf+2\dFs$ mod 6&$\dff$,$\dfs$,$\dft$, $\dFf$, $\dFs$&\\
			\hline
			$61'$&$6_{2\pi/3}1'$&6&$\dff+2\dfs+2\dft$ mod 6&$\dFf+2\dFs+2\dFt$ mod 6&$\dff-\dft$, $\dfs$, $\dff-\dFf$&\\
			&&&&&$\dFf-\dFt$, $\dFs$&\\
			\hline
			$61'$&$6_{\pi}1'$&6&$\dff+2\dfs+2\dft$ mod 6&$2\dFf+2\dFs$ mod 6&$\dff$, $\dfs+\dFs$&\\
			&&&&&$\dFf$, $\dft+\dFs$&\\
			\hline
			$61'$&$6_{4\pi/3}1'$&6&$\dff+2\dfs+2\dft$ mod 6&$\dFf+2\dFs+2\dFt$ mod 6&$\dff-\dft$, $\dfs$, $\dff-\dFf$&\\
			&&&&&$\dFf-\dFt$, $\dFs$&\\
			\hline
			$61'$&$6_{5\pi/3}1'$&6&$\dff+2\dfs+2\dft$ mod 6&$2\dFf+2\dFs$ mod 6&$\dff$,$\dfs$,$\dft$, $\dFf$, $\dFs$&\\
			\hline
			$6'$&$6'$&$3$&$2\dff+3\dfs$ mod 6&$2\dFf+3\dFs$ mod 6&$\dff-\dFf$&$3(\dfs-\dFs)$ mod 6\\
			\hline
			$6'$&$6_{\pi/3}'$&3&$2\dff+3\dfs$ mod 6&$4\dFf$ mod 6&$\dff$, $\dFf$&$3\dfs$ mod 6\\
			\hline
			$6'$&$6_{2\pi/3}'$&3&$2\dff+3\dfs$ mod 6&$4\dFf+3\dFs$ mod 6&$\dff$, $\dFf$&$3(\dfs-\dFs)$ mod 6\\
			\hline
			$6'$&$6_{\pi}'$&3&$2\dff+3\dfs$ mod 6&$2\dFf$ mod 6&$\dff-\dFf$&$3\dfs$ mod 6\\
			\hline
			$6'$&$6_{4\pi/3}'$&3&$2\dff+3\dfs$ mod 6&$4\dFf+3\dFs$ mod 6&$\dff$, $\dFf$&$3(\dfs-\dFs)$ mod 6\\
			\hline
			$6'$&$6_{5\pi/3}'$&3&$2\dff+3\dfs$ mod 6&$4\dFf$ mod 6&$\dff$, $\dFf$&$3\dfs$ mod 6\\
			\hline
			$6mm$&$6mm$&6&$\dff+2\dfs+2\dft$ mod 6&$\dFf+2\dFs+2\dFt$ mod 6&$\delta_1^0-\delta_1^{\Phi/2},\delta_2^0-\delta_2^{\Phi/2},\delta_3^0-\delta_3^{\Phi/2}$&\\
			\hline
			$6mm$&$6_{\pi/3}mm$&6&$\dff+2\dfs+2\dft$ mod 6&$2\dFf+2\dFs$ mod 6&$\dff$,$\dfs$,$\dft$, $\dFf$, $\dFs$&\\
			\hline
			$6mm$&$6_{2\pi/3}mm$&6&$\dff+2\dfs+2\dft$ mod 6&$\dFf+2\dFs+2\dFt$ mod 6&$\dff-\dft$, $\dfs$, $\dff-\dFf$&\\
			&&&&&$\dFf-\dFt$, $\dFs$&\\
			\hline
			$6mm$&$6_{\pi}mm$&6&$\dff+2\dfs+2\dft$ mod 6&$2\dFf+2\dFs$ mod 6&$\dff$, $\dfs+\dFs$&\\
			&&&&&$\dFf$, $\dft+\dFs$&\\
			\hline
			$6mm$&$6_{4\pi/3}mm$&6&$\dff+2\dfs+2\dft$ mod 6&$\dFf+2\dFs+2\dFt$ mod 6&$\dff-\dft$, $\dfs$, $\dff-\dFf$&\\
			&&&&&$\dFf-\dFt$, $\dFs$&\\
			\hline
			$6mm$&$6_{5\pi/3}mm$&6&$\dff+2\dfs+2\dft$ mod 6&$2\dFf+2\dFs$ mod 6&$\dff$,$\dfs$,$\dft$, $\dFf$, $\dFs$&\\
			\hline
			$6mm1'$&$6mm1'$&$6m'm'$&$\dff+2\dfs+2\dft$ mod 6&$\dFf+2\dFs+2\dFt$ mod 6&$\delta_1^0-\delta_1^{\Phi/2},\delta_2^0-\delta_2^{\Phi/2},\delta_3^0-\delta_3^{\Phi/2}$&\\
			\hline
			$6mm1'$&$6_{\pi/3}mm1'$&$6m'm'$&$\dff+2\dfs+2\dft$ mod 6&$2\dFf+2\dFs$ mod 6&$\dff$,$\dfs$,$\dft$, $\dFf$, $\dFs$&\\
			\hline
			$6mm1'$&$6_{2\pi/3}mm1'$&$6m'm'$&$\dff+2\dfs+2\dft$ mod 6&$\dFf+2\dFs+2\dFt$ mod 6&$\dff-\dft$, $\dfs$, $\dff-\dFf$&\\
			&&&&&$\dFf-\dFt$, $\dFs$&\\
			\hline
			$6mm1'$&$6_{\pi}mm1'$&$6m'm'$&$\dff+2\dfs+2\dft$ mod 6&$2\dFf+2\dFs$ mod 6&$\dff$, $\dfs+\dFs$&\\
			&&&&&$\dFf$, $\dft+\dFs$&\\
			\hline
			$6mm1'$&$6_{4\pi/3}mm1'$&$6m'm'$&$\dff+2\dfs+2\dft$ mod 6&$\dFf+2\dFs+2\dFt$ mod 6&$\dff-\dft$, $\dfs$, $\dff-\dFf$&\\
			&&&&&$\dFf-\dFt$, $\dFs$&\\
			\hline
			$6mm1'$&$6_{5\pi/3}mm1'$&$6m'm'$&$\dff+2\dfs+2\dft$ mod 6&$2\dFf+2\dFs$ mod 6&$\dff$,$\dfs$,$\dft$, $\dFf$, $\dFs$&\\
			\hline
			$6'mm'$&$6'mm'$&$3m'$&$2\dff + 3 \dfs$ mod 6&$2\dFf+ 3 \dFt$ mod 6&$\dff-\dFf$&$3\left(\dfs-\dFs\right)$ mod 6\\
			\hline
			$6'mm'$&$6_{\pi/3}'mm'$&$3m'$&$2\dff+ 3 \dfs$ mod 6&$4\dFf$ mod 6&$\dff$, $\dFf$&$3\dfs$ mod 6\\
			\hline
			$6'mm'$&$6_{2\pi/3}'mm'$&$3m'$&$2\dff+ 3 \dfs$ mod 6&$4\dFf + 3 \dFs$ mod 6&$\dff$, $\dFf$&$3\left(\dfs-\dFs\right)$ mod 6\\
			\hline
			$6'mm'$&$6_{\pi}'mm'$&$3m'$&$2\dff+ 3 \dfs$ mod 6&$2\dFf$ mod 6&$\dff-\dFf$&$3\dfs$ mod 6\\
			\hline
			$6'mm'$&$6_{4\pi/3}'mm'$&$3m'$&$2\dff+ 3 \dfs$ mod 6&$4\dFf + 3 \dFs$ mod 6&$\dff$, $\dFf$&$3\left(\dfs-\dFs\right)$ mod 6\\
			\hline
			$6'mm'$&$6_{5\pi/3}'mm'$&$3m'$&$2\dff+ 3 \dfs$ mod 6&$4\dFf$ mod 6&$\dff$, $\dFf$&$3\dfs$ mod 6\\
			\hline
			$6m'm'$&$6m'm'$&6m'm'&$\dff+\dfs+\dft+$ &$\dFf+\dFs+\dFt+$ &$\delta^0_1-\delta^{\Phi}_1,\delta^0_2-\delta^{\Phi}_2,\delta^0_3-\delta^{\Phi}_3,$&\\
						&&&$\delta^0_4+\delta^0_5$ mod 6& $\delta^{\Phi/2}_4+\delta^{\Phi/2}_5$ mod 6 &$\delta^0_4-\delta^{\Phi}_4,\delta^0_5-\delta^{\Phi}_5$&\\
			\hline
		\end{longtable}
	\end{scriptsize}

\subsection{SOC}

 \Tab{tab:RSIsSOC} is a complete list of all RSIs in the conventional 31 2D magnetic PGs and the 51 projective PGs for spin $1/2$ particles.  \Tab{tab:RSIsnoSOC} is calculated using the group theoretical method of \Ref{song2019real} (or equivalently the physical description in \Eq{eq:RSIinduct}). We find that all RSIs groups are products of $\mathds{Z}$ and/or $\mathds{Z}_2$. All projective groups and their irreps can be found at \href{www.cryst.ehu.es/cryst/projectiverepres}{Bilbao Crystallographic Server}.

\begin{center}
	\begin{scriptsize}
		\label{tab:RSIsSOC}
\begin{longtable}{l|lllll}
	\hline
	1&&\\
\hline
$11'$&&\\
\hline
$2$&$\delta_1=-m(\,^{1}\overline{E})+m(\,^{2}\overline{E})$&\\
\hline
$21'$&$\delta_1=+m(\,^{2}\overline{E}^{1}\overline{E})\hspace{0.5cm}\textrm{mod 2}$&\\
\hline
$2_{\pi}1'$&$\delta_1=-m(\,^{1}\overline{E}^{1}\overline{E})+m(\,^{2}\overline{E}^{2}\overline{E})$&\\
\hline
$2'$&$\delta_1=+m(\overline{A})\hspace{0.5cm}\textrm{mod 2}$&\\
\hline
$2_{\pi}'$&&\\
\hline
$m$&$\delta_1=-m(\,^{1}\overline{E})+m(\,^{2}\overline{E})$&\\
\hline
$m1'$&$\delta_1=+m(\,^{2}\overline{E}^{1}\overline{E})\hspace{0.5cm}\textrm{mod 2}$&\\
\hline
$m'$&$\delta_1=+m(\overline{A})\hspace{0.5cm}\textrm{mod 2}$&\\
\hline
$2mm$&&\\
\hline
$2_{\pi}mm$&$\delta_1=-m(\,^{1}\overline{E}_{1})-m(\,^{2}\overline{E}_{1})+m(\,^{1}\overline{E}_{2})+m(\,^{2}\overline{E}_{2})$&\\
\hline
$2mm1'$&$\delta_1=+m(\overline{E})\hspace{0.5cm}\textrm{mod 2}$&\\
\hline
$2_{\pi}mm1'$&$\delta_1=-m(\,^{1}\overline{E}_{1}^{2}\overline{E}_{1})+m(\,^{1}\overline{E}_{2}^{2}\overline{E}_{2})$&\\
\hline
$2'mm'$&$\delta_1=-m(\,^{1}\overline{E})+m(\,^{2}\overline{E})\hspace{0.5cm}\textrm{mod 2}$&\\
\hline
$2_{\pi}'m'm$&&\\
\hline
$2m'm'$&$\delta_1=-m(\,^{1}\overline{E})+m(\,^{2}\overline{E})$&\\
\hline
$4$&$\delta_1=-m(\,^{1}\overline{E}_{1})+m(\,^{1}\overline{E}_{2})$&$\delta_2=-m(\,^{1}\overline{E}_{1})+m(\,^{2}\overline{E}_{2})$\\&$\delta_3=-m(\,^{1}\overline{E}_{1})+m(\,^{2}\overline{E}_{1})$&\\
\hline
$41'$&$\delta_1=-m(\,^{2}\overline{E}_{2}^{1}\overline{E}_{2})+m(\,^{2}\overline{E}_{1}^{1}\overline{E}_{1})$&$\delta_2=+m(\,^{2}\overline{E}_{2}^{1}\overline{E}_{2})\hspace{0.5cm}\textrm{mod 2}$\\
\hline
$4_{\pi/2}1'$&$\delta_1=-m(\,^{2}\overline{E}_{2}^{2}\overline{E}_{2})+m(\,^{2}\overline{E}_{1}^{2}\overline{E}_{1})$&$\delta_2=-2m(\,^{2}\overline{E}_{2}^{2}\overline{E}_{2})+m(\,^{1}\overline{E}_{1}^{1}\overline{E}_{2})$\\
\hline
$4_{\pi}1'$&$\delta_1=-m(\,^{1}\overline{E}_{1}^{2}\overline{E}_{2})+m(\,^{1}\overline{E}_{2}^{2}\overline{E}_{1})$&$\delta_2=+m(\,^{1}\overline{E}_{1}^{2}\overline{E}_{2})\hspace{0.5cm}\textrm{mod 2}$\\
\hline
$4_{3\pi/2}1'$&$\delta_1=-m(\,^{1}\overline{E}_{1}^{1}\overline{E}_{1})+m(\,^{1}\overline{E}_{2}^{1}\overline{E}_{2})$&$\delta_2=-2m(\,^{1}\overline{E}_{1}^{1}\overline{E}_{1})+m(\,^{2}\overline{E}_{1}^{2}\overline{E}_{2})$\\
\hline
$4'$&$\delta_1=+m(\,^{2}\overline{E}^{1}\overline{E})\hspace{0.5cm}\textrm{mod 2}$&\\
\hline
$4_{\pi/2}'$&$\delta_1=+m(\,^{1}\overline{E})-2m(\,^{2}\overline{E}^{2}\overline{E})$&\\
\hline
$4_{\pi}'$&$\delta_1=+m(\,^{1}\overline{E}^{2}\overline{E})\hspace{0.5cm}\textrm{mod 2}$&\\
\hline
$4_{3\pi/2}'$&$\delta_1=+m(\,^{2}\overline{E})-2m(\,^{1}\overline{E}^{1}\overline{E})$&\\
\hline
$4mm$&$\delta_1=+m(\overline{E}_{2})-m(\overline{E}_{1})$&\\
\hline
$4_{\pi/2}mm$&$\delta_1=+m(\,^{2}\overline{E}_{2}')+m(\,^{2}\overline{E}_{2}'')-m(\,^{2}\overline{E}_{1}')-m(\,^{2}\overline{E}_{1}'')$&$\delta_2=+m(\,^{2}\overline{E}_{2}')+m(\,^{2}\overline{E}_{2}'')-m(\,^{1}\overline{E}_{1}^{1}\overline{E}_{2})$\\
\hline
$4_{\pi}mm$&$\delta_1=-m(\,^{1}\overline{E}_{1}^{2}\overline{E}_{2})+m(\,^{1}\overline{E}_{2}^{2}\overline{E}_{1})$&\\
\hline
$4_{3\pi/2}mm$&$\delta_1=-m(\,^{1}\overline{E}_{1}')-m(\,^{1}\overline{E}_{1}'')+m(\,^{1}\overline{E}_{2}')+m(\,^{1}\overline{E}_{2}'')$&$\delta_2=-m(\,^{1}\overline{E}_{1}')-m(\,^{1}\overline{E}_{1}'')+m(\,^{2}\overline{E}_{1}^{2}\overline{E}_{2})$\\
\hline
$4mm1'$&$\delta_1=-m(\overline{E}_{2})+m(\overline{E}_{1})$&$\delta_2=+m(\overline{E}_{2})\hspace{0.5cm}\textrm{mod 2}$\\
\hline
$4_{\pi/2}mm1'$&$\delta_1=-m(\,^{2}\overline{E}_{2}'^{2}\overline{E}_{2}'')+m(\,^{2}\overline{E}_{1}'^{2}\overline{E}_{1}'')$&$\delta_2=-2m(\,^{2}\overline{E}_{2}'^{2}\overline{E}_{2}'')+m(\,^{1}\overline{E}_{1}^{1}\overline{E}_{2})$\\
\hline
$4_{\pi}mm1'$&$\delta_1=-m(\,^{1}\overline{E}_{1}^{2}\overline{E}_{2})+m(\,^{1}\overline{E}_{2}^{2}\overline{E}_{1})$&$\delta_2=+m(\,^{1}\overline{E}_{1}^{2}\overline{E}_{2})\hspace{0.5cm}\textrm{mod 2}$\\
\hline
$4_{3\pi/2}mm1'$&$\delta_1=-m(\,^{1}\overline{E}_{1}'^{1}\overline{E}_{1}'')+m(\,^{1}\overline{E}_{2}'^{1}\overline{E}_{2}'')$&$\delta_2=-2m(\,^{1}\overline{E}_{1}'^{1}\overline{E}_{1}'')+m(\,^{2}\overline{E}_{1}^{2}\overline{E}_{2})$\\
\hline
$4'm'm$&$\delta_1=+m(\overline{E})\hspace{0.5cm}\textrm{mod 2}$&\\
\hline
$4_{\pi/2}'m'm$&$\delta_1=+m(\,^{1}\overline{E}_{1})+m(\,^{2}\overline{E}_{1})-2m(\,^{1}\overline{E}_{2}^{2}\overline{E}_{2})$&\\
\hline
$4_{\pi}'m'm$&$\delta_1=+m(\,^{1}\overline{E}^{2}\overline{E})\hspace{0.5cm}\textrm{mod 2}$&\\
\hline
$4_{3\pi/2}'m'm$&$\delta_1=+m(\,^{1}\overline{E}_{2})+m(\,^{2}\overline{E}_{2})-2m(\,^{1}\overline{E}_{1}^{2}\overline{E}_{1})$&\\
\hline
$4m'm'$&$\delta_1=-m(\,^{1}\overline{E}_{1})+m(\,^{1}\overline{E}_{2})$&$\delta_2=-m(\,^{1}\overline{E}_{1})+m(\,^{2}\overline{E}_{2})$\\&$\delta_3=-m(\,^{1}\overline{E}_{1})+m(\,^{2}\overline{E}_{1})$&\\
\hline
$3$&$\delta_1=+m(\,^{1}\overline{E})-m(\,^{2}\overline{E})$&$\delta_2=+m(\,^{1}\overline{E})-m(\overline{E})$\\
\hline
$31'$&$\delta_1=+m(\,^{1}\overline{E}^{2}\overline{E})-2m(\overline{E}\overline{E})$&\\
\hline
$3_{2\pi/3}1'$&$\delta_1=-2m(\,^{1}\overline{E}^{1}\overline{E})+m(\,^{2}\overline{E}\overline{E})$&\\
\hline
$3_{4\pi/3}1'$&$\delta_1=+2m(\,^{2}\overline{E}^{2}\overline{E})-m(\,^{1}\overline{E}\overline{E})$&\\
\hline
$3m$&$\delta_1=-m(\overline{E}_{1})+m(\,^{2}\overline{E})+m(\,^{1}\overline{E})$&\\
\hline
$3_{2\pi/3}m$&$\delta_1=+m(\,^{1}\overline{E}_{2})+m(\,^{1}\overline{E}_{1})-m(\,^{2}\overline{E}\overline{E})$&\\
\hline
$3_{4\pi/3}m$&$\delta_1=-m(\,^{2}\overline{E}_{1})-m(\,^{2}\overline{E}_{2})+m(\,^{1}\overline{E}\overline{E})$&\\
\hline
$3m1'$&$\delta_1=+m(\overline{E}_{1})-2m(\,^{2}\overline{E}^{1}\overline{E})$&\\
\hline
$3_{2\pi/3}m1'$&$\delta_1=-2m(\,^{1}\overline{E}_{1}^{1}\overline{E}_{2})+m(\,^{2}\overline{E}\overline{E})$&\\
\hline
$3_{4\pi/3}m1'$&$\delta_1=+2m(\,^{2}\overline{E}_{1}^{2}\overline{E}_{2})-m(\,^{1}\overline{E}\overline{E})$&\\
\hline
$3m'$&$\delta_1=+m(\,^{1}\overline{E})-m(\,^{2}\overline{E})$&$\delta_2=+m(\,^{1}\overline{E})-m(\overline{E})$\\
\hline
$6$&$\delta_1=-m(\,^{1}\overline{E}_{3})+m(\,^{2}\overline{E}_{2})$&$\delta_2=-m(\,^{1}\overline{E}_{3})+m(\,^{1}\overline{E}_{2})$\\&$\delta_3=-m(\,^{1}\overline{E}_{3})+m(\,^{2}\overline{E}_{3})$&$\delta_4=-m(\,^{1}\overline{E}_{3})+m(\,^{1}\overline{E}_{1})$\\&$\delta_5=-m(\,^{1}\overline{E}_{3})+m(\,^{2}\overline{E}_{1})$&\\
\hline
$61'$&$\delta_1=-m(\,^{1}\overline{E}_{2}^{2}\overline{E}_{2})+m(\,^{1}\overline{E}_{3}^{2}\overline{E}_{3})$&$\delta_2=-m(\,^{1}\overline{E}_{2}^{2}\overline{E}_{2})+m(\,^{2}\overline{E}_{1}^{1}\overline{E}_{1})$\\&$\delta_3=+m(\,^{1}\overline{E}_{2}^{2}\overline{E}_{2})\hspace{0.5cm}\textrm{mod 2}$&\\
\hline
$6_{\pi/3}1'$&$\delta_1=-m(\,^{1}\overline{E}_{3}^{1}\overline{E}_{3})+m(\,^{2}\overline{E}_{2}^{2}\overline{E}_{2})$&$\delta_2=-2m(\,^{1}\overline{E}_{3}^{1}\overline{E}_{3})+m(\,^{1}\overline{E}_{1}^{1}\overline{E}_{2})$\\&$\delta_3=-2m(\,^{1}\overline{E}_{3}^{1}\overline{E}_{3})+m(\,^{2}\overline{E}_{1}^{2}\overline{E}_{3})$&\\
\hline
$6_{2\pi/3}1'$&$\delta_1=-m(\,^{1}\overline{E}_{2}^{2}\overline{E}_{3})+m(\,^{1}\overline{E}_{1}^{2}\overline{E}_{2})$&$\delta_2=-m(\,^{1}\overline{E}_{2}^{2}\overline{E}_{3})+m(\,^{2}\overline{E}_{1}^{1}\overline{E}_{3})$\\&$\delta_3=+m(\,^{1}\overline{E}_{2}^{2}\overline{E}_{3})\hspace{0.5cm}\textrm{mod 2}$&\\
\hline
$6_{\pi}1'$&$\delta_1=-m(\,^{1}\overline{E}_{1}^{1}\overline{E}_{1})+m(\,^{2}\overline{E}_{1}^{2}\overline{E}_{1})$&$\delta_2=-2m(\,^{1}\overline{E}_{1}^{1}\overline{E}_{1})+m(\,^{1}\overline{E}_{2}^{1}E_{3})$\\&$\delta_3=-2m(\,^{1}\overline{E}_{1}^{1}\overline{E}_{1})+m(\,^{2}\overline{E}_{2}^{2}\overline{E}_{3})$&\\
\hline
$6_{4\pi/3}1'$&$\delta_1=-m(\,^{1}\overline{E}_{3}^{2}\overline{E}_{2})+m(\,^{1}\overline{E}_{2}^{2}\overline{E}_{1})$&$\delta_2=-m(\,^{1}\overline{E}_{3}^{2}\overline{E}_{2})+m(\,^{1}\overline{E}_{1}^{2}\overline{E}_{3})$\\&$\delta_3=+m(\,^{1}\overline{E}_{3}^{2}\overline{E}_{2})\hspace{0.5cm}\textrm{mod 2}$&\\
\hline
$6_{5\pi/3}1'$&$\delta_1=-m(\,^{1}\overline{E}_{2}^{1}\overline{E}_{2})+m(\,^{2}\overline{E}_{3}^{2}\overline{E}_{3})$&$\delta_2=-2m(\,^{1}\overline{E}_{2}^{1}\overline{E}_{2})+m(\,^{1}\overline{E}_{1}^{1}\overline{E}_{3})$\\&$\delta_3=-2m(\,^{1}\overline{E}_{2}^{1}\overline{E}_{2})+m(\,^{2}\overline{E}_{1}^{2}\overline{E}_{2})$&\\
\hline
$6'$&$\delta_1=-m(\,^{1}\overline{E}^{2}\overline{E})+m(\overline{E})$&$\delta_2=+m(\,^{1}\overline{E}^{2}\overline{E})\hspace{0.5cm}\textrm{mod 2}$\\
\hline
$6_{\pi/3}'$&$\delta_1=-2m(\,^{1}\overline{E}^{1}\overline{E})+m(\,^{2}\overline{E}\overline{E})$&\\
\hline
$6_{2\pi/3}'$&$\delta_1=-m(\,^{2}\overline{E})+m(\,^{1}\overline{E}\overline{E})$&$\delta_2=+m(\,^{2}\overline{E})\hspace{0.5cm}\textrm{mod 2}$\\
\hline
$6_{\pi}'$&$\delta_1=+m(\,^{1}\overline{E}^{2}\overline{E})-2m(\overline{E}\overline{E})$&\\
\hline
$6_{4\pi/3}'$&$\delta_1=-m(\,^{1}\overline{E})+m(\,^{2}\overline{E}\overline{E})$&$\delta_2=+m(\,^{1}\overline{E})\hspace{0.5cm}\textrm{mod 2}$\\
\hline
$6_{5\pi/3}'$&$\delta_1=-2m(\,^{2}\overline{E}^{2}\overline{E})+m(\,^{1}\overline{E}\overline{E})$&\\
\hline
$6mm$&$\delta_1=-m(\overline{E}_{2})+m(\overline{E}_{1})$&$\delta_2=-m(\overline{E}_{2})+m(\overline{E}_{3})$\\
\hline
$6_{\pi/3}mm$&$\delta_1=-m(\,^{1}\overline{E}_{3}')-m(\,^{1}\overline{E}_{3}'')+m(\,^{2}\overline{E}_{2}')+m(\,^{2}\overline{E}_{2}'')$&$\delta_2=-m(\,^{1}\overline{E}_{3}')-m(\,^{1}\overline{E}_{3}'')+m(\,^{1}\overline{E}_{1}^{1}\overline{E}_{2})$\\&$\delta_3=-m(\,^{1}\overline{E}_{3}')-m(\,^{1}\overline{E}_{3}'')+m(\,^{2}\overline{E}_{1}^{2}\overline{E}_{3})$&\\
\hline
$6_{2\pi/3}mm$&$\delta_1=-m(\,^{1}\overline{E}_{2}^{2}E_{3})+m(\,^{1}\overline{E}_{1}^{2}\overline{E}_{2})$&$\delta_2=-m(\,^{1}\overline{E}_{2}^{2}E_{3})+m(\,^{2}\overline{E}_{1}^{1}\overline{E}_{3})$\\
\hline
$6_{\pi}mm$&$\delta_1=-m(\,^{1}\overline{E}_{1}')-m(\,^{1}\overline{E}_{1}'')+m(\,^{2}\overline{E}_{1}')+m(\,^{2}\overline{E}_{1}'')$&$\delta_2=-m(\,^{1}\overline{E}_{1}')-m(\,^{1}\overline{E}_{1}'')+m(\,^{1}\overline{E}_{2}^{1}\overline{E}_{3})$\\&$\delta_3=-m(\,^{1}\overline{E}_{1}')-m(\,^{1}\overline{E}_{1}'')+m(\,^{2}\overline{E}_{2}^{2}\overline{E}_{3})$&\\
\hline
$6_{4\pi/3}mm$&$\delta_1=-m(\,^{1}\overline{E}_{3}^{2}E_{2})+m(\,^{1}\overline{E}_{2}^{2}\overline{E}_{1})$&$\delta_2=-m(\,^{1}\overline{E}_{3}^{2}E_{2})+m(\,^{1}\overline{E}_{1}^{2}\overline{E}_{3})$\\
\hline
$6_{5\pi/3}mm$&$\delta_1=-m(\,^{1}\overline{E}_{2}')-m(\,^{1}\overline{E}_{2}'')+m(\,^{2}\overline{E}_{3}')+m(\,^{2}\overline{E}_{3}'')$&$\delta_2=-m(\,^{1}\overline{E}_{2}')-m(\,^{1}\overline{E}_{2}'')+m(\,^{1}\overline{E}_{1}^{1}\overline{E}_{3})$\\&$\delta_3=-m(\,^{1}\overline{E}_{2}')-m(\,^{1}\overline{E}_{2}'')+m(\,^{2}\overline{E}_{1}^{2}\overline{E}_{2})$&\\
\hline
$6mm1'$&$\delta_1=-m(\overline{E}_{2})+m(\overline{E}_{1})$&$\delta_2=-m(\overline{E}_{2})+m(\overline{E}_{3})$\\&$\delta_3=+m(\overline{E}_{2})\hspace{0.5cm}\textrm{mod 2}$&\\
\hline
$6_{\pi/3}mm1'$&$\delta_1=-m(\,^{1}\overline{E}_{3}'^{1}\overline{E}_{3}'')+m(\,^{2}\overline{E}_{2}'^{2}\overline{E}_{2}'')$&$\delta_2=-2m(\,^{1}\overline{E}_{3}'^{1}\overline{E}_{3}'')+m(\,^{1}\overline{E}_{1}^{1}\overline{E}_{2})$\\&$\delta_3=-2m(\,^{1}\overline{E}_{3}'^{1}\overline{E}_{3}'')+m(\,^{2}\overline{E}_{1}^{2}\overline{E}_{3})$&\\
\hline
$6_{2\pi/3}mm1'$&$\delta_1=-m(\,^{1}\overline{E}_{2}^{2}E_{3})+m(\,^{1}\overline{E}_{1}^{2}\overline{E}_{2})$&$\delta_2=-m(\,^{1}\overline{E}_{2}^{2}E_{3})+m(\,^{2}\overline{E}_{1}^{1}\overline{E}_{3})$\\&$\delta_3=+m(\,^{1}\overline{E}_{2}^{2}E_{3})\hspace{0.5cm}\textrm{mod 2}$&\\
\hline
$6_{\pi}mm1'$&$\delta_1=-m(\,^{1}\overline{E}_{1}'^{1}\overline{E}_{1}'')+m(\,^{2}\overline{E}_{1}'^{2}\overline{E}_{1}'')$&$\delta_2=-2m(\,^{1}\overline{E}_{1}'^{1}\overline{E}_{1}'')+m(\,^{1}\overline{E}_{2}^{1}\overline{E}_{3})$\\&$\delta_3=-2m(\,^{1}\overline{E}_{1}'^{1}\overline{E}_{1}'')+m(\,^{2}\overline{E}_{2}^{2}\overline{E}_{3})$&\\
\hline
$6_{4\pi/3}mm1'$&$\delta_1=-m(\,^{1}\overline{E}_{3}^{2}E_{2})+m(\,^{1}\overline{E}_{2}^{2}\overline{E}_{1})$&$\delta_2=-m(\,^{1}\overline{E}_{3}^{2}E_{2})+m(\,^{1}\overline{E}_{1}^{2}\overline{E}_{3})$\\&$\delta_3=+m(\,^{1}\overline{E}_{3}^{2}E_{2})\hspace{0.5cm}\textrm{mod 2}$&\\
\hline
$6_{5\pi/3}mm1'$&$\delta_1=-m(\,^{1}\overline{E}_{2}'^{1}\overline{E}_{2}'')+m(\,^{2}\overline{E}_{3}'^{2}\overline{E}_{3}'')$&$\delta_2=-2m(\,^{1}\overline{E}_{2}'^{1}\overline{E}_{2}'')+m(\,^{1}\overline{E}_{1}^{1}\overline{E}_{3})$\\&$\delta_3=-2m(\,^{1}\overline{E}_{2}'^{1}\overline{E}_{2}'')+m(\,^{2}\overline{E}_{1}^{2}\overline{E}_{2})$&\\
\hline
$6'mm'$&$\delta_1=+m(\overline{E}_{1})-m(\,^{2}\overline{E})-m(\,^{1}\overline{E})$&$\delta_2=-m(\,^{1}\overline{E})+m(\,^{2}\overline{E})\hspace{0.5cm}\textrm{mod 2}$\\
\hline
$6_{\pi/3}'mm'$&$\delta_1=-2m(\,^{1}\overline{E}_{1}^{1}\overline{E}_{2})+m(\,^{2}\overline{E}\overline{E})$&\\
\hline
$6_{2\pi/3}'mm'$&$\delta_1=+m(\,^{2}\overline{E}_{1})+m(\,^{2}\overline{E}_{2})-m(\,^{1}\overline{E}\overline{E})$&$\delta_2=-m(\,^{2}\overline{E}_{1})+m(\,^{2}\overline{E}_{2})\hspace{0.5cm}\textrm{mod 2}$\\
\hline
$6_{\pi}'mm'$&$\delta_1=+m(\overline{E}_{1})-2m(\,^{1}\overline{E}^{2}\overline{E})$&\\
\hline
$6_{4\pi/3}'mm'$&$\delta_1=+m(\,^{2}\overline{E}_{1})+m(\,^{2}\overline{E}_{2})-m(\,^{1}\overline{E}\overline{E})$&$\delta_2=-m(\,^{2}\overline{E}_{1})+m(\,^{2}\overline{E}_{2})\hspace{0.5cm}\textrm{mod 2}$\\
\hline
$6_{5\pi/3}'mm'$&$\delta_1=-2m(\,^{2}\overline{E}_{1}^{2}\overline{E}_{2})+m(\,^{1}\overline{E}\overline{E})$&\\
\hline
$6m'm'$&$\delta_1=-m(\,^{1}\overline{E}_{3})+m(\,^{2}\overline{E}_{2})$&$\delta_2=-m(\,^{1}\overline{E}_{3})+m(\,^{1}\overline{E}_{2})$\\&$\delta_3=-m(\,^{1}\overline{E}_{3})+m(\,^{2}\overline{E}_{3})$&$\delta_4=-m(\,^{1}\overline{E}_{3})+m(\,^{1}\overline{E}_{1})$\\&$\delta_5=-m(\,^{1}\overline{E}_{3})+m(\,^{2}\overline{E}_{1})$&\\
\hline

\end{longtable}
\end{scriptsize}
\end{center}

We now enumerate all symmetry-protected Hofstadter topological invariants in \Tab{tab:HofinvnoSOC} which diagnose Hofstadter SM phases and Hofstadter HOTI phases. We organize the table by listing the crystalline PG at $G^{\phi = 0}$, the (projective) PG at the half-period flux $G^{\phi = \Phi/2}$, and the reduced PG in generic flux $G^{\phi}$. We also calculate the particle number (or total charge) $N$ in terms of the RSIs at $\phi = 0$ and $\phi = \Phi/2$. 
The Hofstadter SM invariants are deduced by requiring compatibility of the $\phi =0$ and $\phi = \Phi$ RSIs when they are reduced in flux to the RSIs of $G^\phi$. If any of the SM invariants are nonzero, a bulk gap closing is enforced. The Hofstadter HOTI indices are given by $N^{\phi=0}-N^{\phi=\Phi/2} \mod n_G$ where $n_G \in \mathbb{N}$ corresponds to the smallest number of particles which can be symmetrically removed from the Wyckoff position. All projective groups and their irreps can be found at \href{www.cryst.ehu.es/cryst/projectiverepres}{Bilbao Crystallographic Server}.

We now give an example of the calculation for PG $G^{\phi=0} = 41'$ with $\gamma = \pi/2$ so that $G^{\phi=\Phi/2} = 4_{\pi/2}1'$, which has a Peierls-indicated SM and HOTI invariant. First we discuss the SM invariant. Both $41'$ and $4_{\pi/2}1'$ are reduced to PG $4$ in generic flux, which breaks their $\mathcal{T}$ and $U\mathcal{T}$ symmetries respectively. The irreps of PG $4$ are ${}^1\overline{E}_1, {}^1\overline{E}_2,{}^2\overline{E}_1, {}^2\overline{E}_2$ which are paired into double irreps by $\mathcal{T}$ or $U\mathcal{T}$. The SM index is protected by the $C_4$ symmetry which is continuous in flux and is Peierls indicated by irrep flow since $\gamma = \pi/2$. We will now see that the SM index $\delta_i^{SM} = \delta_i^{\phi\to0} - \delta_i^{\phi\to\Phi/2}$ diagnoses the the gap closing required by irrep flow. Here $i=1,\dots,3$ since PG $4$ has three RSIs. First we compute the RSI reductions in flux. At $\phi \to 0$, we have ${}^2\overline{E}_2 \, {}^1\overline{E}_2 \to {}^2\overline{E}_2 \oplus {}^1\overline{E}_2$ and ${}^2\overline{E}_1 \, {}^1\overline{E}_1 \to {}^2\overline{E}_1 \oplus {}^1\overline{E}_1$ so that the three RSIs of PG $4$ obey
\bea
\delta_1^{\phi\to0} &= -m({}^1\overline{E}_1) + m({}^1\overline{E}_2) \to - m({}^2\overline{E}_1 \, {}^1\overline{E}_1) + m({}^2\overline{E}_2 \, {}^1\overline{E}_2) =  - \delta^{\phi = 0}_1  \\
\delta_2^{\phi\to0} &= -m({}^1\overline{E}_1) + m({}^2\overline{E}_2) \to = - m({}^2\overline{E}_1 \, {}^1\overline{E}_1) + m({}^2\overline{E}_2 \, {}^1\overline{E}_2) \to - \delta^{\phi = 0}_1  \\
\delta_3^{\phi\to0} &= -m({}^1\overline{E}_1) + m({}^2\overline{E}_1) \to -m( {}^2\overline{E}_1  {}^1\overline{E}_1) + m( {}^2\overline{E}_1  {}^1\overline{E}_1) = 0\ . \\
\eea
At $\phi \to \Phi/2$, we have ${}^2\overline{E}_2 \, {}^2\overline{E}_2 \to {}^2\overline{E}_2 \oplus {}^2\overline{E}_2$, ${}^2\overline{E}_1 \, {}^2\overline{E}_1 \to {}^2\overline{E}_1 \oplus {}^2\overline{E}_1$, and ${}^1\overline{E}_1 \, {}^1\overline{E}_2 \to {}^1\overline{E}_1 \oplus {}^1\overline{E}_2$ so that the three RSIs of PG $4$ obey
\bea
\delta_1^{\phi\to\Phi/2} &= -m({}^1\overline{E}_1) + m({}^1\overline{E}_2) \to -m({}^1\overline{E}_1 \, {}^1\overline{E}_2) + m({}^1\overline{E}_1 \, {}^1\overline{E}_2) = 0 \\
\delta_2^{\phi\to\Phi/2} &= -m({}^1\overline{E}_1) + m({}^2\overline{E}_2) \to  2m({}^2\overline{E}_2{}^2\overline{E}_2) - m({}^1\overline{E}_1{}^1\overline{E}_2) = -\delta_2^{\phi=\Phi/2}  \\
\delta_3^{\phi\to\Phi/2} &= -m({}^1\overline{E}_1) + m({}^2\overline{E}_1) \to 2m(\,^{2}\overline{E}_{1}^{2}\overline{E}_{1}) - m({}^1\overline{E}_1{}^1\overline{E}_2) \\
&= -2m(\,^{2}\overline{E}_{2}^{2}\overline{E}_{2})+2m(\,^{2}\overline{E}_{1}^{2}\overline{E}_{1}) + 2m({}^2\overline{E}_2{}^2\overline{E}_2) - m({}^1\overline{E}_1{}^1\overline{E}_2) =  -\delta_2^{\phi=\Phi/2} +2\delta_1^{\phi = \Phi/2} \ . \\
\eea
Thus we compute the Hofstadter SM indices to be
\bea
\delta_1^{SM} &= \delta_1^{\phi\to0}  - \delta_1^{\phi\to\Phi/2}  =  - \delta^{\phi = 0}_1  - 0 \\
\delta_2^{SM} &= \delta_2^{\phi\to0}  - \delta_2^{\phi\to\Phi/2} =  -\delta^{\phi = 0}_1 +\delta_2^{\phi=\Phi/2} \\
\delta_3^{SM} &= \delta_3^{\phi\to0}  - \delta_3^{\phi\to\Phi/2} = 0 +\delta_2^{\phi=\Phi/2} -2\delta_1^{\phi = \Phi/2} \ . \\
\eea
Note that in \Tab{tab:SMHOTISOC}, we report a different linear combination of $\delta_i^{SM}$ for simplicity, since $\delta^{\phi = 0}_1 = 0, \delta_1^{\phi=\Phi/2} = 0, \delta_2^{\phi=\Phi/2} = 0$ is an equivalent condition for $\delta_i^{SM}= 0 , i = 1,2,3$. We see that since $\delta^{\phi = 0}_1$ is a Hofstadter SM index, there is a Peierls-indicated SM phase corresponding to irrep flow. This is because $\gamma = \pi/2$ interchanges all the irreps, and leads to a gap closing if there is any mismatch in the number of irreps. The $\delta_1^{\phi=\Phi/2},\delta_2^{\phi=\Phi/2}$ RSIs at $\phi = \Phi/2$ also independently diagnose the gap closing. 

We now move on to the Hofstadter HOTI invariant. Firstly, to have a gapped bulk so that the Wannier flow is continuous, we have must $\delta_i^{SM} = 0$, so to define the HOTI index we require that $\delta^{\phi = 0}_1 = \delta_1^{\phi=\Phi/2} =  \delta_2^{\phi=\Phi/2} = 0$. Consulting \Tab{tab:RSIsSOC}, we see that the only RSI allowed to be nonzero is $\delta_2^{\phi=0} = m({}^2\overline{E}_2 \, {}^1\overline{E}_2) \mod 2$. As we will soon find, this RSI diagnoses the HOTI phase. To do the calculation, we observe that the number of particles at the Wyckoff position is adiabatically defined mod $8$ because only 4 Kramers' pairs can be moved offsite. Indeed, we see explicitly that
\bea
N^{\phi =0} &= 2m({}^2\overline{E}_2 \, {}^1\overline{E}_2) + 2m({}^2\overline{E}_1 \, {}^1\overline{E}_1) \!\!\! \mod 8 =  2m({}^2\overline{E}_1 \, {}^1\overline{E}_1) - 2 m({}^2\overline{E}_2 \, {}^1\overline{E}_2)  - 4 m({}^2\overline{E}_2 \, {}^1\overline{E}_2) \!\!\! \mod 8 \\
&= 2 \delta_1^{\phi=0} + 4 \delta_2^{\phi = 0} \mod 8 \\
N^{\phi = \Phi/2} &= 2 m(\,^{2}\overline{E}_{2}^{2}\overline{E}_{2}) + 2 m(\,^{2}\overline{E}_{1}^{2}\overline{E}_{1}) + 2 m(\,^{1}\overline{E}_{1}^{1}\overline{E}_{2}) \mod 8 = 2 \delta_1^{\phi=\Phi/2} + 4 \delta_2^{\phi = \Phi/2} \mod 8
\eea
but enforcing the compatibility conditions $\delta_i^{SM}= 0$ sets $\delta_1^{\phi = 0}   = \delta_1^{\phi = \Phi/2}  = \delta_2^{\phi = \Phi/2}  = 0$, so $N^{\phi = \Phi/2} = 0 \mod 8$. Thus we obtain
\bea
\delta^{HOTI} &= N^{\phi =0} - N^{\phi =\Phi/2} \mod 8 = 4 \delta_2^{\phi = 0} \mod 8
\eea
which is a $\mathds{Z}_2$ invariant. $\delta^{HOTI}  \neq 0$ indicates that as the flux is increased and $\mathcal{T}$ is broken, a quartet of states (as opposed to a quartet of Kramers' pairs) is pumped onto/off of the Wyckoff position. 

In general, we find that the SM invariants are $\mathds{Z}$-valued with rotation symmetries and $\mathds{Z}_2$-valued with $M\mathcal{T}$. In some cases, an SM index depends only on the $\phi = 0$ RSIs and $\gamma$ which determines the projective group at $\phi = \Phi/2$ and hence is Peierls-indicated. The HOTI invariants are all $\mathds{Z}_2$-valued. There are more HOTI indices in the SOC case because $\mathcal{T}^2=-1$ protects Kramers pairs at $\phi =0$ which can be pumped offsite in flux when $\mathcal{T}$ is broken. The value of the HOTI index is the number of charges transferred. 

\begin{center}
	\begin{scriptsize}
			\label{tab:SMHOTISOC}
		\begin{longtable}{|l|l|l|l|l|l|l|}
				\caption{Hofstadter topological invariants with SOC. For brevity, we abbreviate $\delta_i^{\phi=0} \to \delta_i^0$ and $\delta_i^{\phi = \Phi/2}\to \delta_i^{\Phi/2}$, etc.}\\ 
			\hline
			$G^{\phi=0}$&$G^{\phi=\Phi/2}$&$G^{\phi}$&$N^{\phi=0}$&$N^{\phi=\Phi/2}$&SM&HOTI\\
			\hline
$1$&$1$&1&&&&\\ \hline
$11'$&$11'$&1&&&&\\
\hline
$m$&$m$&1&$\dff$ mod 2&$\dFf$ mod 2&&$\dff-\dFf$ mod 2\\
\hline
$m1'$&$m1'$&$m'$&$2\dff$ mod 4&$2\dFf$ mod 4&$\dff-\dFf$ mod 2&\\
\hline
$m'$&$m'$&$m'$&$\dff$ mod 2&$\dFf$ mod 2&$\dff-\dFf$ mod 2&\\
\hline
$2$&$2$&2&$\dff$ mod 2&$\dFf$ mod 2&$\delta_1^{0}-\delta_1^{\Phi}$& \\
\hline
$21'$&$21'$&2&$2\dff$ mod 4&$2\dFf$ mod 4&&$2\dff-2\dFf$ mod 4 \\
\hline
$21'$&$2_{\pi}1'$&2&$2\dff$ mod 4&$2\dFf$ mod 4&$\dFf$&$2\dff$ mod 4\\
\hline
$2'$&$2'$&1&$\dff$ mod 2&$\dFf$ mod 2&&$\dff-\dFf$ mod 2\\
\hline			
$2'$&$2_{\pi}'$&1&$\dff$ mod 2&0 mod 2&&$\dff$ mod 2\\
\hline
$2mm$&$2mm$&2&0 mod 1&0 mod 1&&\\
\hline
$2mm$&$2_{\pi}mm$&2&0 mod 1&$\dFf$ mod 2&&$\dFf$ mod 2\\
\hline
$2mm1'$&$2mm1'$&$2m'm'$&$2\dff$ mod 4&$2\dFf$ mod 4&& $2\dff-2\dFf$ mod 4\\
\hline
$2mm1'$&$2_{\pi}mm1'$&$2m'm'$&$2\dff$ mod 4&$2\dFf$ mod 4&$\dFf$&$2\dff$ mod 4\\
\hline
$2'mm'$&$2'mm'$&$m'$&$\dff$ mod 2&$\dFf$ mod 2&$\dff-\dFf$ mod 2&\\
\hline
$2'mm'$&$2_{\pi}'mm'$&$m'$&$\dff$ mod 2&0 mod 2&$\dff$ mod 2&\\
\hline
			$2m'm'$&$2m'm'$&$2m'm'$&$\dff$ mod 2&$\dFf$ mod 2&$\delta_1^{0}-\delta_1^{\Phi}$& \\
			\hline
			$4$&$4$&4&$\dff+\dfs+\dft$ mod 4&$\dFf+\dFs+\dFt$ mod 4&$\delta_1^{0}-\delta_1^{\Phi},\delta_2^{0}-\delta_2^{\Phi},\delta_3^{0}-\delta_3^{\Phi}$&\\
			\hline

$41'$&$41'$&4&$4\dfs+2\dff$ mod 8&$4\dFs+2\dFf$ mod 8&$\dff+\dFf$&$4(\dfs+\dFs)$ mod 8\\
\hline
$41'$&$4_{\pi/2}1'$&4&$4\dfs+2\dff$ mod 8&$2(\dFs+\dFf)$ mod 8&$\dff$, $\dFf$, $\dFs$&$4\dfs$ mod 8\\
\hline
$41'$&$4_{\pi}1'$&4&$4\dfs+2\dff$ mod 8&$4\dFs+2\dFf$ mod 8&$\dff$, $\dFf$&$4(\dfs-\dFs)$ mod 8\\
\hline
$41'$&$4_{3\pi/2}1'$&4&$4\dfs+2\dff$ mod 8&$2(\dFf+\dFs)$ mod 8&$\dff$, $\dFf$, $\dFs$&$4\dfs$ mod 8\\
\hline
$4'$&$4'$&2&$2\dff$ mod 4&$2\dFf$ mod 4&&$2(\dff-\dFf)$ mod 4\\
\hline
$4'$&$4_{\pi/2}'$&2&$2\dff$ mod 4&$\dFf$ mod 4&$\dFf$&$2\dff$ mod 4\\
\hline
$4'$&$4_{\pi}'$&2&$2\dff$ mod 4&$2\dFf$ mod 4&&$2(\dff-\dFf)$ mod 4\\
\hline
$4'$&$4_{3\pi/2}'$&2&$2\dff$ mod 4&$\dFf$ mod 4&$\dFf$&$2\dff$ mod 4\\
\hline
$4mm$&$4mm$&4&$2\dff$ mod 4&$2\dFf$ mod 4&$\dff-\dFf$&\\
\hline
$4mm$&$4_{\pi/2}mm$&4&$2\dff$ mod 4&$2\dFs-\dFf$ mod 4&$\dff$, $\dFf$, $\dFs$&\\
\hline
$4mm$&$4_{\pi}mm$&4&$2\dff$ mod 4&$2\dFf$ mod 4&$\dff$, $\dFf$&\\
\hline
$4mm$&$4_{3\pi/2}mm$&4&$2\dff$ mod 4&$2\dFs-\dFf$ mod 4&$\dff$, $\dFf$, $\dFs$&\\
\hline
$4mm1'$&$4mm1'$&$4m'm'$&$4\dfs+2\dff$ mod 8&$4\dFs+2\dFf$ mod 8&$\dff+\dFf$&$4(\dfs-\dFs)$ mod 8\\
\hline
$4mm1'$&$4_{\pi/2}mm1'$&$4m'm'$&$4\dfs+2\dff$ mod 8&$2(\dFf+\dFs)$ mod 8&$\dff$, $\dFf$, $\dFs$&$4\dfs$ mod 8\\
\hline
$4mm1'$&$4_{\pi}mm1'$&$4m'm'$&$4\dfs+2\dff$ mod 8&$4\dfs+2\dff$ mod 8&$\dff$, $\dFf$&$4(\dfs-\dFs)$ mod 8\\
\hline
$4mm1'$&$4_{3\pi/2}mm1'$&$4m'm'$&$4\dfs+2\dff$ mod 8&$2(\dFf+\dFs)$ mod 8&$\dff$, $\dFf$, $\dFs$&$4\dfs$ mod 8\\
\hline
$4'm'm$&$4'm'm$&$2m'm'$&$2\dff$ mod 4&$2\dFf$ mod 4&&$2(\dff-\dFf)$ mod 4\\
\hline
$4'm'm$&$4_{\pi/2}'m'm$&$2m'm'$&$2\dff$ mod 4&$\dFf$ mod 4&$\dFf$&$2\dff$ mod 4\\
\hline
$4'm'm$&$4_{\pi}'m'm$&$2m'm'$&$2\dff$ mod 4&$2\dFf$ mod 4&&$2(\dff-\dFf)$ mod 4\\
\hline
$4'm'm$&$4_{3\pi/2}'m'm$&$2m'm'$&$2\dff$ mod 4&$\dFf$ mod 4&$\dFf$&$2\dff$ mod 4\\
\hline
			$4m'm'$&$4m'm'$&$4m'm'$&$\dff+\dfs+\dft$ mod 4&$\dFf+\dFs+\dFt$ mod 4&$\delta_1^{0}-\delta_1^{\Phi},\delta_2^{0}-\delta_2^{\Phi},\delta_3^{0}-\delta_3^{\Phi}$&\\
\hline
			$3$&$3$&3&$\dff+\dfs$ mod 3&$\dFf+\dFs$ mod 3&$\delta_1^{0}-\delta_1^{\Phi},\delta_2^{0}-\delta_2^{\Phi}$&\\
			\hline
$31'$&$31'$&3&$2\dff$ mod 6&$2\dFf$ mod 6&$\dff-\dFf$&\\
\hline
$31'$&$3_{2\pi/3}1'$&3&$2\dff$ mod 6&$2\dFf$ mod 6&$\dff$, $\dFf$&\\
\hline
$31'$&$3_{4\pi/3}1'$&3&$2\dff$ mod 6&$2\dFf$ mod 6&$\dff$, $\dFf$&\\
\hline
$3m$&$3m$&3&$\dff$ mod 3&$\dFf$ mod 3&$\dff-\dFf$&\\
\hline
$3m$&$3_{2\pi/3}m$&3&$\dff$ mod 3&$\dFf$ mod 3&$\dff$, $\dFf$&\\
\hline
$3m$&$3_{4\pi/3}m$&3&$\dff$ mod 3&$\dFf$ mod 3&$\dff$, $\dFf$&\\
\hline
$3m1'$&$3m1'$&$3m'$&$2\dff$ mod 6&$2\dFf$ mod 6&$\dff-\dFf$&\\
\hline
$3m1'$&$3_{2\pi/3}m1'$&$3m'$&$2\dff$ mod 6&$2\dFf$ mod 6&$\dff$, $\dFf$&\\
\hline
$3m1'$&$3_{4\pi/3}m1'$&$3m'$&$2\dff$ mod 6&$2\dFf$ mod 6&$\dff$, $\dFf$&\\
\hline
			$3m'$&$3m'$&$3m'$&$\dff+\dfs$ mod 3&$\dFf+\dFs$ mod 3&$\delta_1^{0}-\delta_1^{\Phi},\delta_2^{0}-\delta_2^{\Phi}$&\\
\hline
			$6$&$6$&6&$\dff+\dfs+\dft+$ &$\dFf+\dFs+\dFt+$ &$\delta^0_1-\delta^{\Phi}_1,\delta^0_2-\delta^{\Phi}_2,\delta^0_3-\delta^{\Phi}_3,$&\\
			&&&$\delta^0_4+\delta^0_5$ mod 6& $\delta^{\Phi/2}_4+\delta^{\Phi/2}_5$ mod 6 &$\delta^0_4-\delta^{\Phi}_4,\delta^0_5-\delta^{\Phi}_5$&\\
			\hline
$61'$&$61'$&6&$2\dff+2\dfs+6\dft$ mod 12&$2\dFf+2\dFs+6\dFt$ mod 12&$\dff-\dFf$, $\dfs-\dFs$&$6(\dft-\dFt)$ mod 12\\
\hline
$61'$&$6_{\pi/3}1'$&6&$2\dff+2\dfs+6\dft$ mod 12&$2(\dFf+\dFs+\dFt)$ mod 12&$\dff$, $\dfs$, $\dFf$, $\dFs$, $\dFt$&$6\dft$ mod 12\\
\hline
$61'$&$6_{2\pi/3}1'$&6&$2\dff+2\dfs+6\dft$ mod 12&$2\dFf+2\dFs+6\dFt$ mod 12&$\dff$, $\dfs$, $\dFf$, $\dFs$&$6(\dft-\dFt)$ mod 12\\
\hline
$61'$&$6_{\pi}1'$&6&$2\dff+2\dfs+6\dft$ mod 12&$2(\dFf+\dFs+\dFt)$ mod 12&$\dff$, $\dFf$, $\dfs+\dFs$, $\dfs+\dFt$&$6(\dfs+\dft)$ mod 12\\
\hline
$61'$&$6_{4\pi/3}1'$&6&$2\dff+2\dfs+6\dft$ mod 12&$2\dFf+2\dFs+6\dFt$ mod 12&$\dff$, $\dfs$, $\dFf$, $\dFs$&$6(\dft-\dFt)$ mod 12\\
\hline
$61'$&$6_{5\pi/3}1'$&6&$2\dff+2\dfs+6\dft$ mod 12&$2(\dFf+\dFs+\dFt)$ mod 12&$\dff$, $\dfs$, $\dFf$, $\dFs$, $\dFt$&$6\dft$ mod 12\\
\hline
$6'$&$6'$&$3$&$\dff+3\dfs$ mod 6&$\dFf+3\dFs$ mod 6&$\dff-\dFf$&$3(\dfs-\dFs)$ mod 6\\
\hline
$6'$&$6_{\pi/3}'$&3&$\dff+3\dfs$ mod 6&$2\dFf$ mod 6&$\dff$, $\dFf$&$3\dfs$ mod 6\\
\hline
$6'$&$6_{2\pi/3}'$&3&$\dff+3\dfs$ mod 6&$2\dFf+3\dFs$ mod 6&$\dff$, $\dFf$&$3(\dfs-\dFs)$ mod 6\\
\hline
$6'$&$6_{\pi}'$&3&$\dff+3\dfs$ mod 6&$2\dFf$ mod 6&$\dff+\dFf$&$3(\dff+\dfs)$ mod 6\\
\hline
$6'$&$6_{4\pi/3}'$&3&$\dff+3\dfs$ mod 6&$2\dFf+3\dFs$ mod 6&$\dff$, $\dFf$&$3(\dfs-\dFs)$ mod 6\\
\hline
$6'$&$6_{5\pi/3}'$&3&$\dff+3\dfs$ mod 6&$2\dFf$ mod 6&$\dff$, $\dFf$&$3\dfs$ mod 6\\
\hline
$6mm$&$6mm$&6&$2(\dff+\dfs)$ mod 6&$2(\dFf+\dFs)$ mod 6&$\dff-\dFf$, $\dfs-\dFs$&\\
\hline
$6mm$&$6_{\pi/3}mm$&6&$2(\dff+\dfs)$ mod 6&$\dFf+2\dFs+2\dFt$ mod 6&$\dff$, $\dfs$, $\dFf$, $\dFs$, $\dFt$&\\
\hline
$6mm$&$6_{2\pi/3}mm$&6&$2(\dff+\dfs)$ mod 6&$2(\dFf+\dFs)$ mod 6&$\dff$, $\dfs$, $\dFf$, $\dFs$&\\
\hline
$6mm$&$6_{\pi}mm$&6&$2(\dff+\dfs)$ mod 6&$\dFf+2\dFs+2\dFt$ mod 6&$\dff$, $\dFf$, $\dFs+\dfs$, $\dFt+\dfs$&\\
\hline
$6mm$&$6_{4\pi/3}mm$&6&$2(\dff+\dfs)$ mod 6&$2(\dFf+\dFs)$ mod 6&$\dff$, $\dfs$, $\dFf$, $\dFs$&\\
\hline
$6mm$&$6_{5\pi/3}mm$&6&$2(\dff+\dfs)$ mod 6&$\dFf+2\dFs+2\dFt$ mod 6&$\dff$, $\dfs$, $\dFf$, $\dFs$, $\dFt$&\\
\hline
$6mm1'$&$6mm1'$&$6m'm'$&$2\dff+2\dfs+6\dft$ mod 12&$2\dFf+2\dFs+6\dFt$ mod 12&$\dff-\dFf$, $\dfs-\dFs$&$6(\dft-\dFt)$ mod 12\\
\hline
$6mm1'$&$6_{\pi/3}mm1'$&$6m'm'$&$2\dff+2\dfs+6\dft$ mod 12&$2(\dFf+\dFs+\dFt)$ mod 12&$\dff$, $\dfs$, $\dFf$, $\dFs$, $\dFt$&$6\dft$ mod 12\\
\hline
$6mm1'$&$6_{2\pi/3}mm1'$&$6m'm'$&$2\dff+2\dfs+6\dft$ mod 12&$2\dFf+2\dFs+6\dFt$ mod 12&$\dff$, $\dfs$, $\dFf$, $\dFs$&$6(\dft-\dFt)$ mod 12\\
\hline
$6mm1'$&$6_{\pi}mm1'$&$6m'm'$&$2\dff+2\dfs+6\dft$ mod 12&$2(\dFf+\dFs+\dFt)$ mod 12&$\dff$, $\dFf$, $\dfs+\dFs$, $\dfs+\dFt$&$6(\dfs+\dft)$ mod 12\\
\hline
$6mm1'$&$6_{4\pi/3}mm1'$&$6m'm'$&$2\dff+2\dfs+6\dft$ mod 12&$2\dFf+2\dFs+6\dFt$ mod 12&$\dff$, $\dfs$, $\dFf$, $\dFs$&$6(\dft-\dFt)$ mod 12\\
\hline
$6mm1'$&$6_{5\pi/3}mm1'$&$6m'm'$&$2\dff+2\dfs+6\dft$ mod 12&$2(\dFf+\dFs+\dFt)$ mod 12&$\dff$, $\dfs$, $\dFf$, $\dFs$, $\dFt$&$6\dft$ mod 12\\
\hline
$6'mm'$&$6'mm'$&$3m'$&$2\dff+3\dfs$ mod 6&$2\dFf+3\dFs$ mod 6&$\dff-\dFf$&$3(\dfs-\dFs)$ mod 6\\
\hline
$6'mm'$&$6_{\pi/3}'mm'$&$3m'$&$2\dff+3\dfs$ mod 6&$2\dFf$&$\dff$, $\dFf$&$3\dfs$ mod 6\\
\hline
$6'mm'$&$6_{2\pi/3}'mm'$&$3m'$&$2\dff+3\dfs$ mod 6&$2\dFf+3\dFs$ mod 6&$\dff$, $\dFf$&$3(\dfs-\dFs)$ mod 6\\
\hline
$6'mm'$&$6_{\pi}'mm'$&$3m'$&$2\dff+3\dfs$ mod 6&$2\dFf$&$\dff-\dFf$&$3\dfs$ mod 6\\
\hline
$6'mm'$&$6_{4\pi/3}'mm'$&$3m'$&$2\dff+3\dfs$ mod 6&$2\dFf+3\dFs$ mod 6&$\dff$, $\dFf$&$3(\dfs-\dFs)$ mod 6\\
\hline
$6'mm'$&$6_{5\pi/3}'mm'$&$3m'$&$2\dff+3\dfs$ mod 6&$2\dFf$&$\dff$, $\dFf$&$3\dfs$ mod 6\\
\hline
			$6m'm'$&$6m'm'$&$6m'm'$&$\dff+\dfs+\dft+$ &$\dFf+\dFs+\dFt+$ &$\delta^0_1-\delta^{\Phi}_1,\delta^0_2-\delta^{\Phi}_2,\delta^0_3-\delta^{\Phi}_3,$&\\
						&&&$\delta^0_4+\delta^0_5$ mod 6& $\delta^{\Phi/2}_4+\delta^{\Phi/2}_5$ mod 6 &$\delta^0_4-\delta^{\Phi}_4,\delta^0_5-\delta^{\Phi}_5$&\\
			\hline
		\end{longtable}
	\end{scriptsize}
\end{center}

\section{Flat-band Hofstadter Hamiltonian with PG $4_\pi1'$}
\label{app:flatbandham}

In this Appendix we build a flat band model with compactly supported Wannier states off the orbital sites. This is a called an obstructed atomic limit. We discuss the 2D topology of the model at filling $1/4$ and $1/2$, and subsequently use Hofstadter topology to prove when there is a protected gap closing at filling $1/4$ but not filling $1/2$. 

\subsection{Four-site Chain}
\label{app:square}

First we provide the details of the 4-site chain that models a single square plaquette with flux through it. We take nearest neighbor and next nearest neighbor hoppings so that at zero-flux, the Hamiltonian is
\bea
\label{eq:Hplaq}
H^{\phi=0}_{4} = \bpm
0 & t  & t' &   t  \\
 t   & 0 &  t  & t' \\
 t' &   t  & 0 &   t  \\
 t   & t' &   t & 0 \\
\epm 
\eea
where the basis is ordered as $s$-orbitals at $(0,0), \mbf{a}_1, \mbf{a}_1+\mbf{a}_2, \mbf{a}_2$ with $\mbf{a}_1 = \hat{x}, \mbf{a}_2 = \hat{y}$. We refer to the center of the unit cell $\mbf{x} = (\mbf{a}_1+\mbf{a}_2)/2$ as the 1b position. This model has $C_4$ (which rotates around $\mbf{x}$) and $\mathcal{T}$ symmetries which generate the point group $41'$. Their representation is
\bea
D[C_4] &= \bpm 0 & 1 & 0 & 0 \\ 0 & 0 & 1 & 0 \\ 0 & 0 & 0 & 1 \\ 1 & 0 & 0 & 0 \\ \epm \\
D[\mathcal{T}] &= K
\eea
where $K$ is complex conjugation. 

We now add flux in the symmetric gauge $\mathcal{A}(\mbf{r}) = \frac{\phi}{2} (\mbf{r}-\mbf{x}) =  \frac{\phi}{2} (-y + 1/2, x-1/2)$ with the Peierls substitution taking all Peierls paths to be straight. We see that the nearest neighbor hoppings become
\bea
t \to t \exp i \int_{\mbf{r}}^{C_4 \mbf{r}} \mathcal{A}(\mbf{r})  \cdot d\mbf{r} = t e^{i \phi/4}
\eea
and the diagonal hoppings become
\bea
t' \to t' \exp i \int_{\mbf{r}}^{C_4^2 \mbf{r}} \mathcal{A}(\mbf{r})  \cdot d\mbf{r} = t'
\eea
where the phase is zero because the path of the integral is straight through the center of the unit cell, which is perpendicular to $\mathcal{A}(\mbf{r})$. We thus obtain the Hofstadter Hamiltonian 
\bea
H^\phi_{4} = \bpm
0 & t e^{-i \phi/4} & t' &   t e^{i \phi/4} \\
 t e^{i \phi/4}  & 0 &  t e^{-i \phi/4}  & t' \\
 t' &   t e^{i \phi/4} & 0 &   t e^{-i \phi/4} \\
 t e^{-i \phi/4}     & t' &   t e^{i \phi/4}  & 0 \\
\epm 
\eea
which still possesses the $D[C_4]$ symmetry, but breaks $\mathcal{T}$ for $\phi \neq 0$. The spectrum is periodic under $\phi \to \phi + \Phi$ where $\Phi = 4\pi = 2 \times 2\pi$ since the smallest closed, due to the diagonal hopping, encloses half the unit cell. This follows from $H^{\phi+\Phi}_4 = U H^\phi_4 U^\dag$ where 
\bea
U \ket{\mbf{r}} = \exp \lp i \int_{\mbf{r}_0}^{\mbf{r}} \mathcal{A}^{\Phi} \cdot d\mbf{r} \rp \ket{\mbf{r}}, \quad \pmb{\nabla} \times \mathcal{A}^{\Phi} = \Phi 
\eea
and we pick $\mbf{r}_0 = (0,0)$ for simplicity. We compute
\bea
U = \bpm 1 & & & \\ & -1 & &  \\ & & 1 & \\ &&& -1 \epm \ .
\eea
We check explicitly that $D[C_4] U = e^{i \gamma} U D[C_4] $ with $\gamma = \pi$. This $\gamma$ can be computed by taking the $C_4$-symmetric closed loop around $\mbf{x}$ to be along the sides of the square, yielding $\gamma = \frac{1}{4} \Phi = \pi$. 

The spectrum of the Hamiltonian is exactly solvable in flux. We compute
\bea
E_n(\phi) &= t' + 2t \cos \frac{\phi}{4}, t' - 2t \cos \frac{\phi}{4}, - t' + 2 t \sin \frac{\phi}{4}, - t' - 2 t \sin \frac{\phi}{4}
\eea
which can be continuously labeled by their $C_4$ eigenvalues (since $C_4$ remains a symmetry at all flux) given in order by $1,-1,i,-i$ which are conventionally called $A, B, {}^1E, {}^2E$. At $\phi =0$, the $\mathcal{T}$ symmetry enhances the point group to $41'$ which enforces the ${}^1E{}^2E$ to be degenerate. At $\phi = \Phi/2 = 2\pi$ where the point group of $\mbf{x}$ is generated by $C_4$ and $U\mathcal{T}$ giving $4_\pi 1'$, we recall from the Main Text that the irreps are $AB$, ${}^1E$, ${}^2E$ which shows the $AB$ irreps must be degenerate. 

\begin{figure}[h]
 \centering
\includegraphics[width=8cm]{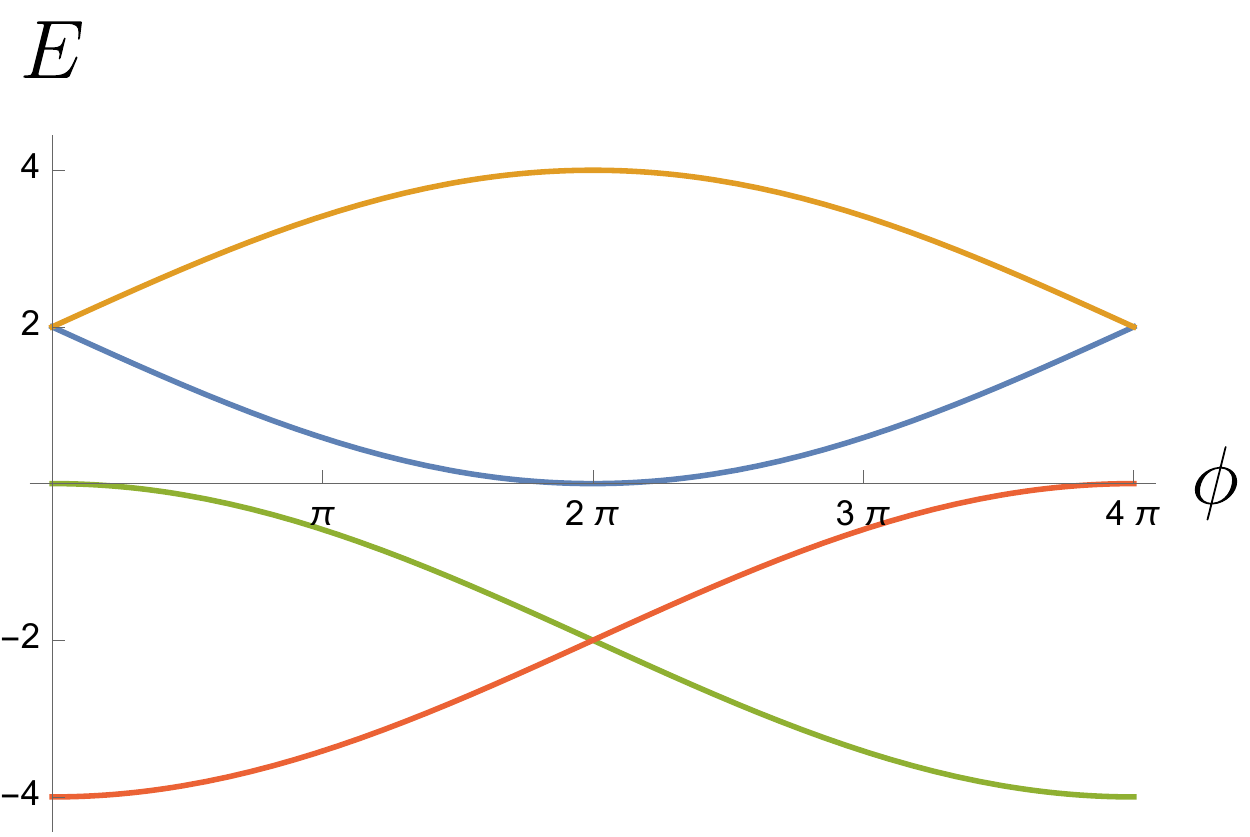} 
\caption{Spectrum of $H_4^\phi$ setting $t = -1, t' = -2$ where red is the $A$ irrep, green is $B$, blue is ${}^1E$, and yellow is ${}^2E$. We see explicitly that the spectrum is $\Phi= 4\pi$ periodic but there is irrep flow between $0$ and $4\pi$ which exchanges the $A,B$ (initially at different energies) and ${}^1E, {}^2E$ irreps (initially at the same energy). Because of the $\mathcal{T}$ symmetry, the gap flowing enforced by irrep flow is also enforced by the degeneracy of the $AB$ irrep of $4_\pi1'$. }
\label{fig_spectrum4}
\end{figure}

We see that with $C_4$ symmetry alone, the gap closing can be diagnosed from the irrep flow $A \to B, B \to A$ between $\phi =0$ and $\phi = \Phi$. Alternatively, with $C_4$ and $U\mathcal{T}$ forming the point group $4_{\pi}1'$ at $\phi = \Phi/2$, the gap closing between $A$ and $B$ is shown explicitly by the double irrep $AB$ of $4_{\pi}1'$. 

\subsection{Construction of the Compact Obstructed Atomic Limit}

\begin{table}[h!]
    \centering
\begin{tabular}{c|ccc}
$41'$ & 1 & $C_4$ & $C_2$  \\
\hline
$A$ & 1 & 1 & 1 \\
$B$ & 1 & $-1$ & 1 \\
$^1\!E^2\!E$ & 2 & $0$ & $-2$ \\
\end{tabular} \
\begin{tabular}{c|ccc}
\rule{0pt}{-2.5ex} $4_{\pi/2}1'$ & 1 & $C_4$ & $C_2$ \\
\hline
\rule{0pt}{2.5ex} \!\!$^1\!EA$ & 2 & $1-i$ & 0 \\
$^2\!EB$ & 2 & $-1+i$ & 0\\
\end{tabular} \
\begin{tabular}{c|ccc}
$4_\pi1'$ & 1 & $C_4$ & $C_2$  \\
\hline
$AB$ & 2 & 0 & 2 \\
$^1\!E$ & 1 & $-i$ & $-1$ \\
$^2\!E$ & 1 & $i$ & $-1$ \\
\end{tabular} 
\caption{We list the (partial) character tables for the irreps of $41'$ and two of its projective representations without SOC. (The irreps of $4_{3\pi/2}1'$ are the complex conjugates of $4_{\pi/2}1'$.) We name the irreps according to their $C_4$ eigenvalues where $A,B,^1\!E,^2\!E$ correspond to $+1,-1,-i,i$ respectively. We do not include the traces of the anti-unitary operators because they are not invariant under unitary transformations. 
\label{tab:4irrepsapp}
 } % title of Table

\end{table}

\begin{figure*}[h]
 \centering
\includegraphics[width=4cm]{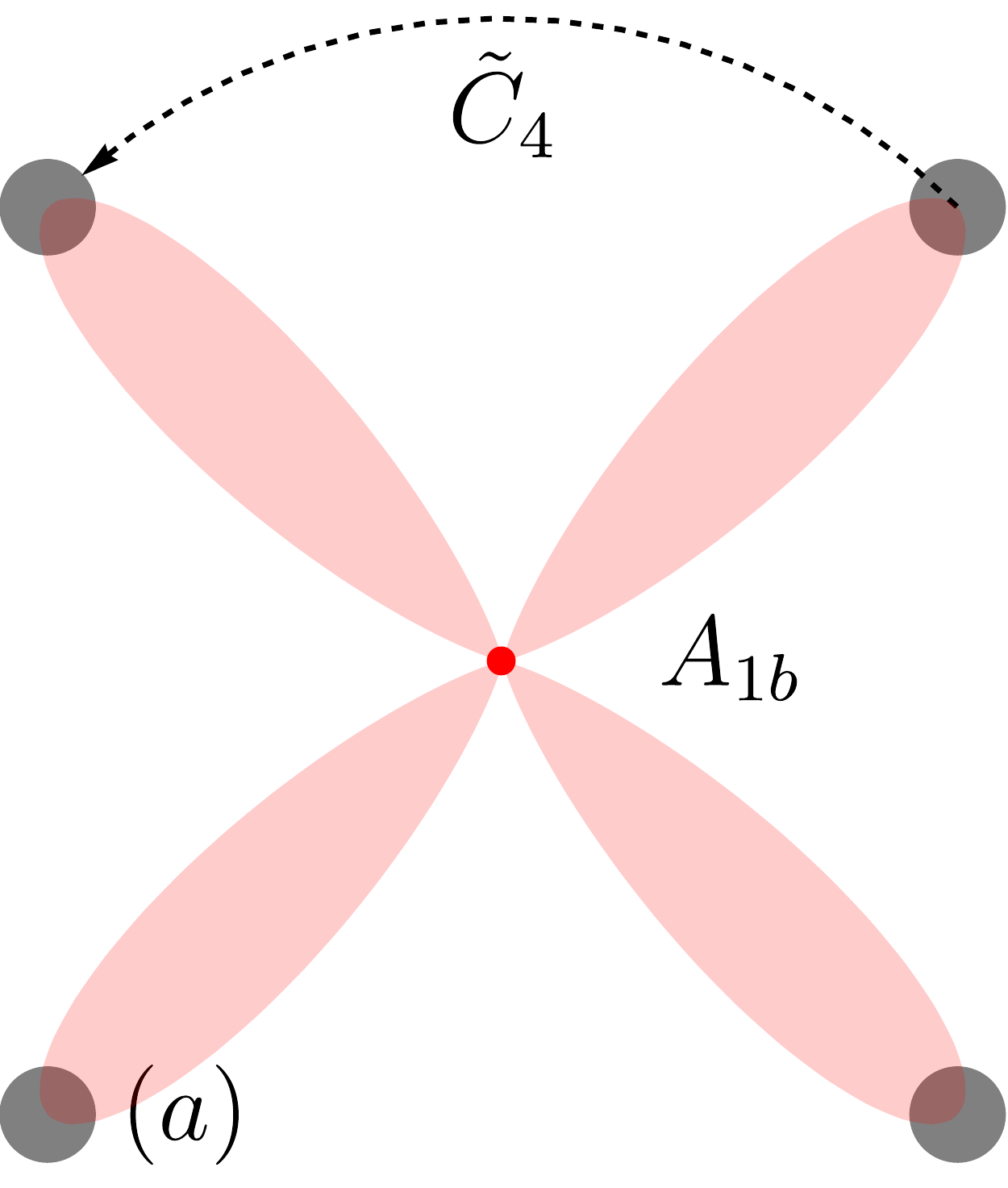}  \qquad
\includegraphics[width=4cm]{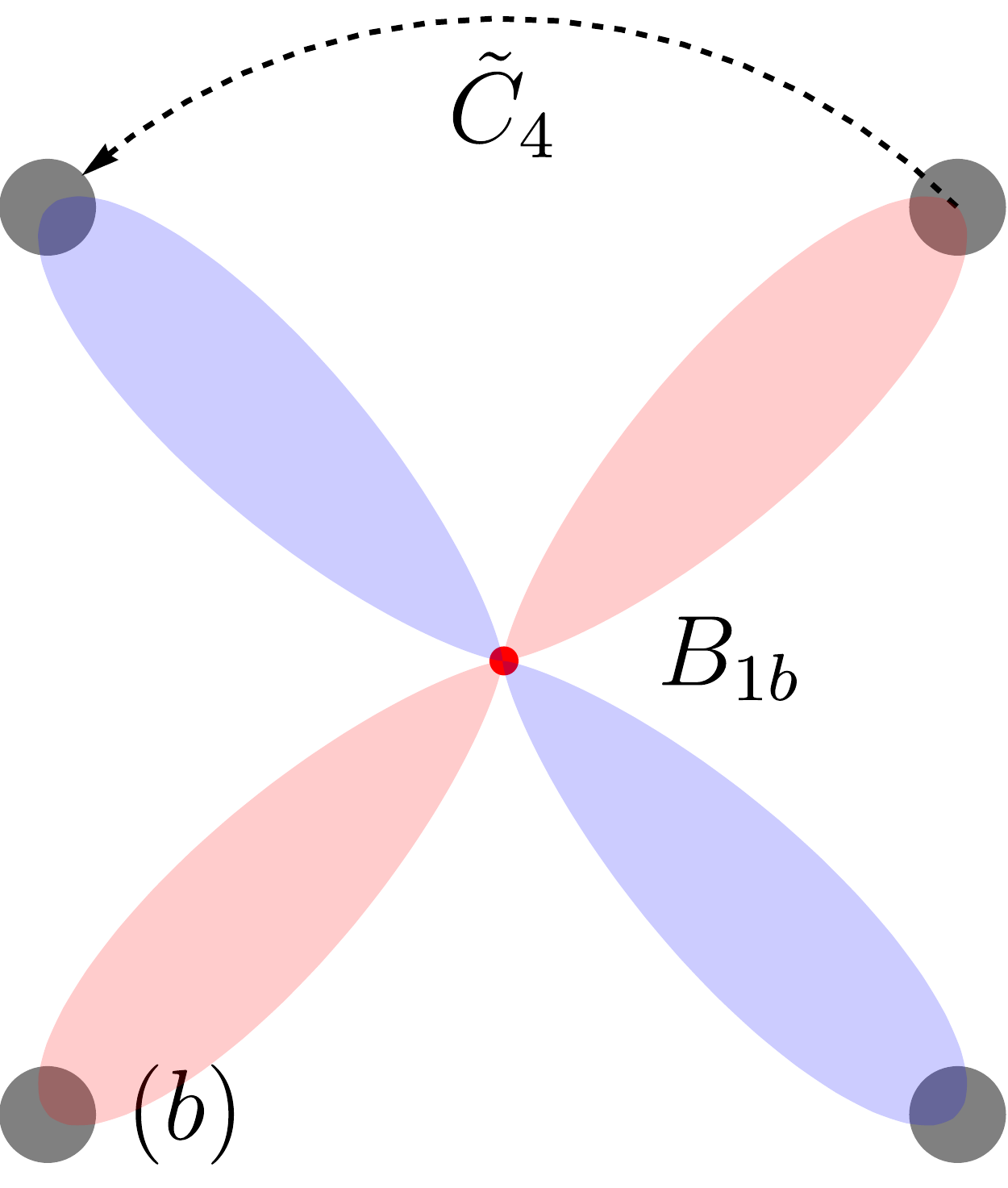} \qquad
\includegraphics[width=4.5cm]{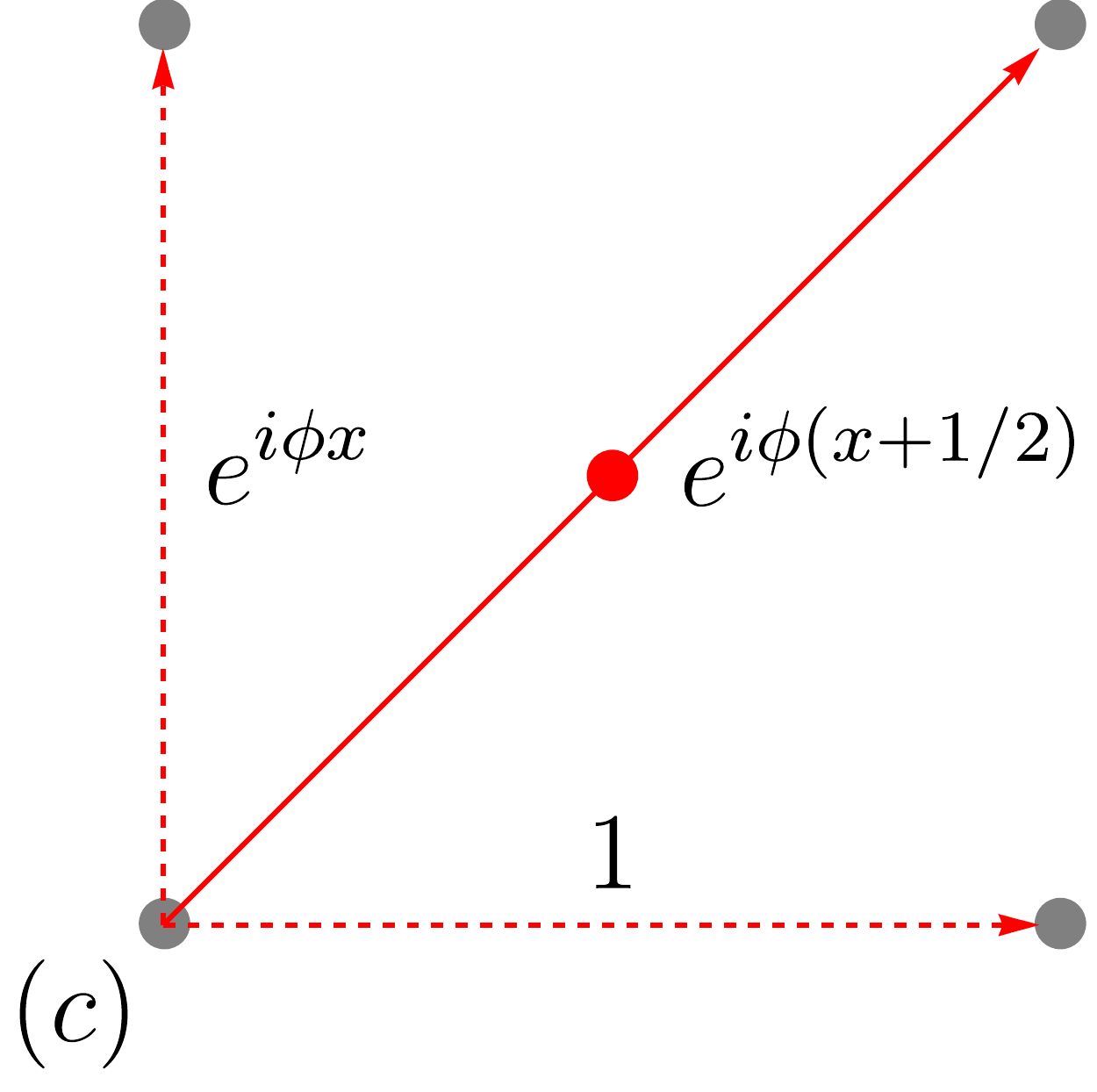} 
\caption{$(a)$ The $A_{1b}$ Wannier function is sketched. Its center is at the 1b position (red dot) and is supported by the orbitals the nearest four atomic sites (1a position, grey dot). $(b)$ The $B_{1b}$ Wannier function is sketched. Its center is at the 1b position (red dot) and is supported by the orbitals the nearest four atomic sites (1a position, grey dot). $(c)$ We show the Peierls paths chosen for the model. For simplicity, the Peierls paths are the same for every orbital. The hoppings are marked by their Peierls phases calculated in the Landau gauge with translation along $x$ broken. } 
\label{fig_PPoal}
\end{figure*}

We construct a flat band model which realizes an obstructed atomic limit (OAL) in space group $p41'$ which has Wyckoff positions 1a (PG $41'$), 1b (PG $41'$), and 2c (PG $41'$) whose irreps are listed in \Tab{tab:4irrepsapp}. We take the lattice vectors to be $\mbf{a}_1 = (1,0), \mbf{a}_2 = (0,1)$. We begin with $A,B,{}^1\!E{}^2\!E$ orbitals (which one can consider as $s$, $d$, $p_x \pm i p_y$ orbitals respectively) which we place at the 1a = $(0,0)$ positions. We define $\tilde{C}_4$ as a rotation around the 1b position so that
\bea
\tilde{C}_4 (0,0) = \mbf{a}_1, \quad \tilde{C}_4 \mbf{a}_1 = \mbf{a}_1 + \mbf{a}_2, \quad \tilde{C}_4 (\mbf{a}_1+\mbf{a}_2) = \mbf{a}_2, \quad \tilde{C}_4 \mbf{a}_2 = (0,0) \ .
\eea
 We will build our model by constructing Wannier functions $\ket{\mbf{R},\rho_{1b}}$ at the 1b = $(1/2,1/2)$ positions according to
\bea
\label{eq:4phi0wann}
\ket{0, A_{1b}} &= \frac{1}{4}\sum_{j =0}^4 \tilde{C}_4^j (\ket{0, A} + \ket{0, B} + \ket{0, {}^1\!E} + \ket{0, {}^2\!E}), \\
\ket{0, B_{1b}} &= \frac{1}{4}\sum_{j =0}^4 (-1)^j \tilde{C}_4^j (\ket{0, A} + \ket{0, B} + \ket{0, {}^1\!E} + \ket{0, {}^2\!E}), \\
\eea
which are supported on the sites $0,\mbf{a}_1, \mbf{a}_2, \mbf{a}_1+\mbf{a}_2$, and we define the Wannier states on other lattice sites through
\bea
\ket{\mbf{R}, \rho_{1b}} &= T_{\mbf{R}} \ket{0, \rho_{1b}}
\eea
and $T_\mbf{R}$ is the translation operator by $\mbf{R}$ unit cells, and $\ket{\mbf{R}, \rho_{1b}}$ denotes the Wannier state at the 1b position $\mbf{R} + (1/2,1/2)$ with irrep $\rho$ (see \Fig{fig_PPoal}a,b). Our focus in on the $A_{1b}$ and $B_{1b}$ irreps, but for completeness we construct the other Wannier states
\bea
\ket{\mbf{R}, {}^1E_{1b}} &= \frac{1}{4} T_{\mbf{R}} \sum_{j =0}^4 (-i)^j\tilde{C}_4^j (\ket{0, A} + \ket{0, B} + \ket{0, {}^1\!E} + \ket{0, {}^2\!E}), \\
\ket{\mbf{R}, {}^2E_{1b}} &= \frac{1}{4} T_{\mbf{R}} \sum_{j =0}^4 i^j \tilde{C}_4^j (\ket{0, A} + \ket{0, B} + \ket{0, {}^1\!E} + \ket{0, {}^2\!E}), \\
\eea
which complete the Hilbert space. The atomic orbitals of the model are denoted $\ket{\mbf{R}, \rho}$ which is the state at the \emph{1a} position carrying irrep $\rho$ in the $\mbf{R}$th unit cell. 

We also need to check that $\ket{\mbf{R}, \rho_{1b}}$ are orthogonalized Wannier functions, e.g. $\braket{\mbf{R}, A_{1b}|\mbf{R}', A_{1b}} = \delta_{\mbf{R},\mbf{R}'}, \braket{\mbf{R}, A_{1b}|\mbf{R}', B_{1b}}  = 0 $. We can do this directly in real space. Because $\ket{0,\rho_{1b}}$ only has nonzero overlaps with neighboring states, there are only finitely many cases to check. First we compute the overlap after translating by $\mbf{a}_1$:
\bea
\braket{0,A_{1b}|\mbf{a}_1, A_{1b}} &= \frac{1}{16} (\bra{\mbf{a}_1,A} - \bra{\mbf{a}_1,B} - i \bra{\mbf{a}_1,{}^1E} + i \bra{\mbf{a}_1,{}^2E})^*(\ket{\mbf{a}_1,A} + \ket{\mbf{a}_1,B} + \ket{\mbf{a}_1,{}^1E} +  \ket{\mbf{a}_1,{}^2E})  \\
& \qquad + \frac{1}{16} (\bra{\mbf{a}_1,A} + \bra{\mbf{a}_1,B} -  \bra{\mbf{a}_1,{}^1E} - \bra{\mbf{a}_1,{}^2E})^*(\ket{\mbf{a}_1,A} - \ket{\mbf{a}_1,B} + i \ket{\mbf{a}_1,{}^1E} - i \ket{\mbf{a}_1,{}^2E}) \\
&= 0 \\
\eea
with the two terms coming from overlaps on the $\mbf{a}_1$ and $\mbf{a}_1+\mbf{a}_2$ sites. The states $\ket{0,A_{1b}}$ and $\ket{\mbf{a}_1 + \mbf{a}_2,A_{1b}}$ overlap on a single corner, and we compute
\bea
\braket{0,A_{1b}|\mbf{a}_1, A_{1b}} &= \frac{1}{16} (\bra{\mbf{a}_1,A} + \bra{\mbf{a}_1,B} - \bra{\mbf{a}_1,{}^1E} - \bra{\mbf{a}_1,{}^2E})^*(\ket{\mbf{a}_1,A} + \ket{\mbf{a}_1,B} + \ket{\mbf{a}_1,{}^1E} +  \ket{\mbf{a}_1,{}^2E})  = 0 \ . \\
\eea
We have computed two of the nearest neighbor overlaps. The other six follow from $C_4$ symmetry. We also only checked the overlaps with the $\mbf{R} = 0$ Wannier states, but all others follow from translation symmetry. It is also direct to check orthogonality in momentum space. By Fourier transforming the Wannier states, we find the orthogonal eigenvectors
\bea
\label{eq:wandef}
u_{A}(\mbf{k}) &= \frac{1}{4} \bpm 1\\ 1\\ 1\\ 1 \epm + \frac{1}{4} e^{- i \mbf{k} \cdot \mbf{a}_1} \bpm 1\\ -1\\ -i\\ i \epm + \frac{1}{4} e^{- i \mbf{k} \cdot (\mbf{a}_1 + \mbf{a}_2)}\bpm 1\\ 1\\ -1\\ -1 \epm + \frac{1}{4} e^{- i \mbf{k} \cdot \mbf{a}_2} \bpm 1\\ -1\\ i\\ -i \epm \\
u_{B}(\mbf{k}) &= \frac{1}{4} \bpm 1\\ 1\\ 1\\ 1 \epm - \frac{1}{4} e^{- i \mbf{k} \cdot \mbf{a}_1} \bpm 1\\ -1\\ -i\\ i \epm +\frac{1}{4}  e^{- i \mbf{k} \cdot (\mbf{a}_1 + \mbf{a}_2)}\bpm 1\\ 1\\ -1\\ -1 \epm - \frac{1}{4} e^{- i \mbf{k} \cdot \mbf{a}_2} \bpm 1\\ -1\\ i\\ -i \epm \\
\eea
in the ordered basis $A,B,{}^1\!E,{}^2\!E$. We can also calculate the Wannier centers $\mbf{x}_c$ directly from the Berry connection (which is $\mbf{k}$ independent):
\bea
\mbf{x}_c = u_{\rho}^\dag(\mbf{k}) \, i \del_\mbf{k} u_{\rho}(\mbf{k}) = \lp \frac{1}{2},\frac{1}{2} \rp , \quad \rho = A,B \\
\eea
which is indeed at the 1b position. Because the Berry potential is independent of $\mbf{k}$, the Berry curvature vanishes identically. 

We now produce a flat band Hamiltonian with eigenstates $u_{A}(\mbf{k}), u_{B}(\mbf{k})$. This is easy to do in terms of the eigenstate projectors:
\bea
h(\mbf{k}) = M_A u_{A}(\mbf{k}) u^\dag_{A}(\mbf{k}) + M_B u_{B}(\mbf{k}) u^\dag_{B}(\mbf{k})
\eea
which puts the band of $A_{1b}$ Wannier states at energy $M_A$ and the $B_{1b}$ states at $M_B$. The ${}^1\!E{}^2\!E$ band is kept at zero energy without loss of generality. Written explicitly, we find
\bea
h(\mbf{k}) &= \frac{M_A+M_B}{4} \left(
\begin{array}{cccc}
 1 & 0 & 0 & 0 \\
 0 & 1 & 0 & 0 \\
 0 & 0 & 1 & 0 \\
 0 & 0 & 0 & 1 \\
\end{array}
\right) +  \frac{M_A+M_B}{16} e^{-i k_x - i k_y} \left(
\begin{array}{cccc}
 1 & 1 & 1 & 1 \\
 1 & 1 & 1 & 1 \\
 -1 & -1 & -1 & -1 \\
 -1 & -1 & -1 & -1 \\
\end{array}
\right)+  \frac{M_A+M_B}{16}  e^{-i k_x + ik_y} \left(
\begin{array}{cccc}
 1 & -1 & -i & i \\
 -1 & 1 & i & -i \\
 -i & i & -1 & 1 \\
 i & -i & 1 & -1 \\
\end{array}
\right)\\
 &\null   + 
\frac{M_A-M_B}{16}  e^{- i k_x} \left(
\begin{array}{cccc}
 2 & 0 & 1-i & 1+i \\
 0 & -2 & -1-i & -1+i \\
 -1-i & 1-i & 0 & -2 i \\
 -1+i & 1+i & 2 i & 0 \\
\end{array}
\right)+  \frac{M_A-M_B}{16}  e^{- i k_y} \
\left(
\begin{array}{cccc}
 2 & 0 & 1+i & 1-i \\
 0 & -2 & -1+i & -1-i \\
 -1+i & 1+i & 0 & 2 i \\
 -1-i & 1-i & -2 i & 0 \\
\end{array}
\right) + h.c. \\
&\equiv M + \frac{M}{4} e^{-i k_x - i k_y} T_{xy} +  \frac{M}{4} e^{-i k_x + i k_y} T_{x\bar{y}} + t e^{-i k_x} T_x + t e^{-i k_y} T_y
\eea
where we have defined $M = (M_A+ M_B)/4$ and $t =  (M_A- M_B)/16$, and identified the hoppings matrices $T_x, T_y, T_{xy},T_{x\bar{y}}$ with the corresponding block matrices in the line above. We set $M_A = -5, M_B = -2$. Filling one band, the $A_{1b}$ irrep in each unit cell is occupied and the RSIs at the 1b position with PG $41'$ are $\delta_1 = m(B) - m(A) = -1, \delta_2 = m(^1\!E^2\!E) - m(A) = -1$. This phase will have a protected bulk gap closing in flux when the Peierls paths yield a nontrivial Schur multiplier $\gamma_{1b} \neq 0$ (see Main Text). Filling two bands, the $A_{1b}$ and $B_{1b}$ orbitals are occupied and the RSIs at the $1b$ position are $\delta_1 = 0, \delta_2 = -1$. When $\gamma_{1b} = \pi$, no gap closing is enforced (see \Tab{tab:HofinvnoSOC} where $\delta_1^{\phi=0}$ is shown to be a SM invariant). We will confirm this by analytically solving the Hofstadter Hamiltonian in the next section. 

To broaden the bands of the model, one can add the nearest-neighbor term $\eps (\cos k_x + \cos k_y) \mathbb{1}_{4\times 4}$ which preserves all symmetries and gives each band a width of $4\eps$. 

%We briefly mention the topology of the \emph{un}occupied bands in this model. At quarter filling, the 3-band conduction manifold is the complement of the occupied $A_{1b}$ orbitals, giving the conduction band representation $A_{1a} \oplus B_{1a} \oplus ({}^1E{}^2E)_{1a} \ominus A_{1b}$, leading to the RSIs $\delta_{1b,1} = 1, \delta_{1b,2} =1$. However, these RSIs do not satisfy the fragile criteria in \Ref{song2019real} and hence the state is not fragile. Indeed, we observe that the RSIs  $\delta_{1b,1} = 1, \delta_{1b,2} =1$ can be obtained from the three-band atomic limit $B_{1b} \oplus ({}^1E{}^2E)_{1b}$. Thus 
%$A_{1a} \oplus B_{1a} \oplus ({}^1E{}^2E)_{1a} \ominus A_{1b}$ and $B_{1b} \oplus ({}^1E{}^2E)_{1b}$ are adiabatically connected. The process that connects them is $A_{1a} \oplus B_{1a} \oplus ({}^1E{}^2E)_{1a} \to A_{1b} \oplus B_{1b} \oplus ({}^1E{}^2E)_{1b}$ where all four states at 1a move off along diagonals and recombine at 1b. 

\subsection{Hofstadter Hamiltonian}

To form the Hofstadter Hamiltonian, we work in the Landau gauge $A = \phi(0,x), \phi = 2\pi \frac{p}{q} $ which preserves $k_y$ as a good quantum number but introduces an extended unit cell along $x$ indexed $\ell = 0,\dots, q-1$ (see \Ref{firstpaper} for extensive details). We take the Peierls paths between nearest neighbors to be straight lines along the bonds of the square lattice and next-nearest neighbors to be diagonal paths, as in \Fig{fig_PPoal}. Then the smallest Peierls path has area $1/2$, so the flux periodicity of the model is $\Phi = 2 \times 2\pi$. To calculate $\gamma_{1b}$, the Schur multiplier of the $\tilde{C}_4$ rotation, we pick a $\tilde{C}_4$-symmetric square path as the boundary of the unit cell, yielding $\gamma_{1b} = \Phi/4 = \pi$. 

The Hofstadter Hamiltonian in the Landau gauge reads
\bea
\mathcal{H}^{\phi = 2\pi p/q}_{\ell, \ell'} &= \delta_{\ell,\ell'} (M/2 + t e^{- i k_y + i \phi \ell } T_y) +  \delta_{\ell, \ell'+1} \lp t e^{-i k_x} T_x + \frac{M}{4} e^{-i k_x - i k_y + i \phi (\ell+ 1/2)} T_{xy} +  \frac{M}{4} e^{-i k_x + i k_y -  i \phi (\ell+ 1/2)} T_{x\bar{y}} \rp + h.c.
\eea
or to include broadening, take $T_x \to T_x + \frac{\eps}{2} \mathbb{1}_{4\times4},T_y \to T_y + \frac{\eps}{2} \mathbb{1}_{4\times4}$ to incorporate the additional nearest-neightbor hopping. $\mathcal{H}^{\phi}$ can be diagonalized to determine the Hofstadter spectrum. As shown in \Fig{fig_OALcross} of the Main Text, there is a gap closing between the $A_{1b}$ and $B_{1b}$ irreps which is protected by $\gamma_{1b} = \pi$. We also see that in the flat band limit at $\phi=0$, the Hofstadter spectrum has zero dispersion for all $\phi$. We now show this analytically using the Wannier basis. 

\begin{figure*}[h]
 \centering
\includegraphics[width=8cm]{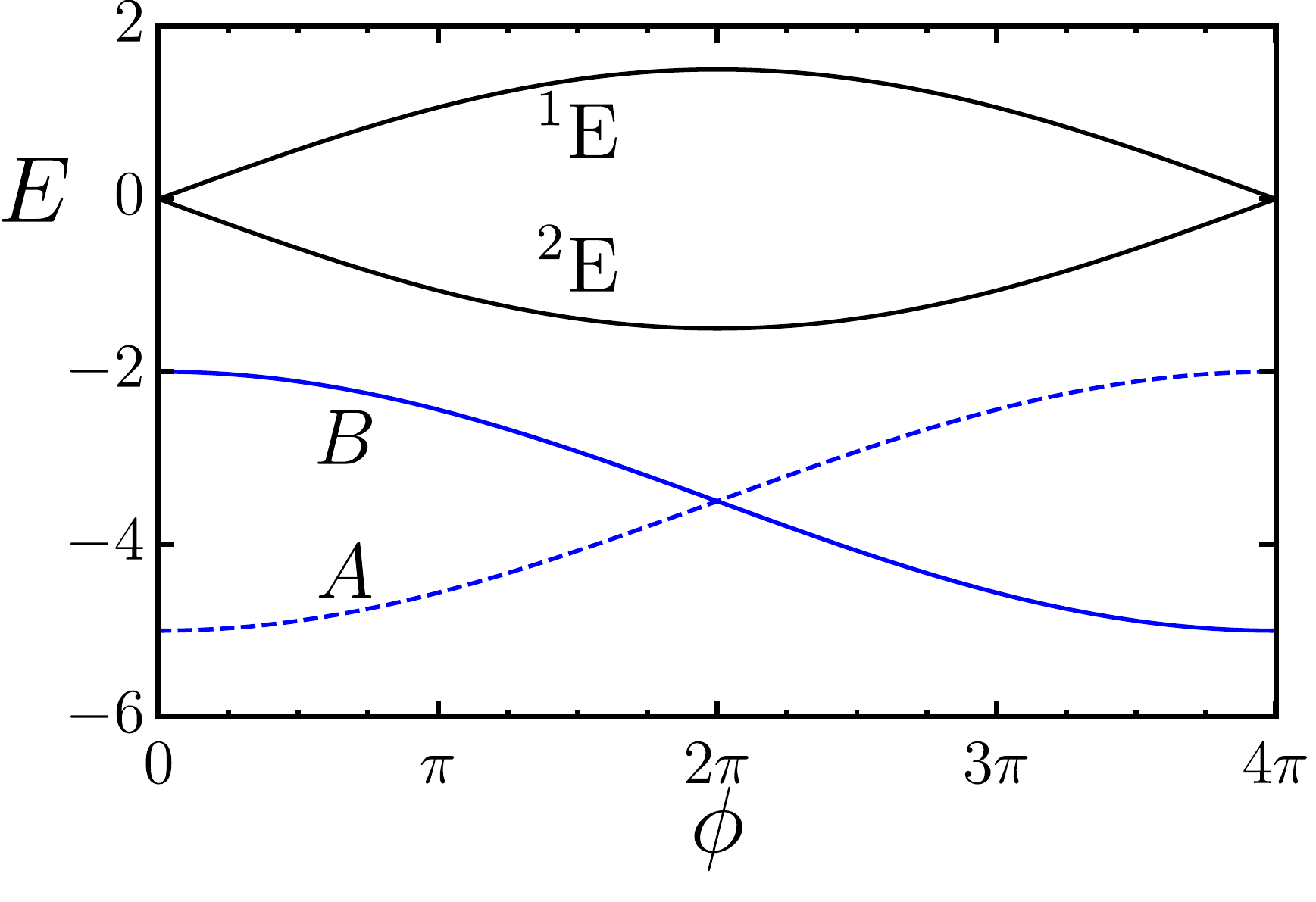}
\caption{We show the spectrum derived in \Eq{eq:OALspecHof} of the flat band OAL at all flux taking $M_A = -5, M_B = -2$. The band crossing at $\Phi/2 = 2\pi$ is protected by the $\gamma_{1b} = \pi$ Schur multiplier from the Peierls paths. } 
\label{fig_41AB}
\end{figure*}

At zero flux, the flat band model is decoupled in the Wannier basis:
\bea
H &=\sum_{\rho,\mbf{R}} E_\rho \ket{\mbf{R},\rho_{1b}} \bra{\mbf{R},\rho_{1b}} 
\eea
where $E_A = M_A, E_B = M_B, E_{{}^1E} = E_{{}^2E}  = 0$. We now want to introduce flux into the model keeping the Wannier state form. We use the fact that the Wannier functions are compactly supported so that
\bea
\ket{0,\rho} &= \sum_{\mbf{t},\al} \tilde{u}_\rho^\al(\mbf{t}) \ket{\mbf{t},\al} \\
\eea
where the sum over lattice vectors $\mbf{t} = 0, \mbf{a}_1, \mbf{a}_2, \mbf{a}_1 + \mbf{a}_2$ is finite and $\al$ sums over the 1a orbitals. From \Eq{eq:wandef}, the coefficients are
\bea
 \tilde{u}_A^\al(0) &= (1,1,1,1), \quad  \tilde{u}_A^\al(\mbf{a}_1) = (1,-1,-i,i), \quad  \tilde{u}_A^\al(\mbf{a}_1+\mbf{a}_2) = (1,1,-1,-1), \quad  \tilde{u}_A^\al(\mbf{a}_2) = (1,-1,i,-i) \\
  \tilde{u}_B^\al(0) &= (1,1,1,1), \quad  \tilde{u}_B^\al(\mbf{a}_1) = -(1,-1,-i,i), \quad  \tilde{u}_B^\al(\mbf{a}_1+\mbf{a}_2) = (1,1,-1,-1), \quad  \tilde{u}_B^\al(\mbf{a}_2) = -(1,-1,i,-i) \ . \\
\eea
Picking the Wannier states at a fixed $\mbf{R}$, the terms in the Hamiltonian $H = \sum_\mbf{R} H_\mbf{R}$ can be written as
\bea
H_{\mbf{R}} &= \sum_\rho E_\rho \ket{\mbf{R},\rho_{1b}} \bra{\mbf{R},\rho_{1b}}  &= \sum_\rho E_\rho \sum_{\mbf{t},\mbf{t}',\al,\be} \ket{\mbf{R}+\mbf{t},\al}  \tilde{u}_n^\al(\mbf{t})  \tilde{u}_n^\be(\mbf{t}')^* \bra{\mbf{R}+\mbf{t}',\be} 
\eea
in terms of the atomic orbitals. Under the Peierls substitution, we have
\bea
\ket{\mbf{R}+\mbf{t},\al} \bra{\mbf{R}+\mbf{t}',\be} \to \exp \lp i \int_{\mbf{R}+\mbf{t}'}^{\mbf{R}+\mbf{t}} \mbf{A} \cdot d\mbf{r}\rp \ket{\mbf{R}+\mbf{t},\al} \bra{\mbf{R}+\mbf{t}',\be} \equiv  \exp \lp i \varphi_{\mbf{t}, \mbf{t}'}(\mbf{R}) \rp \ket{\mbf{R}+\mbf{t},\al} \bra{\mbf{R}+\mbf{t}',\be} 
\eea
which defines the Hofstadter Hamiltonian
\bea
H^\phi &= \sum_\mbf{R} \sum_\rho E_\rho \sum_{\mbf{t},\mbf{t}',\al,\be}   \exp \lp i \varphi_{\mbf{t}, \mbf{t}'}(\mbf{R}) \rp  \ket{\mbf{R}+\mbf{t},\al}  \tilde{u}_n^\al(\mbf{t})  \tilde{u}_n^\be(\mbf{t}')^* \bra{\mbf{R}+\mbf{t}',\be} \ . 
\eea
To solve this Hamiltonian, we first study the $\mbf{R} = 0$ term without loss of generality, $H^\phi_{\mbf{R} = 0}$. $H^\phi_{\mbf{R} = 0}$ is a $16 \times 16$ matrix which acts on the $\ket{\mbf{t},\al}$ basis for $\mbf{t} = 0, \mbf{a}_1, \mbf{a}_2, \mbf{a}_1 + \mbf{a}_2, \al = A, B, {}^1E, {}^2E$. We can explicitly diagonalize $H^\phi_{\mbf{R}=0}$ at all flux. A direct calculation in the $\tilde{C}_4$-symmetric gauge $\mbf{A} = \frac{\phi}{2}(-(y-1/2), x-1/2)$ shows the eigenstates
\bea
\label{eq:OALspecHof}
H^\phi_{\mbf{R} = 0} \ket{0,A_{1b}} &= \frac{1}{2}\lp M_A + M_B + (M_A - M_B) \cos \frac{\phi}{4} \rp  \ket{0,A_{1b}}\\
H^\phi_{\mbf{R} = 0} \ket{0,B_{1b}} &= \frac{1}{2}\lp M_A + M_B - (M_A - M_B) \cos \frac{\phi}{4} \rp  \ket{0,B_{1b}}\\
H^\phi_{\mbf{R} = 0} \ket{0,{}^1E_{1b}} &= \frac{1}{2} (M_A - M_B) \sin \frac{\phi}{4}  \ket{0,{}^1E_{1b}} \\
H^\phi_{\mbf{R} = 0} \ket{0,{}^2E_{1b}} &= -\frac{1}{2} (M_A - M_B) \sin \frac{\phi}{4}  \ket{0,{}^2E_{1b}} \\
\eea
with alll other eigenvalues are zero. Thus $H^\phi_\mbf{R}$ annihilates all Wannier states other than $\ket{0,\rho_{1b}}$, and hence $[H^\phi_\mbf{R}, H^\phi_{\mbf{R}'}] = 0$ follows for all $\mbf{R}$ making use of the magnetic translation operators. This simple structure is what allows the Hofstadter Hamiltonian to have flat bands for all flux. The eigenstates of the full Hamiltonian can be found from this orthogonality. For example, 
\bea
H^\phi \ket{0,\rho_{1b}} = \sum_\mbf{R} H^\phi_\mbf{R}  \ket{0,\rho_{1b}}  = H^\phi_{\mbf{R}=0}  \ket{0,\rho_{1b}} = E_\rho(\phi) \ket{0,\rho_{1b}}\\
\eea
so we have found Wannier state eigenstates of the Hofstadter Hamiltonian. The full spectrum of $H$ is given by \Eq{eq:OALspecHof} and shown in \Fig{fig_41AB}.  Eigenstates at $\mbf{R}=0$ can be obtained by acting with the magnetic translation operators which commute with $H^\phi$. From \Fig{fig_41AB}, we observe, as claimed, that there is a gap closing at $\Phi/2 = 2\pi$ when 1 band is occupied at $\phi =0$, but no gap closing when 2 bands are filled. One can also see the irrep pumping that occurs over $\phi \in (0, \Phi)$ due to $\gamma_{1b} = \pi$, which exchanges the $A$ and $B$ irreps. 

\section{Peierls-indicated Hofstadter HOTI phase from the Quantum Spin Hall Model}
\label{app:qshexample}

To illustrate a an example of the Peierls-indicated Hofstadter HOTI phase protected by $2'_\pi$ at $\phi = \Phi/2$, we consider a variant of the quantum spin Hall model \cite{2006Sci...314.1757B} discussed in \Ref{firstpaper}. Explicitly, the model we consider is
\bea
\label{eq:qshc2zt}
H'''_{QSH}(k_x, k_y) &= (M- \cos k_x  - \cos k_y )I\otimes \tau_3 + \sin k_x \, \sigma_3 \otimes \tau_1 +  \sin k_y \, I \otimes \tau_2  \\
&\qquad + \eps_4 I  \otimes \tau_2 + \eps_5 (\sigma_1 + \sigma_2) \otimes I + \eps_6 (\sigma_1 \otimes \tau_2 + \sigma_2 \otimes \tau_2 + \sigma_1 \otimes \tau_3)
\eea
with the parameters $\eps_4 = .1, \eps_5 = .11, \eps_6 = .05, M = 1.6$. The terms on the second line of \Eq{eq:qshc2zt} break all symmetries of the model except $C_2\mathcal{T}$, and put the model in a fragile topological phase protected by $C_2\mathcal{T}$ \cite{firstpaper}. It can be characterized by pair winding in the Wilson loop, an Euler number of $+1$, a nonzero second Stiefel-Whitney index, or a determinant $-1$ in the nested Wilson loop \cite{PhysRevLett.123.036401,2018arXiv180409719B,2019PhRvX...9b1013A,2018arXiv181002373W}. 

\begin{figure*}
 \centering
\includegraphics[width=8cm]{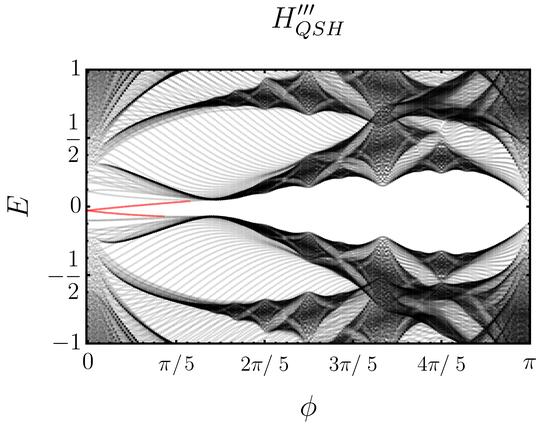} 
\caption{We show the Hofstadter Butterfly for \Eq{eq:qshc2zt} computed on a $25 \times 25$ unit cell lattice with open boundary conditions along both directions. The model realizes a Peierls-indicated Hofstadter HOTI phase and exhibits corner modes (red) that are pumped into the bulk (grey) as $\phi$ is increased.} 
\label{fig:QSHhoti}
\end{figure*}

The hoppings of $H'''_{QSH}$ are all nearest-neighbor on the square lattice. If we choose straight-line Peierls paths, then $\Phi = 2\pi$. As discussed in the main text, at the $C_2\mathcal{T}$-symmetric 1d position $\mbf{x} = (1/2,1/2)$, there is a nonzero Schur multiplier $\gamma_{\mbf{x}} = \frac{1}{2} \Phi = \pi$ which leads to $(UC_2\mathcal{T})^2 = -1$ at $\phi = \Phi/2 = \pi$ flux where $C_2$ is defined as a rotation about the $\mbf{x} = (1/2,1/2)$ position. Now we use the fact that, for the parameters in \Eq{eq:qshc2zt}, the fragile topological ground state is characterized by a quantized nested Wilson loop of $-1$, which indicates an odd numbers of states at the $\mbf{x}$ = 1d Wyckoff position \cite{2018arXiv181002373W}. Thus, because the $C_2\mathcal{T}$-protected RSI of $G^{\phi =0}_{1d} = 2'$ is $\delta^{\phi =0}_{1} = m(A) \mod 2$, we find $\delta^{\phi =0}_{1} = 1$ at the 1d position. Consulting \Tab{tab:HofinvnoSOC}, we see that there is a Peierls indicated HOTI phase for a pumping process between $G^{\phi =0} = 2'$ and $G^{\phi =\Phi/2} = 2_\pi'$. The Hofstadter HOTI invariant is $\delta_{1d}^{HOTI} = \delta^{\phi =0}_1 = 1 \mod 2$ which is nontrivial. In \Fig{fig:QSHhoti}, we show the Hofstadter spectrum computed on open boundary conditions (see \Ref{firstpaper} for an explicit expression for the Hofstadter Hamiltonian) which illustrates how the corner state protected by $N^{\phi =0}_{1d} = 1 \mod 2$ are pumped off the 1d position in flux because $N^{\phi =\Phi/2 = \pi}_{1d} = 0 \mod 2$. Lastly, we remark that \Ref{firstpaper} demonstrated this was a Hofstadter HOTI phase by computing the Wilson loop at $\pi$ flux. However, we were able to predict the Hofstadter HOTI phase from only the zero-flux topology because the phase is Peierls indicated by the projective symmetry group $2_\pi'$ at the 1d position in $\pi$ flux. 

\end{document}